\documentclass[prd,aps,twocolumn,a4paper,showkeys,nofootinbib,showpacs]{revtex4-2}

\usepackage[colorlinks=true,linkcolor=blue,linktocpage=true,anchorcolor=black,citecolor=blue,filecolor=black,menucolor=black,urlcolor=blue]{hyperref}

\usepackage{graphicx,psfrag}
\usepackage{mathrsfs}
\usepackage{amsmath,amsfonts,amssymb}
\usepackage{comment}
\usepackage{ulem}
\usepackage[top=2cm,bottom=2cm,left=1.4cm,right=1.4cm]{geometry}

\usepackage[T1]{fontenc}

\newcommand{\be}{\begin{equation}}
\newcommand{\ee}{\end{equation}}
\newcommand{\bea}{\begin{eqnarray}}
\newcommand{\eea}{\end{eqnarray}}
\newcommand{\bel}{\begin{align}}
\newcommand{\eel}{\end{align}}

\def\GMc2{G M_{\odot} c^{-2}}

\def\lm{{\ell m}}

\def\lm{{\ell m}}

\def\de{\partial}
\def\lm{{\ell m}}

\def\TEOBd{{\texttt{TEOBResumS-Dal\'i}}}

\def\w{{\hat{\omega}}}

\def\hl{{\hat{\ell}}}
\def\hhl{{\hat{\hat{\ell}}}}
\def\hhk{{\hat{\hat{k}}}}

\DeclareSymbolFontAlphabet{\mathrsfs}{rsfs}
\DeclareMathAlphabet{\mathcal}{OMS}{cmsy}{m}{n}
\DeclareSymbolFontAlphabet{\mathrsfs}{rsfs}
\DeclareMathAlphabet\mathbfcal{OMS}{cmsy}{b}{n}

\usepackage{color}
\definecolor{cyan}{rgb}{0,0.9,0.9}
\definecolor{orange}{rgb}{0.9,0.5,0}
\definecolor{magenta}{rgb}{1,0,1}
\definecolor{purple}{rgb}{0.8,0.4,0.8}
\definecolor{gray}{rgb}{0.8242,0.8242,0.8242}
\definecolor{dodgerblue}{rgb}{0.12, 0.56, 1.0}

\begin{document}
        
\title{From the confluent Heun equation to a new factorized and resummed 
gravitational waveform for circularized, nonspinning, compact binaries}
\author{Andrea \surname{Cipriani}${}^{1,2}$}
\author{Alessandro \surname{Nagar}${}^{3,4}$}
\author{Francesco \surname{Fucito}${}^{2}$}
\author{Josè Francisco \surname{Morales}${}^{2}$}

\affiliation{${}^1$Dipartimento di Fisica, Università di Roma Tor Vergata, Via della Ricerca Scientifica 1, 00133 Roma, Italy}
\affiliation{${}^2$INFN Sezione di Roma Tor Vergata, Via della Ricerca Scientifica 1, 00133 Roma, Italy}
\affiliation{${}^3$INFN Sezione di Torino, Via P. Giuria 1, 10125 Torino, Italy}
\affiliation{${}^4$Institut des Hautes Etudes Scientifiques, 91440 Bures-sur-Yvette, France} 

\begin{abstract}
We introduce a new factorized and resummed waveform for circularized, nonspinning, compact binaries that leverages on the solution
of the Teukolsky equation once mapped into a confluent Heun equation. The structure of the solution allows one to identify
new resummed factors that completely absorb all test-mass logarithms and transcendental numbers via 
exponentials and $\Gamma$-functions at any post-Newtonian (PN) order. The corresponding residual relativistic and phase corrections 
are thus polynomial with rational coefficients, that are in fact PN-truncated hypergeometric functions. 
Our approach complements the recent proposal of Ivanov et al.~[Phys.~Rev.~Lett.~135~(2025)~14, 141401], 
notably recovering the corresponding renormalization group scaling of multipole moments from first principles
and fixing the scaling constant. In the test mass limit, our approach (pushed up to 10PN) yields waveforms and
fluxes that are globally more accurate than those obtained using the standard factorized approach of Damour 
et al.~[Phys.~Rev.~D~79 (2009), 064004]. The method generalizes straightforwardly to comparable mass binaries 
implementing the new concept of universal anomalous dimension of multipole moments and might be eventually 
useful to improve current state of the art effective-one-body waveform models for coalescing binaries.
\end{abstract}

\maketitle

\section{Introduction}

The GWTC-4 catalog paper~\cite{LIGOScientific:2025slb}, together with the special 
events papers~\cite{LIGOScientific:2025rsn,LIGOScientific:2025rid,LIGOScientific:2025brd}, collects
all the recently analyzed gravitational wave signals emitted from coalescing compact binary (CBC) 
sources (composed by either black holes or neutron stars) and mirrors our continuously growing 
knowledge of the gravitational wave universe. The analysis of CBC gravitational data, and in particular 
the measurement of the binary intrinsic parameters, relies on gravitational waveform models. 
The effective-one-body (EOB) approach to the general relativistic two-body 
dynamics~\cite{Buonanno:1998gg,Buonanno:2000ef,Damour:2000we,Buonanno:2005xu,
Damour:2009wj,Damour:2016gwp,Damour:2017zjx,Damour:2019lcq,Damour:2025uka}, once informed
by Numerical Relativity (NR) simulations, is currently the only existing analytical method that is 
flexible enough and physically complete so as to allow one to accurately compute CBC waveforms 
for any orbital configuration~\cite{Gamba:2024cvy,Gamba:2025qfg,Chiaramello:2025bhi,Chandra:2025jfc}.
This in particular, The {\tt TEOBResumS} waveform model can deal with both eccentricity and spin-precession at 
the same time~\cite{Gamba:2024cvy}, dynamical capture and scattering~\cite{Nagar:2020xsk},
and tidal interactions~\cite{Gonzalez:2025xba}, although comparisons with NR simulations indicate that the model accuracy 
should be improved in certain corners of the parameter space~\cite{Albanesi:2025txj}.
The EOB approach relies on three building blocks: (i) a Hamiltonian that accounts for the conservative
part of the dynamics; (ii) a radiation reaction force, that accounts for dissipative effect related to the backreaction on the 
system of the emission of gravitational waves and (iii) a waveform computed on the aforementioned dynamics.
Elements (ii) and (iii) have to be consistent, since radiation reaction is build starting from the waveform.
The waveform and thus the radiation reaction relies on a certain factorization and resummation procedure that
was introduced in the nonspinning case by Damour and Nagar~\cite{Damour:2007xr} and 
Damour, Iyer, Nagar~\cite{Damour:2008gu} (DIN hereafter) and then generalized to the case of spin-aligned 
binaries by Pan et al.~\cite{Pan:2010hz}. The crucial new element of this factorization strategy, 
introduced by DIN, was the {\it tail factor}, a closed-form expression that resums all leading order logs
that appear in the PN-expanded waveform phase. This procedure was highly successful  and is pivotal for the 
accuracy of EOB-based waveform models. Although various attempts were pursued to further improve 
the accuracy of the waveform, that mostly focused on additional Pad\'e-type resummations of the residual 
amplitudes~\cite{Nagar:2016ayt,Messina:2018ghh,Nagar:2024dzj,Nagar:2024oyk}, 
no important conceptual improvement occurred since.

Recently, two different investigations appeared that suggest that times might be mature for a new leap forward.
On one hand, Ref.~\cite{Fucito:2023afe} revisited the calculation of the waveform emitted by a test mass
on circular orbits around a Schwarzschild and Kerr black holes using some tools mediated from the CFT/gravity 
correspondence.
In particular, the fact that the Teukolsky equation can be mapped into a confluent Heun equation~\cite{Aminov:2020yma} 
allows one to resum all logs and transcendental numbers into exponential and expressions involving $\Gamma$-functions. 
Despite the idea of having this information encapsulated in closed form is already expressed in Refs.~\cite{Fucito:2023afe,Cipriani:2025ikx}, 
no attempt to compute a new waveform factorization was pursued there. 
By contrast, Ivanov et al.~\cite{Ivanov:2025ozg} (ILPZ hereafter) used an EFT approach to clarify
the role of the renormalized angular momentum of black hole perturbation theory and connect it to
the anomalous dimension of multipole moments. This yielded the introduction of the concept of universal
anomalous dimension of the gravitational multipole moments of a gravitating system in general relativity.
This new concept prompted ILPZ to propose a new factorization of the PN waveform that absorbs 
{\it all} universal logs. Here we combine the results of Refs.~\cite{Fucito:2023afe,Cipriani:2025ikx} 
and ILPZ and propose a new factorized and resummed multipolar waveform that:(i) uses the approach 
of~\cite{Fucito:2023afe,Cipriani:2025ikx} in the test-mass to define closed-form 
analytical expressions (based on exponentials and combinations of $\Gamma$-functions) that incorporate 
all the universal logs and transcendental numbers usually present in PN calculations. 
This allows us to propose a new, factorized and resummed, multipolar waveform where, at given PN order, 
the nonresummed residual information in amplitude and phase in the test-mass limit presents only rational
coefficients;(ii) builds upon the concept of universal anomalous dimension of multipole moments introduced by 
ILPZ to generalize this approach to the comparable mass case, notably factorizing the 4.5PN-accurate 
waveform of Ref.~\cite{Blanchet:2023bwj,Blanchet:2023sbv}. 

The paper is organized as follows. In Sec.~\ref{sec:solution} we report the basics
of black-hole perturbation theory via the Teukolsky formalism. In particular, we present in detail 
the techniques used to analytically solve the Teukolsky equation once its homogeneous form is mapped into a confluent 
Heun equation. This method gives us access to closed-form, non-perturbative analytical expressions 
for the logs (tails) and transcendental numbers that are exploited in Sec.~\ref{sec:circ} once specified to the case of 
circular orbits to introduce the factorized and resummed waveform in the test-mass limit. This
approach is then generalized to the case of comparable mass binaries in Sec.~\ref{sec:comparable_mass}.
In Sec.~\ref{sec:ivanov_et_al} we build on the knowledge gained so far to propose an improved version
of the ILPZ factorization. The quality of the various resummation schemes is evalued in Sec.~\ref{sec:validation},
notably comparing with waveform and fluxes obtained numerically in the test-mass limit. 
Our concluding remarks are collected in Sec.~\ref{sec:end}.
The paper is then completed by various technical Appendixes. In Appendix~\ref{app:PN_exp_hhat} we report the
PN-expanded multipolar waveform in radiative coordinates to the current 3PN accuracy. In Appendix~\ref{sec:AppTails}
we report some useful formulas needed to efficiently expand expressions at high PN order. Appendix~\ref{app:a_expressions}
reports the full expressions, at 10PN accuracy, of the renormalized angular momentum both in the test-mass and
in the comparable mass case, once using the universal anomalous dimension of multipole moments.
In Appendix~\ref{app:rholm_flm_PN},~\ref{app:residual_phases} and~\ref{sec:DIN_ILPZ} we explicitly
list the residual amplitude and phase corrections for different waveform factorizations discussed in
the text.
If not otherwise stated, we used units with $G=c=1$.

\section{Multipolar waveform for a test-particle orbiting around a Schwarzschild black hole}
\label{sec:solution}

\subsection{Teukolsky equation and basic formulas}
\label{sec:teuk}
Let us review the basics of the Teukolsky formalism to describe the gravitational
perturbation of a Schwarzschild black hole\footnote{Another application for the 
determination of the radiated energy from a radially infalling particle released from 
rest in a Schwarzschild spacetime is studied in \cite{Bini:2026cpw}.}. Since our interest here 
is computing the gravitational waveform at infinity, we will work with the Weyl scalar $\Psi_4$ that statisfies the
Teukolsky master equation~\cite{Teukolsky:1973ha} within the Newman-Penrose formalism
using the Kinnersley tetrad~\cite{Fucito:2024wlg}.
This implies that the relation at infinity between the aforementioned Weyl scalar and the strain is
\be 
\label{eq:Psi4_vs_h}
\Psi_4 = -\frac{1}{2} (\ddot{h}_+ - i\ddot{h}_\times) \equiv -\frac{1}{2} \ddot{h} \ ,
\ee
where the dot denotes $d/d t$. 
The Weyl scalar is decomposed in $s=-2$ spin-weighted spherical harmonics
${}_{-2}Y_\lm(\Theta,\Phi)$ as
\be 
\label{eq:Psi4_dec}
\Psi_4 = \frac{1}{R^4} \int \frac{d \omega}{2 \pi} \sum_{\ell, m} \, e^{- i \omega T} \, R_{\lm \omega} (R) \, {}_{-2}Y_\lm(\Theta,\Phi) \ ,
\ee
where $\omega$ is the waveform frequency and $(T,R,\Theta,\Phi)$ are the coordinates 
of the observer at infinity. The spherical harmonics are defined as
\begin{align}
& {}_{s}Y_\lm (\theta, \phi) = e^{i m \phi} \sin^{2 \ell} \left(\frac{\theta}{2}\right) \sqrt{\frac{(2 \ell +1)(\ell-m)! (\ell+m)!}{4 \pi (\ell-s)! (\ell+s)!}} \notag \\
& \sum_{r=0}^{\ell-s} (-)^{\ell+m+r-s} \begin{pmatrix}
\ell -s \\
r
\end{pmatrix} \begin{pmatrix}
\ell + s \\
r+s-m
\end{pmatrix} \cot^{2 r+s-m} \left(\frac{\theta}{2}\right) \ .
\end{align}
The function $R_{\lm \omega}(r)$ entering Eq.~\eqref{eq:Psi4_dec} is here the solution of the 
$s=-2$ version of the Teukolsky equation specialized to the case of a nonspinning black hole 
with mass $M$ in the presence of an external source
\begin{align}
\label{eq:teuk}
& \Delta(r)^2 \dfrac{d}{dr}\left[ \dfrac{1}{\Delta(r)} \dfrac{ d R_{\lm \omega}}{dr}\right]  +\left[\frac{ \omega^2 r^4 +4 i (r-M) \omega r^2 }{\Delta(r)} + \right. \nonumber\\
& \quad\quad\quad\quad \left. -8 i \omega r  - (\ell+2)(\ell-1)  \right]R_{\lm \omega}(r)  =T_{\lm \omega}(r)  \ ,
\end{align}
with $\Delta(r)\equiv r(r-2 M)$.  The solution of this equation is formally constructed following 
three steps: (i) solving the homogeneous equation, Eq.~\eqref{eq:teuk} with $T_{\lm \omega} = 0$, 
with given boundary conditions; (ii) defining the Green function; (iii) computing the complete solution 
using the Green function method. The homogeneous equation has two solutions\footnote{Please note that
from time to time we will drop the indices $(\ell,m)$ to lighten the notation.}, $R_{\rm in}(r)$
and $R_{\rm out}(r)$, obtained imposing ingoing boundary conditions at the horizon and outgoing 
boundary condition at infinity, respectively. From them, one obtains the Wronskian 
(divided by $\Delta(r)$ in order to construct a constant in $r$)
\be
\label{ww}
W=\Delta(r)^{-1}\left\{R_{\text{in}}  \dfrac{d R_{\text{out}}}{dr}- \dfrac{dR_{\text{in}}}{dr}  R_{\text{out}} \right\} \ ,
\ee
and from this the Green function reads
\be
   G(r,r_1)={\dfrac{1}{W}} \left\{
\begin{array}{ccc}
R_{\text{in}} (r_1)   R_{\text{out}} (r)  & ~~~~~~~&  r_1< r \ , \\
R_{\text{in}} (r)   R_{\text{out}} (r_1)  & ~~~~~~~ &  r< r_1 \ , \\
\end{array}
\right.\label{green}
\ee
The solution of the inhomogeneous equation Eq.~\eqref{eq:teuk} reads then
\be \label{eq:inhomogeneous_sol}
R_{\lm \omega} (r) = \int_{2 M}^\infty G(r, r') \frac{T_{\lm \omega}(r')}{\Delta(r')^2} d r' \ .
\ee
Evaluating this function in $r=R \to \infty$, from the asymptotic behaviors of the two homogeneous
solutions in~\eqref{bcpsi}, we obtain the following form of $\Psi_4$ from 
Eq.~\eqref{eq:Psi4_dec} \footnote{Notice that this expression coincides with (6) of~\cite{Pan:2010hz},
once we specialize on circular orbit.}
\be
\Psi_4 \underset{R \to \infty}{\approx} \frac{1}{R} \int \frac{d \omega}{2 \pi} \sum_{\ell,m} e^{-i \omega (T-R_*)} \, Z_{\lm \omega} \,{}_{-2}Y_\lm(\Theta,\Phi)
\ee
where $r_* = r + 2 M \log \left(\frac{r}{2M}-1 \right)$ and
\be
Z_{\lm \omega} \equiv \int_{2M}^\infty \frac{\mathfrak{R}_{\rm{in}}(r') \, T_{\lm \omega}(r')}{\Delta(r')^2} \, d r' \ .
\ee
From \eqref{green}, we realize that the ingoing solution $\mathfrak{R}_{\rm{in}}(r)$ is obtained 
dividing the previous one $R_{\rm{in}}(r)$ by a factor that enters the wronskian $W$. This will
be explicit in Section~\ref{sec:hom}, Eqs.~\eqref{eq:expression_wronskian}-\eqref{eq:def_frak_Rin}.
At this point, from Eq.~\eqref{eq:Psi4_vs_h}, we can obtain the form of the waveform at infinity
\be
h(T, R, \Theta, \Phi) \underset{R \to \infty}{\approx} \frac{1}{R} \int \frac{d \omega}{2 \pi} \sum_{\ell, m} e^{-i \omega (T-R_*)} h_{\lm \omega} {}_{-2}Y_\lm(\Theta,\Phi)
\ee
with the Fourier components of the strain $h_{\lm \omega}$ related to $Z_{\lm \omega}$ as
\be
h_{\ell m\omega} \equiv \frac{2}{\omega^2} Z_{\ell m\omega} = \frac{2}{\omega^2} \int_{2 M}^\infty \frac{\mathfrak{R}_{\text{in}}(r') T_{\lm \omega}(r')}{\Delta(r')^2} d r' \ .
\ee


\subsection{Recasting the homogeneous Teukolsky equation as a confluent Heun equation and its solution}
\label{sec:hom}
Let us thus start by solving the homogeneous Teukolsky equation.
We do so following the procedure of Ref.~\cite{Cipriani:2025ikx} 
that we report here in detail for completeness.
The homogeneous Teukolsky equation can be recast into a 
Confluent Heun Equation (CHE)~\cite{Aminov:2020yma,Bianchi:2021xpr,Bianchi:2021mft,Bonelli:2021uvf,Bonelli:2022ten,Consoli:2022eey} 
that is then solved applying  the method of undetermined coefficients.
This CHE reads
\begin{align}
\label{CHE}
& G_1''(Z){+}G_1'(Z) \left(\frac{1 - m_1 - m_2}{Z - 1} - \frac{2 \left(m_3 - 1\right)}{Z} + \frac{X}{Z^2}\right) + \notag \\
& + G_1(Z) \left(\frac{\frac{1}{4} - u + m_3(m_3 - 1)}{Z^2} + \frac{u + \frac{3}{4} + \delta}{(Z -1) Z}\right)=0 \ ,
\end{align}
and the prime here denotes $d/d Z$. Here, $Z$  is the independent variable of the equation 
while $(X, u, m_1, m_2, m_3)$ are the {\it parameters} of the equation, with
the shorthand $\delta \equiv m_1 m_2 + m_1 m_3 + m_2 m_3 - m_1 - m_2 - m_3$.
For a general CHE these parameters are independent among themselves. However,
the identification with the homogeneous Teukolsky equation makes them linearly
dependent, as we will see below.
The identification with Eq.~\eqref{eq:teuk} stems from
\begin{align}
\label{eq:Z}
& Z\equiv \dfrac{2 M}{r} \ , \\
\label{eq:RvsG}
& R^{\rm hom}_{\lm \omega} (Z) =  G_1(Z) e^{-{\frac{X}{2 Z}}} (1-Z)^{1-\frac{m_1+m_2}{2}} \left(\frac{X}{Z}\right)^{1+m_3} \ ,
\end{align}
and the physical range $r \in (2 M, \infty)$ is thus mapped into $Z \in (0,1)$. 
Finally, one has to connect the parameters of the CHE, $(X,u,m_1,m_2,m_3)$, to the
other two physical quantities appearing in Eq.~\eqref{eq:teuk}, i.e. $(\ell, M \omega)$.
This is done using the following dictionary between parameters
\begin{align} 
\label{eq:X}
&X=4 i M \omega \ ,\\
\label{eq:m1}
&m_1 = \frac{X}{2} \ , \\
\label{eq:m2}
&m_2 =  -2 +  \frac{X}{2} \ , \\
\label{eq:m3}
&m_3 = -2 -  \frac{X}{2}  \ ,\\
\label{eq:u}
&u= \left(\ell+\frac{1}{2}\right)^2+\frac{5 X}{2} +\frac{X^2}{4}  \ .
\end{align}
It can be shown that Eq.~\eqref{CHE} has two regular singular points in $Z = \{1, \infty\}$ 
and an irregular one in $Z = 0$. 
The $u$-dependence of the solution $G_1(Z)$ of Eq.~\eqref{CHE} turns out to be simpler
once $u$ is replaced by a different quantity, called $a$, that we assume to be related to $u$
as
\be \label{eq:u_vs_a}
u(a) = a^2 + \sum_{i=1}^{i_{{\rm max},a}} u_i(a) X^i \ .
\ee
Note that the upper limit of the series, $i_{{\rm max},a}$ is a priori infinite, 
but in practice we will always consider a finite truncation. Once $i_{{\rm max}, a}$ 
is fixed, the coefficients $u_i$ are determined recursively as described 
in Sec.~\ref{sbs:get_a} below.

Equation~\eqref{CHE} can be solved all over the full $Z$-domain under the condition 
$X = 4 i M \omega \ll 1$ (\textit{soft limit}), i.e. working perturbatively in $X$. 
The two independent solutions can be obtained recursively in terms of hypergeometric functions. 
More precisely, at leading order in $X$, i.e. when the term $X/Z^2$ in Eq.~\eqref{CHE} is neglected, 
the two solutions are called $H^1_\alpha(Z)$, with $\alpha=\pm$, and 
read\footnote{Note that the equation obtained neglecting the term $X/Z^2$
in fact only depends on the parameters $(m_1,m_2,m_3,a)$ and so do the solutions.
However, in writing Eq.~\eqref{eq:hyper_function1} we have rewritten the $m_i$'s parameters 
in term of $X$ using Eqs.~\eqref{eq:m1}-\eqref{eq:m3}.}
\begin{align} 
\label{eq:hyper_function1}
H^1_\alpha(Z) &= Z^{-\frac{5}{2} + \alpha a - \frac{X}{2}} \times \nonumber\\
                        &\times \,_2 F_1 \left( \frac{1}{2} + \alpha a - \frac{X}{2}, \frac{5}{2} + \alpha a - \frac{X}{2},1 + 2\alpha a; Z \right) \ .
\end{align}
Following Ref.~\cite{Cipriani:2025ikx}, the generalization of this leading-order solution to higher order in $X$ is 
obtained by assuming it to have the following structure
\be
\label{eq:ansatz_G1}
G^1_\alpha(Z) = P_1(Z) H^1_\alpha(Z) + \hat{P}_1(Z) Z H^{1'}_\alpha(Z) \ ,
\ee
where $H^{1'}_\alpha(Z) \equiv d H^1_\alpha(Z) / d Z$ are their first derivatives 
and $P_1$ and $\hat{P}_1$ are polynomials in $X$ and $1/Z$, with $a$-dependent 
coefficients. Their structure reads
\begin{align}
\label{eq:p1}
&P_1(Z) = 1 + \sum_{j=1}^{j_{\rm max}} \sum_{i=0}^{j-1} d_{ij} X^j Z^{i-j}  \ ,\\
\label{eq:p1hat}
& \hat{P}_1(Z) = \sum_{j=1}^{j_{\rm max}} \sum_{i=0}^{j} \hat{d}_{ij} X^j Z^{i-j} \ ,
\end{align}
where $j_{\rm max}$ is the order of  truncation of our solutions in powers  of $X$.
The leading-in-$X$ solution, Eq.~\eqref{eq:hyper_function1}, corresponds to $j_{\rm max}=0$ (also for what said before).

By inserting Eqs.~\eqref{eq:ansatz_G1}-\eqref{eq:p1hat} together 
with Eq.~\eqref{eq:u_vs_a} into Eq.~\eqref{CHE},
one finally finds the coefficients $(d_{ij},\hat{d}_{ij},u_i(a))$ that can be 
expressed\footnote{Evidently, these coefficients are determined as functions 
of $(m_1,m_2,m_3,a)$, but  one uses Eqs.~\eqref{eq:X}-\eqref{eq:m3} to have them
as functions of $X$ or $\omega$.} as explicit functions of $(M \omega,a)$.
In particular, for $j_{\rm max}=1$  one finds
\begin{align}
& d_{01} = \frac{1}{2} + \frac{12 + 20 i M \omega - 8 (M \omega)^2}{1-4a^2} \ ,\\
&\hat{d}_{01} = - \hat{d}_{11} = \frac{4+4 i M \omega}{1-4 a^2} \ , \\
& u_1 = \frac{5+30i M \omega + 32 i (M \omega)^3 + a^2 (8 i M \omega-20)}{8a^2-2} \ .
\end{align}
Notice how the monodromy of the solutions coming from Eq.~\eqref{eq:hyper_function1} 
around the irregular singular point $Z=0$ is codified by the quantity $a$, as it can be seen under the rotation $Z \to e^{2 \pi i} Z$. 

The solution $G_1(Z)$ is determined perturbatively in $X$ keeping $Z\in (0,1)$, that corresponds to
the physical domain of the original Teukolsky equation. Because of this, the quantity $Y \equiv X/Z = 2 i \omega r$ 
is also very small. The physical meaning of this is that $G_1(Z)$ is in fact the PN representation of 
the solution of the CHE, 
being exact in all the separations but perturbative in the velocities.
Since our aim is to eventually compute the gravitational waveform at infinity,
it is useful to work with the Post Minkowskian representation of the  same 
CHE solution, i.e. a solution which is exact for all the velocities but it is perturbative in $1/r$.
This means that this solution is determined perturbatively in $(X,Z)$, while $Y$ is kept finite.
This solution is obtained from Eq.~\eqref{CHE}, replacing $Z = X/Y$ and denoting the corresponding solution as $G_0(Y)$
\begin{align}
\label{CHE_Y}
& G_0''(Y){+}G_0'(Y) \left(\frac{1 {-} m_1 {-} m_2}{Y {-} X} + \frac{m_1 {+} m_2 {+} 2 m_3 {-} 1}{Y} {-}1 \right) + \notag \\
& + G_0(Y) \left(\frac{u + \frac{3}{4} + \delta}{(X-Y) Y} + \frac{1 + \delta + m_3 (m_3-1)}{Y^2}\right)=0 \ .
\end{align}
In the following we will thus only use the two independent solutions of this, denoted as $G^0_\alpha(Y)$,
with $\alpha=\pm$. Following the same reasoning as discussed for $G^1_\alpha(Z)$, at leading order 
in $X$ (i.e.,  neglecting the $X$-dependence in the coefficients of Eq.~\eqref{CHE_Y}), the two solutions 
are now called $H^0_\alpha(Y)$ and read
\be
\label{eq:hyper_function0}
H^0_\alpha(Y) = Y^{\frac{5}{2} - \alpha a + \frac{X}{2}} \,_1 F_1 \left(\frac{5}{2} - \alpha a + \frac{X}{2}, 1 - 2\alpha a; Y\right) \ .
\ee
To go beyond the leading order and include the full $X$-dependence (perturbatively) 
we proceed as above for the other solution, assuming thus the following ansatz 
\be
\label{eq:ansatz_G0}
G^0_\alpha(Y) = P_0(Y) H^0_\alpha(Y) + \hat{P}_0(Y) Y H^{0'}_\alpha(Y) \ ,
\ee
with $H^{0'}_\alpha(Y)\equiv d H^0_\alpha(Y)/d Y$ their first derivatives and $P_0$ and $\hat{P}_0$ 
are polynomials in $X$ and $1/Y$ with $a$-dependent coefficients. In particular they are
\begin{align}
\label{eq:p0}
&P_0(Y) = 1 + \sum_{i=1}^{i_{\rm max}} \sum_{j=0}^{i-1} c_{ij} X^i Y^{j-i}  \ ,\\
\label{eq:p0hat}
& \hat{P}_0(Y) = \sum_{i=1}^{i_{\rm max}} \sum_{j=0}^{i-1} \hat{c}_{ij} X^i Y^{j-i} \ ,
\end{align}
where $i_{\rm max}$ is the order of  truncation of our solutions in powers 
of $X$. The leading-in-$X$ solution, Eq.~\eqref{eq:hyper_function0}, corresponds to $i_{\rm max}=0$.
By plugging Eqs.~\eqref{eq:ansatz_G0}-\eqref{eq:p0hat} together with Eq.~\eqref{eq:u_vs_a} into Eq.~\eqref{CHE_Y},
one finally finds the coefficients $(c_{ij},\hat{c}_{ij},u_i(a))$, that can be expressed as explicit functions of $(M \omega,a)$.
In particular, for $i_{\rm max}=1$ one finds
\begin{align}
& c_{10} = \frac{5+30i M \omega + 32 i (M \omega)^3 + a^2 (8 i M \omega-20)}{8a^2-2} = u_1 \ ,\\
&\hat{c}_{10} = \frac{1-4a^2+16 (M \omega)^2 +16 i M \omega}{8a^2-2} \ .
\end{align}
Notice how the monodromy of the solutions coming from Eq.~\eqref{eq:hyper_function0} 
around the irregular singular point $Y=\infty$ is codified by the quantity $a$, 
as it can be checked under the rotation $Y \to e^{-2 \pi i} Y$.

Since the functions $G^0_\alpha(Y)$ and $G^1_\alpha(Z)$ are solutions of the 
same differential equation (just written in two different variables) 
and display the same monodromy around the irregular singular point, 
they have to be proportional to each other. Their ratios are defined as
\be
 \label{eq:ratios}
g_\alpha (X) = \frac{G^1_\alpha \left(\frac{X}{Y}\right)}{G^0_\alpha(Y)} \ ,
\ee
and turn out to depend only on $X$. For example, if we truncate the two couples of solutions $G^0_\alpha$ and $G^1_\alpha$ at the first order in $X$, it turns out that
\begin{align} \label{eq:expression_galpha}
g_\alpha(X) & = X^{\alpha a+m_3 - \frac{1}{2}} \left[1 + X \left(\frac{1}{4} + \frac{8 \alpha m_1 m_2 m_3}{(4 a^2-1)^2} + \right. \right. \notag \\
& \left. \left. - \frac{m_3 (m_1+m_2+ 2 m_3 - 2) - m_1 m_2}{4 a^2 -1} \right) + \mathcal{O}\left(X^2\right)\right] \ ,
\end{align}
and then using the dictionary in Eqs.~\eqref{eq:m1}-\eqref{eq:m3} everything becomes a function of $X$, $a$
and of the order of truncation $i_{\rm max}$.
The value of $i_{\rm max}$ is determined by requiring a certain PN-accuracy of the waveform once 
PN-expanded\footnote{As anticipated, our method is such to obtain some parts of the solution in
resummed, nonexpaned, form. \label{footnote_general}}. For example, as we will see below, it is necessary to fix $i_{\rm max}=13$ 
in order to compute a waveform that,  once PN expanded, is correct at 10PN accuracy.

\subsubsection{Computation of $a$}
\label{sbs:get_a}
The solutions $G_\alpha^0(Y)$ depend on two separate quantities, $X$ and $a$.
It turns out that also $a$ can be expressed as function of $X$. This can
be done exploiting Eq.~\eqref{eq:u_vs_a}, that can be written as
\be
\label{eq:fora}
u = a^2 + \sum_{i=1}^{i_{{\rm max},a}} u_i(a) X^i = \left(\ell+\frac{1}{2}\right)^2+\frac{5 X}{2} +\frac{X^2}{4} \ ,
\ee
and that can be inverted order by order in $X$ using the expressions of the coefficients $u_i = c_{i,i-1}$ 
obtained above. Although this can be done using standard algebraic manipulators, the 
usual inversion of the series can be computationally inefficient, especially when $i_{{\rm max},a}$
increases so to go to high orders in $X$. An alternative route is then possible.
In Ref.~\cite{Poghosyan:2020zzg} it was shown that $a$ satisfies the following 
equation 
 \be
 \label{fractionequality}
 \frac{ X \, \mathfrak{M}(a+1)}{P(a+1) -\frac{ X \, \mathfrak{M}(a+2)}{P(a+2)-\ldots
 }}
 +\frac{ X \, \mathfrak{M}(a)}{P(a-1)-\frac{ X \, \mathfrak{M}(a-1)}{P(a-2)-\ldots
 }}-P(a)=0 \ ,
 \ee
 with
 \begin{align}
 &P(a)=a^2+a X+2X -\frac{3}{4}X^2-\left(\ell+\frac{1}{2}\right)^2 \ ,  \\
 &\mathfrak{M}(a)= \left( a-\frac{X}{2}-\frac{1}{2}\right) \left( a-\frac{X}{2}+\frac{3}{2}\right) \left( a+\frac{X}{2}+\frac{3}{2} \right) \ .
 \end{align}
At this point\footnote{The two equations are actually rewritten using the dictionary from
\begin{align}
P(a)&=a^2 - u +x\left(a+\dfrac{1}{2}-m_1-m_2-m_3\right) \ , \\
\mathfrak{M}(a)&=\prod_{i=1}^3\left(a-m_i-\dfrac{1}{2}\right) \ ,
\end{align}
to perform the actual calculation.} 
Eq.~\eqref{fractionequality} can be solved for $a$ order by order in $X$.
In particular, we use the following expression for $a$
\be \label{eq:expression_a_general}
a = \ell + \frac{1}{2} + \sum_{i=1}^{i_{{\rm max},a}} a_i X^i \, ,
\ee
and we set $i_{{\rm max},a}$ to the order in $X$ at which we want to determine $a$. For example, for $\ell=2,3$ one finds,
by setting $i_{{\rm max},a}=10$ and $X= 4 i M\omega \equiv 4 i \w$
 \begin{widetext}
 \begin{align}\label{acycleexp}
a(2) & =\frac{5}{2}-\frac{214 \w^2}{105}-\frac{3390466 \w^4}{1157625}-\frac{153440219802466 \w^6}{15021833990625}-\frac{71638806585865707261481 \w^8}{1520451676706008921875} + \notag \\
& -\frac{270360664939833821554899493653643 \w^{10}}{1099244369724415858768042968750} + \mathcal{O}(\w^{12}) \, \\
\nonumber\\
a(3) & =\frac{7}{2}-\frac{26 \w^2}{21}-\frac{21842 \w^4}{33957}-\frac{381415329076 \w^6}{481821815475}-\frac{47254211021655226 \w^8}{35059764403038375}+\notag \\
&-\frac{225004388212297377065114 \w^{10}}{80542621464278043695625} + \mathcal{O}(\w^{12}) \ .
\end{align}
\end{widetext}
The quantity $a$ that we are computing here has an interpretation both within the gauge theory context 
and in the Effective Field Theory (EFT) one as well. For the former, it turns out that the CHE 
Eq.~\eqref{CHE_Y} can be seen as the quantum version of the Seiberg-Witten
(SW) curve of the $\mathcal{N}=2$ supersymmetric extension of QCD, SQCD, with gauge
group SU(2) and with three massive fundamental flavors (whose masses are exactly 
$m_1,m_2,m_3$ that appear in Eq.~\eqref{CHE_Y}) and $a$ is the quantum period of this curve~\cite{Nekrasov:2002qd,Nekrasov:2003af,Nekrasov:2003rj,Nekrasov:2009rc}. 
In other words, for the gauge theory we are considering, the classical SW curve is a torus 
and the quantity $a$ is the period of a specific 1-form, the SW-differential, along one of the 
two non-trivial homological 1-cycles of the torus itself.
The higher-order corrections that characterize it, as reported in Eq.~\eqref{acycleexp}, 
are computed following the procedure we depicted above once the classical SW curve 
is promoted to the quantum one~\cite{Flume:2002az,Bruzzo:2002xf,Poghossian:2010pn,Fucito:2011pn,Aminov:2020yma,Bianchi:2021mft}.
For what concerns the connection to the EFT side, we are going to see it in Section \ref{sec:circ}, inspired
also from the results of ~\cite{Ivanov:2025ozg}.

\subsection{Solving the homogeneous Teukolsky equation} 
\label{sec:hom_solution_Teuk}
Let us go back now to the determination of the functions $R_{\text{in,out}}(Y)$,
solutions of the homogeneous equation. To obtain them we proceed as follows.
From Eq.~\eqref{eq:RvsG}, that defines the relation between the solution of
the Heun equation and the solution of the homogeneous Teukolsky equation,
we have
\be 
\label{eq:Ra_G1a}
R^{\rm hom}_{\pm} (Z) =  G^1_\pm(Z) \, e^{-{\frac{X}{2 Z}}} (1-Z)^{1-\frac{m_1+m_2}{2}} \left(\frac{X}{Z}\right)^{1+m_3} \, ,
\ee
(where we dropped the $\ell, m ,\omega$  indices in $R^{\rm hom}_{\lm \omega, \pm}$ to ease the notation). 
For what discussed in the previous Sec.~\ref{sec:hom},
it is convenient to work with the $G^0_\pm(Y)$. For this reason we are going to use as independent variables $X$ and $Y$ and, remembering the proportionality relation between $G^0_\pm(Y)$ and $G^1_\pm(X/Y)$ in Eq.~\eqref{eq:ratios} and 
the dictionary in Eqs.~\eqref{eq:m1}-\eqref{eq:m3}, the relation Eq.~\eqref{eq:Ra_G1a} becomes
\begin{align}
\label{eq:R_alpha}
R^{\rm hom}_{\pm} (Y) & =  g_\pm(X) G^0_\pm(Y) e^{-\frac{Y}{2}} \left(1-\frac{X}{Y}\right)^{2-\frac{X}{2}} Y^{-1-\frac{X}{2}} \notag \\
&  \equiv g_\pm(X) R_\pm(Y) \, .
\end{align}

In this way the incoming and outgoing solutions are linear combinations of these two last ones
\be
\label{eq:Rin_Rup}
R_{\text{in,out}}(Y) = \sum_{\alpha=+,-} c^{\text{in,out}}_\alpha g_\alpha(X) R_\alpha(Y) \ ,
\ee
where $c^{\text{in,out}}_\pm$ are coefficients that are determined by imposing incoming boundary 
conditions at the horizon and outgoing ones at infinity, as we explain next. 
In order to reach the boundaries and establish the behavior of $R_{\rm in,out}$, we need to use the standard asymptotic formulas 
of the hypergeometric functions that enter $R_\alpha(Y)$ as expressed in Eq.~\eqref{eq:R_alpha}, that is 
(see e.g. Ref.~\cite{Cipriani:2025ikx} and references therein)
\begin{align}
H^0_\alpha (Y) & \underset{Y\to \infty}{\approx} \sum_{\beta=\pm}  B_{\alpha \beta}  Y^{\left(2+{X\over 2}\right)(1+\beta ) } 
e^{ Y (1+\beta)\over 2} \ ,\\
  H^1_\alpha(Z) & \underset{Z\to 1}{\approx} \sum_{\beta=\pm}  F_{\alpha\beta}  (1{-}z)^{  \left(\frac{X}{2}{-}1 \right)
   \left(1{+}\beta \right)} \ ,
\label{fusion0}
\end{align}
with  
\begin{align}
& F_{\alpha \beta} = \frac{\Gamma(1+2 \alpha a) \Gamma(2 \beta (1- 4 i \w))}{\Gamma \left(\frac{1}{2} + \alpha a - \beta \, 2 i \w \right) \Gamma \left(\frac{1}{2} + \alpha a + \beta (2-2 i \w) \right)} \label{eq:fusion_matrix} \ , \\
& B_{\alpha\beta} = \frac{e^{\frac{\pi}{4} (\beta-1)(4 \w + 2 i \alpha a - 5 i)} \Gamma(1-2 \alpha a)}{\Gamma \left(\frac{1}{2} - \alpha a + \beta(2 + 2 i \w)\right)} \label{eq:braiding_matrix}\ .
\end{align}
The behavior of the two functions $R_{\rm in,out}(r)$ for $r \to \infty$ and $r \to 2 M$ is 
\begin{widetext}
\begin{align} 
R_{\text{in}} (r)  & \approx 
\left\{
\begin{array}{lll}
B^{\text{in}}_{-} \, e^{- i \omega r}  \left(\frac{2 M}{r}\right)^{1+2 i M \omega}    +  B^{\rm in}_{+} \, e^{i \omega r} \left(\frac{2 M}{r}\right)^{-3- 2 i M \omega} &~~~~~~~   &\quad r \to \infty   \\
D^{\text{in}}_{-}  \, e^{- i \omega r} \left(1- \frac{2 M }{r}\right)^{2-2 i M \omega}   +  D^{\text{in}}_{+} \, e^{i \omega r} \left(1- \frac{2 M }{r}\right)^{2 i M \omega}&   & \quad
	r \to 2M    \\
\end{array}
\right.   \\
R_{\text{out}} (r)  & \approx 
\left\{
\begin{array}{lll}
   B^{\text{out}}_{+} \, e^{i \omega r}  \left(\frac{2 M}{r}\right)^{-3-2 i M \omega} + B^{\text{out}}_{-}  \, e^{-i \omega r}  \left(\frac{2 M}{r}\right)^{1+ 2 i M \omega} &~~~~~~~   &\quad r \to \infty   \\
D^{\text{out}}_{-} \, e^{- i \omega r} \left(1- \frac{2 M}{r}\right)^{2-2 i M \omega}  +D^{\text{out}}_{+} \, e^{i \omega r} \left(1-\frac{2 M}{r}\right)^{2 i M \omega}  &   & \quad
	r \to 2M   \\
\end{array}
\right.  \label{bcpsi}
\end{align}
\end{widetext}
where $B^{\text{in,out}}_{\pm}, D^{\text{in,out}}_{\pm}$ are some coefficients to be specified. 
At this point, the incoming condition at the horizon $r = 2M$ for $R_{\text{in}}$ 
and the outgoing one at infinity for $R_{\text{out}}$ are respectively equivalent to impose 
\begin{align}
D^{\text{in}}_+  &= 0 \ , \label{din}\\
 B^{\text{out}}_- &= 0 \ , \label{bout}
\end{align} 
which eventually yield
\begin{align}
\frac{c^{\text{in}}_+}{c^{\text{in}}_-} &= - \frac{F_{-+}}{F_{++}} \ , \label{eq:ratio_in} \\
\frac{c^{\text{out}}_-}{c^{\text{out}}_+} &= -  \frac{g_+(X)}{g_-(X)}  \frac{B_{+-}}{B_{--}} \label{eq:ratio_out}\ .
\end{align} 
All the details concerning the asymptotic expressions of the hypergeometric functions and how we obtain
the ratios \eqref{eq:ratio_in}-\eqref{eq:ratio_out} through the computation of the coefficients 
$B^{\rm out}_-$ and $D^{\rm in}_+$ are reported in Appendix~\ref{app:conn_formulas}. 
Note that the boundary conditions determine only the ratios of the coefficients $c^{\text{in,out}}_\pm$, 
while the overall normalization of $R_{\text{in,out}}$ in Eq.~\eqref{eq:Rin_Rup} is left undetermined.
It turns out that the ratio $g_+/g_-$ entering \eqref{eq:ratio_out} can be easily evaluated order by order in $X$. We write 
\be
\frac{g_+(X)}{g_-(X)} = X^{2a} \lambda_{\text{inst}}^{\rm NS}(X) \ ,
\ee 
 with
\be
\label{eq:gamma_inst}
\lambda_{\text{inst}}^{\rm NS} (a,X) \equiv e^{-\partial_a \mathcal{F}_{\text{inst}}^{\rm NS}(a,X)} \ ,
\ee
and
\be
\label{eq:FNS}
\mathcal{F}_{\text{inst}}^{\rm NS}(a,X) = - \sum_{j=1}^{i_{{\rm max},{\cal F}}} \frac{u_j(a)}{j} X^j \ .
\ee
Likewise the case of $(i_{\rm max},i_{{\rm max},a})$, the value of $i_{{\rm max},{\cal F}}$
is arbitrary and it is chosen accordingly to the PN-accuracy that
one desires to recover once the solution is fully PN-expanded.
The quantity defined here as  $\mathcal{F}_{\text{inst}}^{\rm NS}$ is interpreted 
as the instantonic  part of the Nekrasov-Shatashvili (NS) prepotential~\cite{Cipriani:2025ikx}
of a ${\cal N} = 2$ supersymmetric SU(2) gauge theory with three massive fundamental 
flavors~\cite{Nekrasov:2002qd, Flume:2002az, Bruzzo:2002xf, Nekrasov:2003af, Nekrasov:2003rj, Nekrasov:2009rc, Poghossian:2010pn, Fucito:2011pn}.
In this framework, $a$ is the scalar vacuum expectation value (i.e., the quantum SW period),
$u$ the Coulomb branch parameter and Eq.~\eqref{eq:FNS} the quantum version of the Matone relation \cite{Matone:1995rx,Flume:2004rp}.
The function $\lambda_{\rm inst}^{\rm NS}$ is given by the exponential of a complicated rational 
function of $a$ and $\hat{\omega}$. Formally it can be written as
\be
\label{eq:lambda_exp}
\lambda_{\rm inst}^{\rm NS}=e^{\de_a u_1 X + \frac{1}{2}\de_a u_2 X^2 + \dots} \ ,
\ee
and the first two contributions, i.e. $i_{{\rm max},{\cal F}}=2$, explicitly read
\begin{align}
\de_a u_1 X &= -\frac{512 a \, \w^2 (1+\w^2)}{(1-4 a^2)^2} \ ,
\nonumber\\
 \frac{1}{2}\de_a u_2 X^2 & = \frac{a(3871-6392 a^2 + 496 a^4) \w^2}{8(1-4a^2)^2(a^2-1)^2} + \notag \\
 & - \frac{p_1(a) \w^4 + p_2(a) \w^6 + p_3(a) \w^8}{(1-4a^2)^4 (a^2-1)^2} \ ,
\end{align}
where $p_1(a),p_2(a)$ and $p_3(a)$ are the following polynomials in $a$
\begin{align}
& p_1(a) = 6 a(-9471 - 4112 a^2 + 20576 a^4 - 256 a^6 + 256 a^8) \ , \\
& p_2(a) = 96 a (-1199 - 368 a^2 + 2128 a^4 + 384 a^6) \ , \\
& p_3(a) = 1536 a (-37 - 16 a^2 + 80 a^4)  \label{eq:p3_example}\ .
\end{align}
We stress that the overall rescalings of the homogenous solutions are irrelevant, since they 
 cancel against the rescaling of $W$ in the definition of the Green function in Eq.~\eqref{green}. 
 Using the asymptotic behaviors of Eq.~\eqref{bcpsi} one can evaluate the constant $W$ at $r = \infty$. 
 It reads
\be \label{eq:expression_wronskian}
W = \frac{X}{8 M^3} B^{\rm in}_- B^{\rm out}_+ \, .
\ee
 We find convenient to normalize incoming and outgoing solutions such that $W=1$. This can be done by taking
 \begin{align}
& \mathfrak{R}_{\text{in}}(Y) \equiv \frac{R_{\rm in} (Y)}{X B^{\rm in}_-} \ , \label{eq:def_frak_Rin}\\
& \mathfrak{R}_{\text{out}}(Y) \equiv \frac{8 M^3 \, R_{\rm out} (Y)}{B^{\rm out}_+} \, . \label{eq:def_frak_Rout}
\end{align}
that are invariant under rescalings of the solutions.

In the end, we get the following expressions for $\mathfrak{R}_{\text{in}}(Y)$ and $\mathfrak{R}_{\text{out}}(Y)$
\begin{widetext}
\begin{align}
& \mathfrak{R}_{\text{in}}(Y) = \frac{X^{\frac{X}{2}}}{B_{--}} \frac{1}{1- X^{2a} \lambda_{\text{inst}}^{\rm NS}(X) \frac{F_{-+} B_{+-}}{F_{++} B_{--}}} \left[R_-(Y) - X^{2a} \lambda_{\text{inst}}^{\rm NS}(X) \frac{F_{-+}}{F_{++}} R_+(Y) \right] \label{eq:R_in_fin} \ , \\
& \mathfrak{R}_{\text{out}}(Y) = \frac{8 \,  e^{\partial_{m_3} \mathcal{F}^{\rm NS}_{\text{inst}}|_{m_3=-2-\frac{X}{2}}}}{X^{3+\frac{X}{2}} B_{++}} \frac{1}{1- \frac{B_{-+} B_{+-}}{B_{++} B_{--}}} \left[R_+(Y) - \frac{B_{+-}}{B_{--}} R_-(Y) \right] \label{eq:R_out_fin}\ .
\end{align}
\end{widetext}
We notice, in passing, that $X^{2a}=e^{-\de_a{\cal F}_{\rm tree}^{\rm NS}}$ and $\frac{F_{-+} B_{+-}}{F_{++} B_{--}}=e^{-\de_a{\cal F}_{\rm 1-loop}^{\rm NS}}$,
where ${\cal F}_{\rm tree, 1-loop}^{\rm NS}$ respectively correspond to the tree level and 1-loop contributions to the same NS prepotential mentioned above.
Let us stress that functional dependence on $a$ of Eqs.~\eqref{eq:R_in_fin} and \eqref{eq:R_out_fin} is \textit{exact}. By contrast, 
the quantum period $a$, the polynomials $(P_0,\hat{P}_0)$ from Eqs.~\eqref{eq:p0}-\eqref{eq:p0hat} entering the definitions 
of $R_{\pm}(Y)$ as well as the NS prepotential ${\cal F}^{\rm NS}_{\rm inst}$ are computed perturbatively. 
In other words, if $(a,P_0,\hat{P}_0,{\cal F}^{\rm NS}_{\rm inst})$ were known with infinite accuracy, 
Eqs.~\eqref{eq:R_in_fin} and \eqref{eq:R_out_fin} would formally look the same.

\subsection{Solving the Teukolsky equation in the presence of a point-particle source}
\label{sec:teuk_part}
Let us finally present the solution of the Teukolsky equation in the presence of a point-particle source, 
solution that is formally given by Eq.~\eqref{eq:inhomogeneous_sol}. Since waves are extracted far 
away from the BH, Eq.~\eqref{eq:inhomogeneous_sol} can be approximated as
\begin{align}
R_{\ell m\omega} (r) & \underset{r \to \infty}{\approx}  \int_{2 M}^\infty \mathfrak{R}_{\text{in}} (r_1) \mathfrak{R}_{\text{out}}(r) \frac{T_{\ell m\omega}(r_1)}{\Delta(r_1)^2} d r_1 = \notag \\
& =  \mathfrak{R}_{\text{out}}(r) \int_{2 M}^\infty \frac{\mathfrak{R}_{\text{in}} (r_1) T_{\ell m\omega}(r_1) }{\Delta(r_1)^2} d r_1 \ ,
\end{align}
and further using the asymptotic behavior of $\mathfrak{R}_{\text{out}}(r)$ when $r \to \infty$ from Eq.~\eqref{bcpsi}, 
i.e.
\be
\mathfrak{R}_{\rm out} (r) \underset{r \to \infty}{\approx} r^3 \, e^{i \omega r_*} \ ,
\ee
we obtain that
\be \label{eq:final_solution}
R_{\ell m\omega} (r) \underset{r \to \infty}{\approx} r^3 \, e^{i \omega r_*}  Z_{\ell m\omega} \ ,
\ee
where $Z_{\ell m}$ is the following integral
\be 
\label{eq:Z_lm}
Z_{\ell m\omega} = \int_{2 M}^\infty \frac{\mathfrak{R}_{\text{in}} (r_1) T_{\ell m\omega}(r_1) }{\Delta(r_1)^2} d r_1 \ .
\ee
Once we specialized the harmonic components of the stress-energy tensor $T_{\lm\omega}$ to be those 
of a point-particle source, described by a $\delta$-function, and moving along a trajectory characterized
by $r(t)$ and dimensionless conserved specific energy $\hat{E}\equiv E/\mu$ and angular momentum 
$p_\varphi=P_\varphi/(M \mu)$ (with $\mu$ the mass of the particle)
we have
\be
\label{eq:Zlmw}
Z_{ \ell m\omega} = \int \frac{d t}{r(t)^2} e^{i \omega t - i m \phi(t)} \sum_{n=0}^2 b^n_{\lm} \mathcal{L}^n [\mathfrak{R}_{\rm in} (r(t))] \ ,
\ee
where  the coefficients $b^n_{\lm}$ are defined as
\begin{align} 
& b^0_{\ell m} = \pi \sqrt{(\ell -1) \ell (\ell +1) (\ell+2)} \,\, {}_{0}S_\lm \left(\frac{\pi}{2}\right) \label{eq:b0_lm} \ ,\\
& b^1_{\ell m} = 2 \pi \sqrt{(\ell -1) (\ell+2)} \,\, {}_{-1}S_\lm \left(\frac{\pi}{2}\right) \label{eq:b1_lm} \ ,\\
& b^2_{\ell m} = 2 \pi \,\, {}_{-2}S_\lm \left(\frac{\pi}{2}\right) \label{eq:b2_lm} \ .
\end{align}
The functions ${}_{s}S_\lm (\theta)$ are related to the spherical harmonics via 
${}_{s}Y_\lm (\theta,\phi) = e^{i m \phi} {}_{s}S_\lm (\theta)$ and we are assuming 
that the particle's motion takes place on the equatorial plane. 
For what concerns the $\mathcal{L}^i$, these are the following multiplicative and differential operators
\begin{align}
&  \mathcal{L}^0 = \frac{\hat{E}}{A} \left(1 + \frac{\dot{r}}{A}\right)^2 , \label{eq:L0}\\
&  \mathcal{L}^1 = \frac{i M p_\varphi}{r} \left(1 + \frac{\dot{r}}{A} \right) \left(2 - r \partial_r + \frac{i \omega r}{A}\right) , \label{eq:L1}\\
& \begin{aligned}
    \mathcal{L}^2 = & \frac{M^2 p_\varphi^2 A}{\hat{E} \, r^2} \left[A^{-2} \left(- i \omega r + \frac{1}{2} (\omega r)^2 + i M \omega \right) + \right. \\
    & \left. + \left(1 + \frac{i \omega r}{A}\right) r \partial_r - \frac{1}{2} r^2 \partial^2_r \right] \label{eq:L2} \ ,
  \end{aligned}
\end{align}
where $A(r) \equiv 1 - \frac{2 M}{r}$.

\section{Multipolar waveform for circular orbits and its factorized form}
\label{sec:circ}
\subsection{Structure of the strain waveform}
Let us now specialize the general formulas derived above to the case of circular orbits. These
orbits are defined by the condition $\dot{r}=0$, the phase is $\phi(t) = \Omega t$, $\Omega$ is
the orbital frequency and $r\equiv r_\Omega=(M\Omega)^{-2/3}$ is the orbital radius.
The integral in Eq.~\eqref{eq:Zlmw} thus becomes trivial, as it amounts in identifying 
$\omega=m \Omega$. Equation~\eqref{eq:Zlmw} is the recast as
\be 
\label{eq:final_form_Z_lm}
Z_{\ell m\omega} = \frac{1}{r^2} \left. \left[C_0 \mathfrak{R}_{\text{in}}+ C_1 r \partial_r \mathfrak{R}_{\text{in}} + C_2 r^2 \partial^2_r \mathfrak{R}_{\text{in}}\right]\right|_{r=r_\Omega} \ ,
\ee
that is related to the waveform Fourier modes as
\be
\label{eq:hlm}
h_{\lm \omega}=\dfrac{2}{\omega^2}Z_{\lm \omega} \ .
\ee
In Eq.~\eqref{eq:final_form_Z_lm} it is intended that the derivatives of 
$\mathfrak{R}_{\text{in}}$ are evaluated 
at $r=r_\Omega$ after differentiation and the coefficients $(C_0,C_1,C_2)$ read
\begin{align}
& \begin{aligned} 
    C_0 = & \frac{\hat{E}_{\rm circ}}{A} b^0_{\ell m} + \frac{2 i M p_\varphi^{\rm circ}}{r} \left(1 + \frac{i m\Omega r}{2 A}\right) b^1_{\ell m} + \\
    & - p_\varphi^{\rm circ} M \Omega \frac{i m\Omega r}{A^2} \left(1 - \frac{M}{r} + \frac{i m\Omega r}{2}\right) b^2_{\ell m} \ ,
  \end{aligned}\\
  \label{eq:A1} 
& C_1 = \left(- \frac{i M }{r} b^1_{\ell m} + M \Omega \left(1 + \frac{i m\Omega r}{A}\right) b^2_{\ell m}\right)p_\varphi^{\rm circ} \ ,\\
& C_2 = -\frac{1}{2} p_\varphi^{\rm circ} M \Omega b^2_{\ell m} \label{eq:A2} \ ,
\end{align}
where $(\hat{E}_{\rm circ},p_\varphi^{\rm circ})$ are the energy and angular moment of a test-particle on 
circular orbits that expressed in terms of the frequency parameter $x\equiv (M\Omega)^{2/3}=r_{\rm \Omega}^{-1}$ 
read
\begin{align}
\hat{E}_{\rm circ}      &= \frac{1-2x}{\sqrt{1-3x}} \ ,\\
p_{\varphi}^{\rm circ}&= \frac{1}{\sqrt{x (1-3x)}} \ .
\end{align}
The structure of the solution in Eq.~\eqref{eq:final_form_Z_lm} is standard, see e.g. Eq.~(2.11) of 
Ref.~\cite{Fujita:2010xj} (and references therein), that is completely equivalent to ours once the 
notations are matched. However, Ref.~\cite{Fujita:2010xj,Fujita:2012cm} solved the Teukolsky 
equation using the MST method, that relies on hypergeometric and Coulomb functions, 
and expressed the solution in fully PN-expanded form.
Our approach based on the solution of the confluent Heun equation will allow us to
identify new knowledge in the form of closed-form, non-PN expanded, analytical structures 
that account, by themselves, of all logarithmic terms and transcendental numbers appearing 
in the PN-expanded waveform of Ref.~\cite{Fujita:2010xj}. Let us remember in fact that, although 
the MST procedure can be easily pushed to high PN order (e.g. 22PN~\cite{Fujita:2012cm}), 
it is then crucial to resum the PN-expanded results in order to make them reliable and predictive 
in the strong-field, fast velocity regime. At a conceptual level, Refs.~\cite{Damour:2007xr,Damour:2008gu} 
(see also~\cite{Damour:1997ub}) pointed out that it is useful to recast the PN-expanded waveform 
multipoles in a special factorized and resummed form, notably factorizing out closed-form analytical expression. 
More precisely, Ref.~\cite{Damour:2008gu} identified these closed-form factors to be the {\it source} of
the field (i.e. the energy or the angular momentum, depending on the parity of the waveform mode) 
and the {\it tail factor} that resums the infinite series of leading logarithms in the waveform phase.
The solution of the Teukolsky equation as CHE will allow us to identify {\it new}, closed-form,
resummed structures that incorporate all logs (both leading and subleading, in either the 
waveform amplitude and phase) as well as the transcendental numbers. In practice, a wise
use of the CHE allows one to reduce the residual PN-expanded information to just polynomials
with rational coefficients.
Our starting point to identify these analytical structures is given by the solution $\mathfrak{R}_{\text{in}}$
in Eq.~\eqref{eq:R_in_fin}. Since $Y = 2 i \omega r$ and $X=4 i \w$, it rewrites as
\begin{widetext}
\begin{align}
\mathfrak{R}_{\text{in}} (r)& = \frac{(4 i \w)^{2 i \w} \, e^{- i \pi \left(\frac{5}{2} +a + 2 i \w \right)} \Gamma \left(a - 2 i \w - \frac{3}{2}\right)}{\Gamma \left(1+2 a \right)} \frac{1}{1- (4 i \w)^{2a} \lambda_{\text{inst}}^{\rm NS} \frac{e^{- 2 i a \pi} \Gamma(1-2a)^2 \Gamma \left(\frac{5}{2} + a - 2 i \w \right) \Gamma \left(\frac{1}{2} + a - 2 i \w \right) \Gamma \left(-\frac{3}{2} + a - 2 i \w \right)}{\Gamma(1+2a)^2 \Gamma \left(\frac{5}{2} - a - 2 i \w \right) \Gamma \left(\frac{1}{2} - a - 2 i \w \right) \Gamma \left(-\frac{3}{2} - a - 2 i \w \right)}} \times \notag \\
&  \times \left[R_-(r) - (4 i \w)^{2a} \lambda_{\text{inst}}^{\rm NS}\frac{\Gamma(1-2a) \Gamma \left(\frac{5}{2} + a - 2 i \w \right) \Gamma \left(\frac{1}{2} + a - 2 i \w \right)}{\Gamma(1+2a) \Gamma \left(\frac{5}{2} - a - 2 i \w \right) \Gamma \left(\frac{1}{2} - a - 2 i \w \right)} R_+(r) \right] \ .
\end{align}
\end{widetext}
The functions $R_{\pm}(r)$ contain, respectively, the factors $(2 i \omega r)^{-1-2 i \w} (2 i \omega r)^{\frac{5}{2}\mp a + 2 i \w} = (2 i \omega r)^{\frac{3}{2} \mp a}$ (see Eq.~\eqref{eq:R_alpha}). These factors, once PN-expanded, generate  logarithmic and transcendental contributions. 
It is thus useful to separate this contribution from the rest and define the function {\it without} it as
\be
\label{eq:Spm}
\mathcal{S}_\pm(r) \equiv (2 i \omega r)^{-\frac{3}{2} \pm a } R_{\pm}(r) \ .
\ee
In this way these functions contain only (truncated) hypergeometric functions. 
At this point we collect the factor $(2 i \omega r)^{\frac{3}{2} + a}$ in 
front of the square brackets, so to recast the above expression as
\begin{widetext}
\begin{align}
\label{eq:R_in_fin_explicit}
\mathfrak{R}_{\text{in}} (r)& = (4 \w)^{2 i \w} (2 \omega r)^{\frac{3}{2}+a} \frac{e^{\pi \w} e^{- i \frac{\pi}{2} \left(\frac{7}{2} +a\right)} \Gamma \left(a - 2 i \w - \frac{3}{2}\right)}{\Gamma \left(1+2 a \right)} \frac{1}{1-  \lambda_{\text{inst}}^{\rm NS} (4 \w)^{2a} \frac{e^{-i a \pi} \Gamma(1-2a)^2 \Gamma \left(\frac{5}{2} + a - 2 i \w \right) \Gamma \left(\frac{1}{2} + a - 2 i \w \right) \Gamma \left(-\frac{3}{2} + a - 2 i \w \right)}{\Gamma(1+2a)^2 \Gamma \left(\frac{5}{2} - a - 2 i \w \right) \Gamma \left(\frac{1}{2} - a - 2 i \w \right) \Gamma \left(-\frac{3}{2} - a - 2 i \w \right)}} \notag \\
& \times \left[\mathcal{S}_-(r) - \lambda_{\text{inst}}^{\rm NS}\textcolor{black}{\left(\frac{2 M}{r}\right)^{2a}}  \textcolor{black} {\frac{\Gamma(1-2a) \Gamma \left(\frac{5}{2} + a - 2 i \w \right) \Gamma \left(\frac{1}{2} + a - 2 i \w \right)}{\Gamma(1+2a) \Gamma \left(\frac{5}{2} - a - 2 i \w \right) \Gamma \left(\frac{1}{2} - a - 2 i \w \right)}} \mathcal{S}_+(r) \right] \ .
\end{align}
\end{widetext}
Note that, although this expression formally looks to be in closed form, in fact it intrinsically involves
the truncation indices $(i_{\rm max},i_{{\rm max},a},i_{{\rm max},{\cal F}})$
needed to determine $({\cal S}_{\pm},a,\lambda_{\rm inst}^{\rm NS})$ recursively.
As a reminder, the functions ${\cal S}_\pm$, Eq.~\eqref{eq:Spm}, are related to $R_\pm$ via Eq.~\eqref{eq:R_alpha}, 
that in turn are connected to $G^0_\pm$, Eq.~\eqref{eq:ansatz_G0} where the polynomials $(P_0,\hat{P}_0)$ 
defined in Eqs.~\eqref{eq:p0}-\eqref{eq:p0hat} appear with their explicitly dependence on the truncation index $i_{\rm max}$.
Similarly, $a$ comes from Eq.~\eqref{eq:fora} and $\lambda_{\rm inst}^{\rm NS}$ from Eq.~\eqref{eq:FNS}.
Assuming that we want the solution above to be accurate at a given PN order, 
labeled with $N_{\rm PN}$, we have
\begin{align}
i_{\rm max} &= N_{\rm PN}+3 \ , \label{eq:rule_imax}\\
i_{{\rm max},a} &= i_{{\rm max}, {\cal F}} = \left \lfloor{2i_{\rm max}/3}\right\rfloor \ ,
\end{align}
where $\lfloor \rfloor$ denotes the integer part. The reason for this is that $a$ involves 
only even powers of  $\hat{\omega}$, implying that it contributes every three PN orders,
starting from 3PN (i.e., 3PN, 6PN, 9PN etc). Note however that, given the modular structure
of the solution, in principle the three orders of truncation can be {\it independent} and one
can inject more (or less) analytical information in each of the three building blocks 
$({\cal S}_{\pm},a,\lambda_{\rm inst}^{\rm NS})$ depending on the impact of each one on
the accuracy of $\mathfrak{R}_{\text{in}}$ when compared to numerical results 
(see in particular Sec.~\ref{sec:studyh22} below).

By taking two derivatives with respect to $r$ of Eq.~\eqref{eq:R_in_fin_explicit} one finally gets the strain from 
Eqs.~\eqref{eq:final_form_Z_lm} and \eqref{eq:hlm}. Note that in $\mathfrak{R}_{\text{in}}$ we kept 
explicit the dependence on $r$ so to avoid the introduction of mistakes in getting the waveform, 
because one has {\it first} to derive with respect to $r$ and {\it then} impose the Kepler's relation between 
radius and frequency holding on circular orbits and have $r\to r_\Omega$ so that effectively
the final waveform only depends on the frequency parameter $x$.
Inspecting the analytical structure of $\mathfrak{R}_{\text{in}}$, Eq.~\eqref{eq:R_in_fin_explicit}, 
one sees that the final solution is given by a complicated combination of transcendental and 
algebraic pieces.  As it is well known (see e.g. Ref.~\cite{Pan:2010hz}) the PN-expanded waveform, 
obtained ab-initio with a different method, presents logs, transcendental  numbers 
(e.g. $\pi$, $\gamma_E$ etc) together with rational coefficients.  The origin of these different contributions
stems from the analytical, closed-form, expressions of Eq.~\eqref{eq:R_in_fin_explicit}, that remain unchanged
under the action of the $r$-derivatives necessary to obtain the waveform strain. More precisely:
\begin{itemize}
\item[(i)]{}All logs come from terms like $(4\hat{\omega})^{2i\hat{\omega}}$, i.e. schematically
of the from $\hat{\omega}^{P^n_0(\hat{\omega})}$ where $P^{n}_0$ generically indicates a polynomial
in $\hat{\omega}$ of order $n$. Once PN-expanded these term bring $\omega^p (\log(\hat{\omega}))^q$-type
corrections in both the waveform amplitude and phase. These type of corrections are generically denoted
as tails.
\item[(ii)]The $\pi$'s come from the $e^{\pi P^m_0(\hat{\omega})}$ and $e^{i\pi P^n_0(\hat{\omega})}$ 
functions, while the $\gamma_E$ from the
expansion of the many $\Gamma$ functions present (see Sec.~\ref{sec:PN_comparison} below).
\item[(iii)]all the other functions present in Eq.~\eqref{eq:R_in_fin_explicit}, eventually lead to 
purely rational contributions. 
\end{itemize}
The strain depends on the quantity $a$, that, as we mentioned above, can be interpreted as
the quantum period of the Seiberg-Witten curve of a certain supersymmetric gauge theory.
Actually, within the current context of black hole perturbation theory (BHPT), Refs.~\cite{Fucito:2023afe,Bautista:2023sdf}
pointed out that $a$ is related to the so-called renormalized angular momentum, introduced 
in Ref.~\cite{Mano:1996mf} (see also Refs.~\cite{Mano:1996gn}), which depends on the black 
hole mass $M$ and its dimensionless spin $\chi\equiv J/M^2$. 
Here we address this quantity as $\hat{\ell}$ 
(instead of the commonly used $\nu$ notation\footnote{to avoid confusion with the symmetric mass
ratio $\nu\equiv m_1 m_2/(m_1+m_2)^2$}) and it is related to $a$ 
\begin{align}
\hl &= a - \dfrac{1}{2} \nonumber\\
     &\simeq \ell -\dfrac{15\ell^2(\ell+1)^2+13\ell(\ell+1)+24}{(2\ell+1)\ell(\ell+1)(4\ell(\ell+1)-3)}\hat{\omega}^2+O(\hat{\omega}^3) \ .
\end{align}
The explicit expression of $\hl$, in PN-expanded form up to $\hat{\omega}^{10}$ are 
listed in Appendix~\ref{app:a_expressions}. The physical meaning of $\hl$ is to capture 
the scale dependence of the BH multipole moments in the near zone.  Note that the computation of
$a$ we performed solving Eqs.~\eqref{eq:fora} (or Eq.~\eqref{fractionequality}) is one of the three 
possible methods of having access to this quantity, the other being the MST recursive relation 
or the Monodromy matrix method~\cite{Castro:2013lba,Nasipak:2024icb}. See also
Ref.~\cite{Bautista:2023sdf} or the supplemental material of Ref.~\cite{Ivanov:2025ozg} for 
additional information. More recent developments in the study of this quantity for several 
geometries in the eikonal limit are reported in Ref.~\cite{Bini:2025ltr, Bini:2025bll}. In this respect, it is important to mention
that Ref.~\cite{Ivanov:2025ozg} showed that $\hat{\ell}$ precisely determines the anomalous 
dimension of BH multipole moments, $\gamma_\lm^{\rm BH}$, defined as
\be
\label{eq:gamma_lm_BH}
\gamma_\lm^{\rm BH}=\hat{\ell}(\hat{\omega},\chi)-\ell \ .
\ee
This quantity will be important, following the insights of Ref.~\cite{Ivanov:2025ozg}, to modify 
$\hl$ into a more general quantity to be used for comparable mass binaries. We will come back 
to this in Sec.~\ref{sec:comparable_mass} below. From now on, the $a$-dependence in all formulas
mentioned above has actually to be intended as $\hat{\ell}$-dependence.

\subsection{Factorized multipolar waveform}
\label{sec:h_factorized}

Let us turn now to presenting a new factorization of the waveform stemming from the analytic
structure of $\mathfrak{R}_{\rm{in}}$ so to improve the original DIN proposal~\cite{Damour:2008gu}.
The multipolar waveform is first factorized in a Newtonian prefactor and a post-Newtonian 
correction $\hat{h}_\lm^{(\epsilon)}$
\be
h_\lm = h_\lm^{(N,\epsilon)}\hat{h}_\lm^{(\epsilon)} \ .
\ee
Note that in the test-mass limit, $h_\lm^{(N,\epsilon)}$ is simply given by the test-mass reduction
of Eq.~(4) of DIN. We propose to factorize the PN correction as
\be
\label{eq:hhat_fac}
\hat{h}_{\lm}^{(\epsilon)}=\hat{S}^{(\epsilon)}T_{\ell m}z_\lm \left[1+ \tilde{T}_\lm \left(-\frac{\ell+1}{\ell}\right)^\epsilon \tilde{z}_\lm\right] \ ,
\ee
where $\epsilon$ indicates the parity of $\ell+m$, so that $\epsilon=0$ for $\ell+m$ even, while $\epsilon=1$ 
for $\ell+m$ odd. The source of the field, $\hat{S}^{(\epsilon)}$,  corresponds to either the energy 
($\epsilon=0$) or the Newton-normalized angular momentum ($\epsilon=1$)
\begin{align} 
\hat{S}^{(0)} &= \hat{E}_{\text{circ}} \ , \label{eq:source_energy_test_mass}\\
\hat{S}^{(1)} &= \sqrt{x}p^{\text{circ}}_\varphi  \ , \label{eq:source_angmom_test_mass}
\end{align}
We will refer to the functions $T_\lm$ and $\tilde{T}_\lm$ as {\it tail factors}, i.e. take 
into account logarithmic as well as transcendental terms, while $(z_\lm,\tilde{z}_\lm)$ 
are the remainder  (purely rational and complex) contributions. 
Note that the internal tail, $\tilde{T}_\lm$, starts contributing at 5PN order, as we will explicitly see below. 
The two tail factors can be written in a convenient form that separates the 
contributions according to their behavior under PN-expansion, as already hinted above 
in the discussion around Eq.~\eqref{eq:R_in_fin_explicit}. In particular, for each tail term, 
one can separate factors that generate only the logarithmic terms from factors 
that generate the transcendental numbers and no logs. More precisely, we write
\begin{align}
& T_{\ell m} = \frac{\Gamma (2 \ell + 2)}{\Gamma (\ell - 1)}\frac{T_{\ell m}^{1,\text{log}} \, \, T_{\ell m}^{1,\text{trnsc}}}{1- \lambda_{\text{inst}}^{\rm NS}T_{\ell m}^{2,\text{log}}\, \, T_{\ell m}^{2,\text{trnsc}}} \ , \label{eq:ext_tail_def}\\
& \widetilde{T}_{\ell m} = - \lambda_{\text{inst}}^{\rm NS} \tilde{T}_{\ell m}^{ \text{log}} \tilde{T}_{\ell m}^{\text{trnsc}} \label{eq:int_tail_def} \ .
\end{align}
Here, the superscript log refers to the fact that the corresponding term produces, once PN expanded, 
logarithmic contributions, while by trnsc we mean that the factor generates transcendental numbers. 
The functions $T_{\ell m}^{2, \text{log}}, \tilde{T}_{\ell m}^{\text{log}},T_{\ell m}^{2, \text{Tr}}, \tilde{T}_{\ell m}^{\text{trnsc}}$ 
can be read from Eq.~\eqref{eq:R_in_fin_explicit} (setting explicitly $r=r_\Omega$), 
while $T_{\ell m}^{1, \text{log}}$ and $T_{\ell m}^{1, \text{trnsc}}$  are obtained from the 
same expression after multiplying  for some factors in such a way that $T_{\ell m}$ starts with 1. 
The tail contributions are thus given by
\begin{align}
\label{eq:Ta1}
T_{\ell m}^{1,\text{log}} &= (2\omega r_\Omega)^{\hat{\ell}-\ell} (4 \w)^{2 i \w}= e^{(\hl-\ell)\log(2 \omega r_{\Omega})} e^{2 i \w \log \left(4 \w \right)} \ ,\\
\label{eq:Ta2}
T_{\ell m}^{2,\text{log}}&=(4 \w)^{2\hl +1} =  e^{(2\hl +1) \log (4 \w)}\ ,\\
\tilde{T}_{\ell m}^{\text{log}} &= \left(\frac{2 M}{r_{\rm \Omega}}\right)^{2\hl + 1} = e^{(2\hl +1) \log \left(\frac{2M}{r_{\Omega}}\right)}\ , 
\end{align}
while the transcendental contributions read
\begin{align}
&T_{\ell m}^{1,\text{trnsc}} = \frac{\Gamma \left(\hl - 1 - 2 i \w \right)}{\Gamma \left(2 \hl+ 2\right)} e^{\pi \w} e^{-i\frac{\pi}{2} \left(\hl- \ell \right)} \label{eq:T1_trscn}\ ,\\
&T_{\ell m}^{2,\text{trnsc}}  = e^{-i \left(\hl + \frac{1}{2}\right) \pi} \nonumber\\ 
&\!\!\!\times\frac{ \Gamma(-2\hl)^2 \Gamma \left(\hl + 3 - 2 i \w \right) \Gamma \left(\hl + 1 - 2 i \w \right) \Gamma \left(\hl - 1 - 2 i \w \right)}{\Gamma(2\hl+2)^2 \Gamma \left(2 - \hl - 2 i \w \right) \Gamma \left(-\hl - 2 i \w \right) \Gamma \left(-2 - \hl - 2 i \w \right)} \ ,\\
&\label{eq:tTlmTr}\tilde{T}_{\ell m}^{\text{trnsc}}  = \frac{\Gamma(-2\hl) \Gamma \left(\hl + 3 - 2 i \w \right) \Gamma \left(\hl + 1 - 2 i \w \right)}{\Gamma(2 \hl + 2) \Gamma \left(2 - \hl - 2 i \w \right) \Gamma \left(-\hl - 2 i \w \right)}  \ . 
\end{align}
Let us make here the following remark. The two $T^{1,{\rm Ta}}_\lm$ and 
$T^{1,{\rm trnsc}}_\lm$ factors precisely show the functional dependence on $\hat{\ell}-\ell$ 
in the amplitude and phase predicted by Ivanov et al.~\cite{Ivanov:2025ozg} 
(once specified to the test-mass limit) by running renormalization group evolution of 
the multipoles up to the orbital scale $1/r_\Omega$, see their Eqs.~(32) and (33).
It is interesting to note that the analysis in the test-mass limit predicts the presence
of the 2 in the first log of Eq.~\eqref{eq:Ta1}, a number that could not be obtained by
the RG evolution of Ref.~\cite{Ivanov:2025ozg}. We will also see that the presence of this
$2$ and the precise combination of the $\Gamma$-functions given by
Eq.~\eqref{eq:ext_tail_def} are important ingredients also to obtain a factorized waveform
for comparable mass binaries that improves the proposal of Ref.~\cite{Ivanov:2025ozg}.

Let us finally turn to discuss the structure of the remainder functions $(z_\lm,\tilde{z}_\lm)$. 
In the current context they are represented in closed form as infinite power series 
with rational (complex) coefficients, after being PN expanded. They read
\begin{align}
\label{eq:flm}
z_\lm &= \frac{1}{\hat{S}^{(\epsilon)} b^{\epsilon}_{\ell m} \left(- i \ell x^{\frac{1}{2}}\right)^\epsilon} \sum_{n=0}^2 C_n \mathcal{R}^-_n \ ,\\
\label{eq:tflm}
 \tilde{z}_{\ell m} & = \left(-\frac{\ell+1}{\ell}\right)^{-\epsilon} \frac{\sum_{n=0}^2 C_n \mathcal{R}^+_n}{\sum_{n=0}^2 C_n \mathcal{R}^-_n}  \ ,
\end{align}
with
\be
\label{eq:Ralphan}
 \mathcal{R}^\alpha_n = \left. \left(2 i \omega r\right)^{- \left(\frac{3}{2}-\alpha a\right)} r^n \partial_{r}^n R_\alpha(r) \right|_{r=r_\Omega, \omega = m \Omega, a = \hl + \frac{1}{2}}  \ ,
\ee
and $R_{\pm}(r)$ are the functions we obtained above stemming from the solution of the CHE.
The term $\left(-\frac{\ell+1}{\ell}\right)^\epsilon$ is factorized from $\tilde{z}_\lm$ in such a way that,
once PN-expanded when $\ell+m$ is odd, this quantity starts with 1. 
We stress that Eqs.~\eqref{eq:flm}-\eqref{eq:tflm} as they are give access to the complete analytic
representation of the multipolar waveform for a test-mass orbiting a Schwarzschild black hole around
circular orbits. This solution needs, however, that we specify a certain order of truncation for
the recursion that allows us to determine the quantity $a$ from Eq.~\eqref{eq:fora}. In practice, to compute
the waveform analytically one has to: (i) solve Eq.~\eqref{eq:fora} at a given order, determining thus $\lambda_{\rm inst}^{\rm NS}$
from Eqs.~\eqref{eq:gamma_inst}; and (ii) explicitly compute $(z_\lm,\tilde{z}_\lm)$ from the ${\cal R}_n^{\pm}$ 
of Eq.~\eqref{eq:Ralphan}. This approach gives an analytical solution to the problem with an accuracy that can 
be arbitrarily high depending on the PN order of computation of $\hat{\ell}$. This fact was already pointed out in 
Ref.~\cite{Cipriani:2025ikx} by pointwise comparisons with the energy flux computed numerically.

Here we want to follow a different approach and work instead with PN-expansions of $(z_\lm,\tilde{z}_\lm)$,
retaining up to 10PN order. The hope is that the fact that all the logarithmic and transcendental contributions 
are packaged in the new tail factors will eventually allow for an analytical representation of the waveform 
(and of the related energy flux) that is more accurate than the standard approach of 
DIN~\cite{Damour:2008gu} at the same PN order. It is convenient to write the remainders as amplitude and phase as
\begin{align}
z_\lm&=f_\lm e^{i\delta_\lm} \equiv (\rho_\lm)^\ell e^{i\delta_\lm}\ ,\\
\tilde{z}_\lm &= \tilde{f}_\lm e^{i\tilde{\delta}_\lm} \ .
\end{align}
Note that the current $f_\lm$ coincide, at 1PN accuracy, with the corresponding ones obtained by DIN~\cite{Damour:2008gu}..
As such, the same reasoning of DIN holds and it is thus useful to $1/\ell$ resum the residual amplitudes.
From Eq.~\eqref{eq:hhat_fac} the final factorized form reads then
\be
\label{eq:hhat_final}
\hat{h}_{\lm}^{(\epsilon)}=\hat{S}^{(\epsilon)}T_{\ell m}(\rho_\lm)^\ell e^{i\delta_\lm} \left[1+ \tilde{T}_\lm \left(-\frac{\ell+1}{\ell}\right)^\epsilon \tilde{f}_\lm e^{i\tilde{\delta}_\lm}\right] \ ,
\ee
where it is intended that the amplitude corrections $(\rho_\lm,\tilde{f}_\lm)$ are polynomials 
in $x$, the phase corrections $(\delta_\lm,\tilde{\delta}_\lm)$ are polynomials in  $x$
and all involve only rational coefficients. Expanding $(z_\lm,\tilde{z}_\lm)$ up to 10PN, 
as mentioned above, for $\rho_{22}$ we obtain
\begin{widetext}
\begin{align} \label{eq:rho22_test_mass}
\rho_{22} & = 1-\frac{43}{42}x-\frac{20555}{10584}x^2-\frac{4296031 bv}{4889808}x^3+\frac{9228174993589}{800950550400}x^4-\frac{8938613036677}{2116091577600}x^5+ \notag \\
& -\frac{1060700697798333909671}{24231643979185843200} x^6+\frac{3567168919606240724303840051}{43991338062012939037440000}x^7+\frac{8339316227220569285816625738049}{279101750471556914871889920000}x^8+ \notag \\
& -\frac{522338057689474511990262498143822507399}{857097472947610731676894961786880000}x^9+\frac{1523513000214555169284583871085138536795675131}{1333729377653777059562416250036563968000000}x^{10} \ ,
\end{align}
for $\tilde{f}_{22}$ we have
\begin{align} \label{eq:ftilde_22}
\tilde{f}_{22} & = 1+\frac{4391}{2247} x+\frac{53185}{2646}x^2+\frac{17096210}{305613}x^3 +\frac{4747421406107252}{71641272277575} x^4 +\frac{8197825650198689}{18747248820300} x^5+ \notag \\
& +\frac{93413981315288045717}{265033606022345160}x^6 - \frac{84886593520942215307406177173}{115729970007349756213155000} x^7+\frac{12091990099120207716578842317287}{2316762577156478297276430000}  x^8 + \notag \\
& -\frac{18745458158059179828839098304527937}{5506639808719744112850116685000} x^9 -\frac{12907954629421965590241825710607690689624960837}{465999378807314866788638416951557033375000} x^{10}  \ ,
\end{align}
\end{widetext}
while for the residual phases we obtain
\begin{align}
 \label{eq:delta22_test_mass}
& \delta_{22} = {-\frac{17}{3} x^{\frac{3}{2}} - \frac{259}{81} x^{\frac{9}{2}} - \frac{58940243}{3539025} x^{\frac{15}{2}}} \ , \\
\nonumber\\
\label{eq:delta_tilde_22_test_mass}
& \tilde{\delta}_{22} = {\frac{25}{3} x^{\frac{3}{2}} + \frac{12077}{567} x^{\frac{9}{2}} + \frac{159283133}{694575} x^{\frac{15}{2}}}  \ .
\end{align}
This procedure is then applied to all multipoles up to $\ell=8$, where each residual function is truncated at 10PN accuracy.
The complete analytical information is reported in Appendices~\ref{app:rholm_flm_PN} and~\ref{app:residual_phases}.
We recall that, as already said in Section \ref{sec:hom}, in order to determine the expressions of 
$\rho_\lm$ and $\tilde{f}_\lm$ at 10 PN order, we have to set at least $i_{\rm max} = 13$ inside the polynomials 
$P_0$ and $\hat{P}_0$ in \eqref{eq:p0}. In this way the results that come out are consistent at that PN order.

\subsection{Comparison with the PN-expanded waveform}
\label{sec:PN_comparison}
We have mentioned above that each factor $T_{\lm}^{i,k}$ with $i=1,2$ and $k=({\rm log,trnsc})$ gives different 
logarithmic or transcendental contribution entering at various PN order. For illustrative purposes, let us consider
only the $\ell=m=2$ case and show here explicitly the PN-expansion of the various factors
\begin{widetext}
\begin{align} \label{eq:PN_T1Ta_22}
T_{22}^{1,\text{log}} & = 1+ 6 i \left[2 \log(2) + \log(x)\right] x^{3/2} - \frac{2}{105} \left[856 \log(2) + 3780 \log(2)^2 + 214 \log(x) + 3780 \log(2) \log(x) + \right. \notag \\
& \left. + 945 \log(x)^2 \right] x^3 - \frac{4 i}{35} \left[1712 \log(2)^2 + 2520 \log(2)^3 + 1284 \log(2) \log(x) + 3780 \log(2)^2 \log(x) + 214 \log(x)^2 + \notag \right. \\
&\left. +1890 \log(2) \log(x)^2 +315 \log(x)^3\right] x^{9/2} + \mathcal{O}\left(x^6\right) \ , 
\end{align}
\begin{align}
T_{22}^{2,\text{log}} & = 32768 x^{15/2} - \frac{28049408}{35} \left[2 \log(2) + \log(x)\right] x^{21/2} + \frac{1048576}{385875} \left[-3390466 \log(2) + 14425740 \log(2)^2 + \right. \notag \\
& \left. - 1695233 \log(x) +14425740 \log(2) \log(x) + 3606435 \log(x)^2\right] x^{27/2} + \mathcal{O}\left(x^{33/2}\right) \ , \\
\nonumber \\
\tilde{T}_{22}^{\text{log}} & = 32 x^5 - \frac{54784}{105} \left[\log(2) + \log(x)\right] x^8 + \frac{2048}{1157625} \left[-1695233 \log(2) + 2404290 \log(2)^2 + \right. \notag \\
& \left. - 1695233 \log(x) +4808580 \log(2) \log(x) + 2404290 \log(x)^2\right] x^{11} + \mathcal{O}\left(x^{14}\right) \ , \\
\nonumber\\
T_{22}^{1,\text{trnsc}} & =\frac{1}{120}+\frac{1}{60} (2 i \gamma_E +\pi ) x^{3/2}+\frac{\left[29318-6300 \gamma_E^2+\gamma_E  (-6420+6300 i \pi )+3210 i \pi +525 \pi^2\right] x^3}{94500}+ \notag \\
& + \frac{1}{47250}\left[-4200 i \gamma_E^3+29318 \pi +5350 i \pi^2-525 \pi^3-60 \gamma_E^2 (214 i+105 \pi )+2 i \gamma_E  \left(29318+6420 i \pi + \right. \right. \notag \\
& \left. \left. + 525 \pi^2\right)+4200 i \varPsi^{(2)}(1)\right] x^{9/2} +\mathcal{O}\left(x^6\right) \ , \\
\nonumber \\
T_{22}^{2,\text{trnsc}} & =\frac{49}{732736 \, x^{3/2}}+\frac{7 i}{17120}+\left(\frac{121258243}{17640619200}-\frac{7 \gamma_E}{6420}+\frac{7 i \pi }{12840}\right) x^{3/2} + \left(\frac{423318901 \, i}{8655444000}-\frac{i \gamma_E}{150}-\frac{\pi }{300} +\frac{7 i \pi^2}{3210} \right) x^3+ \notag \\
& + \left(\frac{2423925578234843}{5722804300059000}-\frac{569063 \gamma_E}{4808580}+\frac{2 \gamma_E^2}{225}+\frac{569063 i \pi}{9617160}-\frac{2 i \gamma_E  \pi}{225} -\frac{7 \pi ^2}{450}+\frac{14 \, \varPsi ^{(2)}(1)}{535}\right) x^{9/2}+\mathcal{O}\left(x^{11/2}\right) \ ,\\
\nonumber \\
\tilde{T}_{22}^{\text{trnsc}} & = -\frac{7}{214}+\frac{149 i}{1070} x^{3/2} - \frac{5956078}{3606435} x^3 + \left(\frac{240034898 \, i}{30053625}-\frac{32 i \pi^2}{45} \right) x^{9/2} + \mathcal{O}\left(x^{6}\right)  \label{eq:PN_TtTr_22} \ .
\end{align}
It is also useful to report the PN expansion of the quantity $\lambda_{\rm inst}^{\rm NS}$, 
which turns out to be a polynomial in $x$, but it is important in order to understand
how the denominator of Eq.~\eqref{eq:ext_tail_def} and \eqref{eq:int_tail_def} start. 
Hence we always consider the (2,2) mode
\be
\lambda_{{\rm inst},\ell=m=2}^{\rm NS} = 1 - \frac{500}{49} x^3 - \frac{45098078528}{318087567} x^6 - \frac{2024036734584885289408}{953484714476442225} x^9 + \mathcal{O}(x^{12}) \ .
\ee
\end{widetext}
From these expressions we clearly see how, for this mode, the denominator of the
external tail \eqref{eq:ext_tail_def} starts to contribute from 6PN, while the internal
one \eqref{eq:int_tail_def} becomes relevant from 5PN. Notice also that $T_{22}$
in \eqref{eq:ext_tail_def} starts with 1 because $\Gamma(6)=120$ which cancels
the $1/120$ that appears inside $T^{1,\text{trnsc}}_{22}$.

\subsection{Refactorizing existing analytical information}
\label{sec:fromDIN}
Our approach based on the mapping of the Teukolsky equation into a CHE proposes a new way of
computing the circularized waveform emitted by a test-mass on a Schwarzschild black hole that can fully
replace the standard PN expansion in the MST formalism since certain expression (notably what we called the
tail and transcendental parts) are not PN-expanded ab initio, but are kept in closed form during the full procedure.
The functions $(z_\lm,\tilde{z}_\lm)$ can be obtained at (a priori) arbitrary PN accuracy (or even evaluated in closed 
form in terms of hypergeometric functions),
eventually yielding more analytical information than the one available with the standard MST approach. In this respect,
let us recall that the test-mass waveform (and fluxes) is currently known at {\it global} 22PN order~\cite{Fujita:2012cm} 
for the  Schwarzschild case\footnote{For the Kerr case we are limited to 11PN global accuracy in full analytic 
form~\cite{Fujita:2014eta}, although the knowledge was pushed up to 21.5PN with high-order analytical terms 
extracted from high-precision numerical data~\cite{Johnson-McDaniel:2015vva}.}.

By contrast, the well-known factorization approach to the waveform of DIN~\cite{Damour:2008gu} 
(see also Sec.~\ref{sec:DIN} below for a reminder) is conceptually different with respect to the one we 
discussed so far in that it: (i) starts from the PN-expanded waveform at a given global 
PN accuracy and (ii) factorizes closed form functions, like the tail factor and the source,
so to obtain residual amplitudes and phases for each waveform mode.
Since PN-results are available in the DIN-factorized form at high-order, i.e. 22PN for Schwarzschild~\cite{Fujita:2012cm}
and 11PN for Kerr~\cite{Fujita:2014eta} it is meaningful to wonder whether it is possible to simply 
recast them in the new factorized format, without the need of recomputing everything from
the beginning. Note in this respect that this information is available in DIN factorized form, with
the $(\rho_\lm^{\rm DIN},\delta_\lm^{\rm DIN})$ of Refs.~\cite{Fujita:2012cm,Fujita:2014eta}
freely available in electronic form \cite{Fujita:BHPC}.
It is possible to map these results in our form by simply equating the elements of the factorized waveforms
as
\be
\label{eq:main}
T^{\rm DIN}_\lm (\rho_\lm^{\rm DIN})^\ell e^{i\delta_\lm^{\rm DIN}}= T_\lm z_\lm\left[1+\tilde{T}_\lm\left(-\dfrac{\ell+1}{\ell}\right)^\epsilon\tilde{z}_\lm\right]
\ee
under the hypothesis that the $(z_\lm,\tilde{z}_\lm)$ are complex polynomials of the form
\begin{align}
z_{\lm} &= \sum_{n=0}^{N_z} b_n i^n \, x^{\frac{n}{2}} \ , \\
\tilde{z}_{\lm} &= \sum_{n=0}^{N_{\tilde{z}}} d_n i^n \, x^{\frac{n}{2}} \ ,
\end{align}
where $(N_z,N_{\tilde{z}})$ are the respective PN orders of the two functions (that a priori may differ)
and all coefficients $\{b_m, d_m \}$ are real. 
For pedagogical purposes, let us focus only on the $\ell=m=2$ mode. To determine $\tilde{z}_{22}$
at, say, 2PN order we need to start from DIN factorized results at 10PN order because the internal 
tail, $\tilde{T}_{22}$ starts at 5PN. As a result, $z_{22}$ can be fully determined at 7PN order. 
So, starting with DIN results at 10PN yields $N_z=14$ and $N_{\tilde{z}}=4$ because of the structure 
of the PN-expansions of $(T_{22},\tilde{T}_{22})$ explicitly shown above \footnote{Technically speaking, the system of equations in the coefficients $(b_m,d_m)$ can be solved only if $N_z = N_{\tilde{z}} = 14$, but at the end only the coefficients that impact at global 7 PN order must be retained, i.e. all the $d_m$ with $m=0,\dots,4$.}. By expanding Eq.~\eqref{eq:main} 
at 10PN one can  solve for the coefficients $(b_m,d_m)$ and eventually obtain

\begin{widetext}
\begin{align}
z_{22} & = 1-\frac{43}{21}x-\frac{17}{3} i x^{3/2}-\frac{536 x^2}{189}+\frac{731}{63} i x^{5/2}-\frac{201356 x^3}{14553}+\frac{9112}{567} i x^{7/2}+\frac{6107384702 x^4}{99324225}+\frac{635344 i x^{9/2}}{43659}+ \notag \\
& + \frac{1617025696474 x^5}{95649228675}-\frac{64867320634 i x^{11/2}}{297972675}-\frac{257475652476904 x^6}{1926228312855}+\frac{24471767655142 i x^{13/2}}{286947686025} + \notag \\
& -\frac{4258865347928184866512 x^7}{16235937670515688125} \ , \\
& \notag\\
\tilde{z}_{22} & = 1 + \frac{4391 x}{2247} + \frac{25}{3} i x^{3/2} + \frac{53185}{2646} x^2 \ ,
\end{align}
\end{widetext} 
that, once written as amplitude and phase, precisely coincide with Eqs.~\eqref{eq:rho22_test_mass}-\eqref{eq:delta22_test_mass}
at 7PN and Eqs.~\eqref{eq:ftilde_22} and \eqref{eq:delta_tilde_22_test_mass} at 2PN.
This is a practical algorithmic procedure to recast already available analytic information in the new 
factorized and resummed form without the need of ab-initio calculations. Although in this paper we will content 
ourselves to keep $(z_\lm,\tilde{z}_\lm)$ at 10PN-accuracy, that is enough for our purposes, if the needs come to go to 
higher order one can exploit already calculated information instead of resorting to the  full solution of the Heun equation.

\section{Comparable-mass binaries}
\label{sec:comparable_mass}
Building upon the knowledge gained in the test-mass case we can now propose a new
factorized and resummed EOB waveform that is valid also for comparable mass binaries.
We have that $M=m_1+m_2$ is the total mass of the system and we adopt the
convention that $m_1\geq m_2$. We define the reduced mass of the system as
$\mu\equiv m_1 m_2/M$ and the symmetric mass ratio as $\nu\equiv \mu/M$.
We use some dimensionless phase-space variables, $p_\varphi\equiv P_\varphi/(\mu M)$ 
and $u\equiv M/r$, where $r$ is the relative separation between the two objects.
To construct this new analytic proposal we blend together the (i) new tail and transcendental 
factors defined above with (ii) the factorized structure of the waveform of DIN for comparable 
mass binaries and (iii) the finding of ILPZ~\cite{Ivanov:2025ozg} 
that the renormalized angular momentum $\hat{\ell}$ can be generalized to comparable 
mass binaries using the new concept of universal anomalous dimension of the gravitational 
multipoles. Let us remember that ILPZ pointed out that from the anomalous dimension of BH
multipole moments introduced above, $\gamma_\lm^{\rm BH}$, Eq.~\eqref{eq:gamma_lm_BH}
one can compute an {\it universal anomalous dimension} $\gamma_\lm^{\rm univ}$ that 
is valid for a general system (e.g., like the case of a binary with comparable masses) that
is obtained suitably expanding the BH anomalous dimension around $\chi=0$ and then
replacing the mass of the black hole, $M$, with the energy of the system ${\cal E}$ and the
dimensionless black hole spin, $\chi$, with ${\cal J}\equiv P_\varphi/{\cal E}^2$, where 
$P_\varphi$ is addressing the orbital angular momentum of the system. 
The universal dimension (note that we reinserted the gravitational constant $G$) is then given by
\be
\label{eq:gamma_univ}
\gamma_\lm^{\rm univ}=\left[\gamma_{\lm}^{\rm BH}(G{\cal E}\omega,0)+\partial_{\chi}\gamma_\lm^{\rm BH}(G{\cal E}\omega,0){\cal J}\right]_{G^{2\ell +1}} \ ,
\ee
where $[\dots]_{G^{n}}$ denotes an expansion through order $n$.
Following in spirit the proposal of Ref.~\cite{Ivanov:2025ozg} we will then 
use the so-computed universal anomalous dimension to define the quantity
\be
\hat{\hat{\ell}}=\ell +\gamma_\lm^{\rm univ} \ .
\ee
We then propose to generalize the $\hl$ dependence of test-mass formulas
discussed above to the comparable mass case by the simple replacement
\be
\hat{\ell}\to \hat{\hat{\ell}} \ 
\ee
in the various tail (and transcendental) factors introduced above for the 
test-mass case, Eqs.~\eqref{eq:ext_tail_def}-\eqref{eq:tTlmTr}.

Let us introduce the various building blocks of the factorized waveform that
now incorporates all available corrections in $\nu$. It is intended that in 
the $\nu\to 0$ limit our formulas reduce to the test-mass ones discussed 
above. The PN correction to the multipolar waveform is factorized as
\be
\label{eq:hlm_com}
\hat{h}_\lm = \hat{S}^{(\epsilon)}_{\rm eff} T_\lm (\rho_\lm)^\ell e^{i\delta_\lm}\left[1+\tilde{T}_{\lm} \left(-\frac{\ell+1}{\ell}\right)^\epsilon \tilde{f}_\lm e^{i\tilde{\delta}_\lm}\right] \ ,
\ee
where  $\hat{S}^{(\epsilon)}_{\rm eff}$ is the {\it effective} source computed along a 
sequence of EOB circular orbits, i.e. either the effective EOB energy, when $\ell+m$ 
is even, $\epsilon=0$, or the Newton-normalized orbital angular momentum 
of the binary, when $\ell+m$ is odd, $\epsilon=1$. The effective EOB Hamiltonian for circular
orbits is defined as $\hat{H}_{\rm eff}=H_{\rm eff}/\mu=\sqrt{A(1+p_\varphi^2 u^2)}$, 
where $A$ is the full EOB (resummed) interaction potential that is taken at a given PN accuracy. 
The effective Hamiltonian is related to the real Hamiltonian of the system as
\be
H_{\rm real}=M\sqrt{1+2\nu(\hat{H}_{\rm eff}-1)} \ .
\ee
Circular orbits are determined  by the condition $\de_u[A(u)(1+p_\varphi^2 u^2)]=0$, 
which leads to the following representation of the squared angular momentum 
\be
\label{eq:pph_circ}
\left[p_{\varphi}^{\rm circ}(u)\right]^2= -\dfrac{A'(u)}{(u^2 A(u))'} \ ,
\ee
where the prime stands for $d/du$. From one of the Hamilton's equations we obtain 
the orbital frequency for circular orbits as function of $u$
\be
\label{eq:Omg_u}
M\Omega(u) = \dfrac{M A(u) p_{\varphi}^{\rm circ}(u) u^2}{E_{\rm real}\hat{E}_{\rm eff}} \ ,
\ee
where with $E_{\rm real}$ and $\hat{E}_{\rm eff}$ we indicate the real and effective
energies obtained evaluating the Hamiltonians $H_{\rm real}$ and $\hat{H}_{\rm eff}$ 
along the sequence of circular orbits characterized by the angular momentum 
of Eq.~\eqref{eq:pph_circ}. 
Along circular orbits we have $x\equiv (M\Omega)^{2/3}$ and from
Kepler's constraints, $\Omega^2 r^3=M$,  we have $x \equiv (M\Omega)^{2/3} = M/r_\Omega$,
where we indicate with $r_\Omega$ the orbital radius as related to the orbital frequency.
One can then invert numerically Eq.~\eqref{eq:Omg_u} so to obtain a parametric
relation between $x$ and $u$ and have $\hat{S}^{(\epsilon)}_{\rm eff}(x)$.
The tail factors $T_\lm$ and $\tilde{T}_\lm$ are now given by
\begin{align} 
& T_{\ell m} = \frac{\Gamma (2 \ell + 2)}{\Gamma (\ell - 1)}\frac{{T_{\ell m}^{1,\text{log}}} {T_{\ell m}^{1,\text{trnsc}}}}{1- \lambda_{\text{inst}}^{\rm NS}{T_{\ell m}^{2,\text{log}}} T_{\ell m}^{2,\text{trnsc}}} \ ,\label{eq:T_comp_mass}\\
& \tilde{T}_{\ell m} = - \lambda_{\text{inst}}^{\rm NS}{\tilde{T}_{\ell m}^{ \text{log}}}{\tilde{T}_{\ell m}^{\text{trnsc}}} \ ,
\end{align}
but with the $\hat{\ell}$'s replaced by $\hat{\hat{\ell}}$. To do so, we need the
universal anomalous dimension of multipole moments $\gamma_\lm^{\rm univ}$,
that can be obtained from the renormalized angular momentum (e.g. as obtained
using the MST method) using Eq.~\eqref{eq:gamma_univ} above.
Specifically, for the quadrupole we have~\cite{Ivanov:2025ozg}
\be 
\label{eq:gamma_2m_univ}
\gamma^{\rm univ}_{2m}=-\dfrac{214}{105}\hhk\,{\!^2}+ \dfrac{2m{\cal J}}{3}\hhk\,{\!^3}-\dfrac{3390466}{1157625}\hhk\,{\!^4} +\dfrac{381863m{\cal J}}{99225}\hhk\,{\!^5}\ ,
\ee
where ${\cal J}\equiv p_\varphi^{\rm circ}/E_{\rm real}^2$ and $\hhk\equiv E_{\rm real} m \Omega$, 
using here the notation introduced in Ref.~\cite{Damour:2008gu}. 
The various tail factors now read
\begin{align}
{T_{\ell m}^{1,\text{log}}} &= e^{(\hhl-\ell)\log(2 k r_{\Omega})} e^{2 i \hhk \log \left(4 M k\right)} \ ,\\
{T_{\ell m}^{2,\text{log}}} &=e^{(2\hhl +1)\log(4 M k)} \ ,\\
{\tilde{T}_{\ell m}^{\text{log}}} &= e^{(2\hhl +1)\log\left(\frac{2 M}{r_{\Omega}}\right)} \ ,
\end{align}
where we defined $k\equiv m\Omega$. Let us recall that Ref.~\cite{Damour:2008gu} introduced the 
use of $\hhk$ to promote the tail factor found in the test-mass to the comparable mass case 
in order to take into account tail effects linked to the propagation of the multipolar wave in a 
Schwarzschild background of mass $M_{\rm ADM}=E_{\rm real}$~\cite{Damour:2008gu,Faye:2014fra}. 
In the equations above we follow precisely the same rationale to rise our new test-mass tail
factor to the comparable mass case. Similarly, the test-mass transcendental factors now read
\begin{align}
&\textcolor{black}{T_{\ell m}^{1,\text{trnsc}}} = {\frac{\Gamma \left(\hhl - 1 - 2 i \hat{\hat{k}} \right)}{\Gamma \left(2 \hhl+ 2\right)} e^{\pi \hat{\hat{k}}} e^{-i\frac{\pi}{2} \left(\hhl -\ell\right)}} \ ,\\
&{T_{\ell m}^{2,\text{trnsc}}}  ={e^{-i \left(\hhl + \frac{1}{2}\right) \pi}} \times \\ 
&{\times\frac{ \Gamma(-2\hhl)^2 \Gamma \left(\hhl + 3 - 2 i \hat{\hat{k}} \right) \Gamma \left(\hhl + 1 - 2 i \hat{\hat{k}} \right) \Gamma \left(\hhl - 1 - 2 i \hat{\hat{k}} \right)}{\Gamma(2\hhl+2)^2 \Gamma \left(2 - \hhl - 2 i \hhk \right) \Gamma \left(-\hhl - 2 i \hhk \right) \Gamma \left(-2 - \hhl - 2 i \hhk \right)}} \ ,\nonumber\\
& {\tilde{T}_{\ell m}^{\text{trnsc}}}  =   {\frac{\Gamma\left(-2\hhl\right) \Gamma \left(\hhl + 3 - 2 i \hhk \right) \Gamma \left(\hhl + 1 - 2 i \hhk \right)}{\Gamma(2 \hhl + 2) \Gamma \left(2 - \hhl - 2 i \hat{\hat{k}}\right) \Gamma \left(-\hhl - 2 i \hhk \right)}}  \ .
\end{align}
We are now left to determine the residual amplitudes and phases in Eq.~\eqref{eq:hlm_com}. 
We do so by factoring out the PN-expanded waveform multipole by multipole. For the $\ell=m=2$ multipole
it is fully known at 4PN accuracy~\cite{Marchand:2020fpt,Larrouturou:2021dma,Larrouturou:2021gqo,Trestini:2022tot,Blanchet:2022vsm,Trestini:2023wwg,Blanchet:2023sbv,Blanchet:2023bwj}, while subdominant modes are known up to (global) 3.5PN accuracy~\cite{Blanchet:2008je,Faye:2014fra,Henry:2021cek,Henry:2022ccf}. 
For convenience the analytically known PN-expanded $\hat{h}_\lm$'s are listed in Appendix~\ref{app:PN_exp_hhat}.
Because of the global 4PN accuracy of the train waveform, in the factorization we only have to take 
into account the external tail $T_\lm$, since the internal one starts at 5PN order. 
Furthermore, since the denominator of Eq.~\eqref{eq:T_comp_mass} contributes from 6PN, we don't need to consider 
it, so in the following we are going to denote as $T_{\lm}$ only $\Gamma(2 \ell +2)/\Gamma(\ell-1) T^{1,\rm log}_{\lm} T^{1,\rm trnsc}_{\lm}$. 
We report here explicitly the factorization of the $\ell=m=2$  mode, while all others, up to the currently 
known $\ell=5$ ones, are listed in Appendix~\ref{app:rholm_flm_PN}. 
The $\ell=m=2$ PN-correcting factor at 4PN accuracy, expressed in radiative 
coordinates~\cite{Blanchet:2022vsm,Blanchet:2023sbv,Blanchet:2023bwj}, reads\footnote{Note that following~\cite{Blanchet:2023bwj} the argument 
$x$ should be half the gravitational wave frequency, while, as in Ref.~\cite{Nagar:2024dzj} we approximate it with the orbital frequency, neglecting
hereditary effects (see Ref.~\cite{Trestini:2025nzr} for details). We made this choice for consistency with our previous work. This choice does not 
change conceptually the results presented here, though the EOB waveform models of Refs.~\cite{Nagar:2024dzj,Nagar:2024oyk} will have to be 
correctly updated accordingly.}
\begin{widetext}
\begin{align}
\label{eq:h224PN}
\hat{h}_{22}^{\rm 4PN}&= 1 + \left(-\dfrac{107}{42}+\dfrac{55}{42}\nu\right)x + 2\pi x^{3/2} + \left(-\dfrac{2173}{1512}-\dfrac{1069}{216}\nu+\dfrac{2047}{1512}\nu^2\right)x^2
+\left[-\dfrac{107\pi}{21}+\left(\dfrac{34\pi}{21}-24i\right)\nu\right]x^{5/2} + \nonumber \\
&+\left[\dfrac{27027409}{646800} - \dfrac{856 \gamma_E}{105}+\dfrac{428 i \pi}{105} + \dfrac{2 \pi^2}{3} - \left(\dfrac{278185}{33264} - \frac{41 \pi^2}{96}\right) \nu  - \dfrac{20261}{2772} \nu^2 + \dfrac{114635}{99792} \nu^3 - \dfrac{428}{105} \log(16 x)\right]x^3 + \nonumber\\
&+\left[- \dfrac{2173\pi}{756} + \left(- \dfrac{2495\pi}{378}+ 
     \dfrac{14333i}{162}\right) \nu + \left(\dfrac{40\pi}{27} - \dfrac{4066 i}{945} \right)\nu^2\right]x^{7/2} + \left[- \dfrac{846557506853}{12713500800} + \dfrac{45796 \gamma_E}{2205} - \dfrac{22898 i \pi}{2205} + \right. \nonumber \\
& - \dfrac{107 \pi^2}{63} + \left(\dfrac{256450291}{7413120}  - \dfrac{1025 \pi^2}{1008}\right) \nu^2 - \dfrac{81579187}{15567552} \nu^3 + \dfrac{26251249}{31135104} \nu^4 + \dfrac{22898}{2205} \log(16 x) + \left(-\dfrac{336005827477}{4237833600} + \right.\nonumber \\
& \left. \left. + \dfrac{15284 \gamma_E}{441} - \dfrac{219314 i \pi}{2205} - \dfrac{9755 \pi^2}{32256} + \dfrac{7642}{441} \log(16 x)\right) \nu \right] x^4 \ .
\end{align}
\end{widetext}
We use the following 4PN accurate energy and 4PN accurate angular momentum along circular orbits
\begin{widetext}
\begin{align}
\dfrac{E^{\text{4PN}}_{\rm real}}{M}& = 1-\frac{1}{2} \nu  x \left\{1 + \left(-\frac{3}{4}-\frac{\nu}{12}\right) x + \left(-\frac{27}{8} + \frac{19}{8} \nu - \frac{\nu^2}{24} \right) x^2 + \left[-\frac{675}{64} + \left(\frac{34445}{576} - \frac{205}{96}\pi^2\right) \nu - \frac{155}{96} \nu^2 + \right. \right. \notag \\
& \left. \left. - \frac{35}{5184} \nu^3 \right] x^3 + \left[-\frac{3969}{128} + \left(-\frac{123671}{5760} + \frac{9037}{1536} \pi^2 + \frac{896}{15} \gamma_E + \frac{448}{15} \log(16 x)\right)\nu + \left(-\frac{498449}{3456} + \frac{3157}{576} \pi^2\right) \nu^2 +\right. \right. \notag \\
& \left. \left. + \frac{301}{1728} \nu^3 + \frac{77}{31104} \nu^4\right] x^4\right\} \ ,
\end{align}
\begin{align}
p_{\varphi,{\rm circ}}^{\rm 4PN}& = \frac{1}{\sqrt{x}} \left\{1+ \left(\frac{3}{2} + \frac{\nu}{6}\right) x + \left(\frac{27}{8} - \frac{19}{8} \nu + \frac{\nu^2}{24}\right) x^2 + \left[\frac{135}{16} + \left(-\frac{6889}{144} + \frac{41}{24}\pi^2\right) \nu + \frac{31}{24} \nu^2 + \right. \right. \notag \\
& \left. \left. + \frac{7}{1296} \nu^3 \right] x^3 + \left[\frac{2835}{128} + \left(\frac{98869}{5760} - \frac{128}{3} \gamma_E - \frac{6455}{1536} \pi^2 - \frac{64}{3} \log(16 x)\right)\nu + \left(\frac{356035}{3456} - \frac{2255}{576} \pi^2\right) \nu^2 + \right. \right. \notag \\
& \left. \left. - \frac{215}{1728} \nu^3 - \frac{55}{31104} \nu^4\right] x^4\right\} \ ,
\end{align}
\end{widetext}
and from them we compute $\hat{S}^{(0)}_{\rm eff}=E_{\rm eff}^{\rm 4PN}$ and $\hat{S}^{(1)}_{\rm eff}=\sqrt{x}p_{\varphi,{\rm circ}}^{\rm 4PN}$
to be factored out. For what concerns the tail, there is one important subtlety we need to face correctly,
i.e. the fact that $\hat{h}_{\lm}^{\rm 4PN}$ is expressed using radiative coordinates, while the test-mass 
expressions presented above were obtained in Schwarzschild coordinates.
These two coordinate systems differ for the choice of the origin of time and 
this translates into a difference in the overall phase of the waveform.
In radiative coordinates\footnote{Note that the definition of the Newtonian factor $h_\lm^{N, {\rm radiative}}$ 
follows Ref.~\cite{Faye:2014fra} that omits the factor $e^{-im\varphi}$.} we have
\be
h_\lm = h_\lm^{N, {\rm radiative}} \, \hat{h}_\lm^{\rm radiative} \, e^{-i m \psi} \ ,
\ee
and, according to~\cite{Faye:2014fra}, the change from the Schwarzschild coordinates to the radiative ones 
implies that the phase variable $\psi$ is related to the actual orbital phase of the binary $\varphi$ by
\be
\psi = \varphi - 2 E_{\rm real} \Omega \log\left(\dfrac{\Omega}{\Omega_0}\right) \ ,
\ee
where $\Omega_0$ is a constant frequency related to a length scale in a way we are about to see. 
$\hat{h}_{\lm}^{\rm radiative}$ is the normalized waveform in radiative coordinates. By 
inserting the term $e^{- i m \varphi}$ inside the Newtonian-part in radiative coordinates and 
$2 E_{\rm real} \Omega \log\left(\dfrac{\Omega}{\Omega_0}\right)$ inside the PN-expanded waveform, 
we obtain the ingredients in the Schwarzschild coordinates of the previous Section
\be
h_\lm = h_\lm^{N, {\rm Schwarzschild}} \, \hat{h}_\lm^{\rm Schwarzschild} \ ,
\ee
meaning that
\be
\hat{h}_\lm^{\rm Schwarzschild} = \hat{h}_\lm^{\rm radiative} e^{2 i E_{\rm real} m \Omega \log \left(\Omega/\Omega_0\right)} \ .
\ee
In order to match the PN results of $\hat{h}_\lm^{\rm Schwarzschild}$ and $\hat{h}_\lm^{\rm radiative}$ 
known in literature, we have to fix
\be \label{eq:omega_0_value}
\Omega_0 = \dfrac{e^{\frac{11}{12}-\gamma_E}}{4 r_0} \ ,
\ee
where, following Ref.~\cite{Fujita:2010xj,Pan:2010hz}, the scale $r_0$ is fixed to $r_0 = 2 M / \sqrt{e}$.
At this point, using the phase factor $e^{2 i E_{\rm real} \Omega \log \left(\Omega/\Omega_0\right)}$ 
in front the tail part, we can write
\be \label{eq:transf_tail_factors}
T_{\lm}^{\rm Schwarzschild} = T_{\lm}^{\rm radiative} e^{2 i E_{\rm real} m \Omega \log \left(\Omega/\Omega_0\right)} \ .
\ee
This is valid for all the factorization proposals we are going to investigate also in the following
Sections. From all this reasoning it is clear that the residual phase $\delta_\lm$ is the same in
the two coordinates systems. In our scenario, using for $T_{\lm}^{\rm Schwarzschild}$ what
is written in \eqref{eq:T_comp_mass} with all the simplifications due to the fact we are at 4PN, i.e.
\begin{align}
T_{\lm}^{\rm Schwarzschild} =&=\dfrac{\Gamma(2\ell+2)}{\Gamma(\ell-1)}\dfrac{\Gamma(\hhl-1-2i\hhk)}{\Gamma(2\hhl+2)}\dfrac{}{}e^{\pi\hhk}(2k r_{\Omega})^{\hhl - \ell}\nonumber\\
&\times e^{2i\hhk\log(4 M m \Omega)} e^{-i\frac{\pi}{2}(\hhl-\ell)} \ ,
\end{align}
we obtain that
\begin{align} \label{eq:our_tail_comp_mass}
T^{\rm radiative}_\lm &=\dfrac{\Gamma(2\ell+2)}{\Gamma(\ell-1)}\dfrac{\Gamma(\hhl-1-2i\hhk)}{\Gamma(2\hhl+2)}\dfrac{}{}e^{\pi\hhk}(2k r_{\Omega})^{\hhl - \ell}\nonumber\\
&\times e^{2i\hhk\log(4 M m \Omega_0)} e^{-i\frac{\pi}{2}(\hhl-\ell)} \ ,
\end{align}
and using the expression of $\Omega_0$ coming from \eqref{eq:omega_0_value}, we have
\begin{align}
\label{eq:Tlm_ext}
T^{\rm radiative}_\lm &=\dfrac{\Gamma(2\ell+2)}{\Gamma(\ell-1)}\dfrac{\Gamma(\hhl-1-2i\hhk)}{\Gamma(2\hhl+2)}\dfrac{}{}e^{\pi\hhk}(2k r_{\Omega})^{\hhl - \ell}\nonumber\\
&\times e^{2i\hhk\log(2m \phi_0)}e^{-i\frac{\pi}{2}(\hhl-\ell)} \ ,
\end{align}
where the value of $\phi_0$ is
\be \label{eq:new_phase}
\phi_0=\dfrac{e^{\frac{17}{12}-\gamma_E}}{4} \ .
\ee
With this choice, the residual phase extracted from the (radiative-coordinates) 
4PN waveform coincides with the one obtained in the test-mass limit that we computed above.
The residual amplitude at 4PN then reads
\begin{widetext}
\begin{align}
\label{eq:rho22}
\rho_{22} & = 1+\left(-\frac{43}{42}+\frac{55 \nu }{84}\right)x+\left(-\frac{20555}{10584}-\frac{33025 \nu }{21168}+\frac{19583}{42336}\nu^2\right)x^2+ \Bigg[-\frac{4296031}{4889808}+\left(\frac{41 \pi^2}{192}-\frac{48993925}{9779616}\right) \nu \nonumber\\
&-\frac{6292061}{3259872}\nu^2+\frac{10620745}{39118464}\nu^3\Bigg]x^3+x^4 \Bigg[\frac{9228174993589}{800950550400}+\nu \left(-\frac{2487107795131}{145627372800}-\frac{9953\pi ^2}{21504}+\dfrac{464}{35}{\rm eulerlog}_2(x)\right) \notag \\
& +\left(\frac{10815863492353}{640760440320}-\frac{3485 \pi ^2}{5376}\right) \nu^2-\frac{2088847783}{11650189824} \nu^3 +\frac{70134663541}{512608352256}\nu^4\Bigg] \ ,
\end{align}
\end{widetext}
where $\text{eulerlog}_m(x)\equiv \gamma_E + \log(2m\sqrt{x})$, and the residual phase is
\begin{align}
\label{eq:delta22}
\delta_{22}&=-\dfrac{17}{3}y^{3/2}-24\nu y^{5/2} \nonumber\\
&+ \left(\dfrac{30995}{1134}\nu+\dfrac{962}{135}\nu^2\right)y^{7/2} -\nu\dfrac{4976}{105}\pi y^4\ ,
\end{align}
with $y=(E_{\rm real}\Omega)^{2/3}$. The $\nu=0$ limit of this function coincides with 
Eq.~\eqref{eq:rho22_test_mass}-\eqref{eq:delta22_test_mass}. Note that $\nu$-dependent logs are still present in 
Eq.~\eqref{eq:rho22} as well as $\nu$-dependent terms that are proportional to $\pi^2$.
As pointed out in Ref.~\cite{Ivanov:2025ozg}, the residual log come from tail of memory 
effect that are not universal and not present in the test-mass case.

Let us finally stress that our procedure, though inspired by ILPZ, is technically different as it stems
from the $\hat{\ell}$-dependence found in the test-mass solution. In ILPZ, they proposed to replace 
all $\ell$'s appearing in the prefactor resumming the infinite number of leading (infrared) logarithms into $\hhl$. 
In our case, instead, we systematically promote each $\hat{\ell}$ to $\hhl$, as it seems a rather natural
procedure, without modifying the $\Gamma(2\ell +2)/\Gamma(\ell-1)$ factor in Eq.~\eqref{eq:T_comp_mass}.
In this respect, in Eq.~\eqref{eq:Tlm_ext} we could have used the combination $\Gamma(\hl-1-2i\hhk)/\Gamma(2\hl+2)$
instead. When this is done, one finds that all terms in $\rho_{22}$ are unchanged except for the
\be
\dfrac{464}{35}\text{eulerlog}_2(x)=\dfrac{464}{35}\left(\gamma_E+2\log(2)+\dfrac{1}{2}\log(x)\right) \ ,
\ee 
term, that is instead replaced  by
\be
\dfrac{52}{3}\gamma_E + \dfrac{928}{35}\log(2) + \dfrac{232}{35}\log(x) \ .
\ee 
This results suggests that the replacement $\hl\to\hhl$ in the $\Gamma$-dependent factor
seems a good practice, because it yields  a coefficient of $\gamma_E$ such to reconstruct 
the full eulerlog$_2(x)$ function.

\subsection{Comparison with Damour-Iyer-Nagar}
\label{sec:DIN}
To have a better understanding of the effect of the new factorized terms it is useful to compare $(\rho_{22},\delta_{22})$
of above with those obtained with the standard DIN factorization~\cite{Damour:2007xr,Damour:2008gu,Fujita:2010xj,Pan:2010hz}.
The result of this calculation was reported in Ref.~\cite{Nagar:2024dzj} since the 4PN-accurate $\rho_{22}^{\rm DIN}$ is
already incorporated within the state-of-the-art waveform model {\tt TEOBResumS-Dal\'i}~\cite{Nagar:2024dzj,Nagar:2024oyk}
as well as in the LEOB-model~\cite{Damour:2025uka} for coalescing black hole binaries.
Here we repeat in detail the calculation of~\cite{Nagar:2024dzj} for completeness.
The DIN factorized waveform is formally written as
\be
\label{eq:hlmDIN}
\hat{h}_\lm=\hat{S}_{\rm eff}^{(\epsilon)}T_{\ell m}^{\rm DIN}(\rho_{\ell m}^{\rm DIN})^\ell e^{i\delta^{\rm DIN}_\lm} \ ,
\ee
where the superscript DIN allows us to distinguish these functions from the new ones. In particular, the tail factor
$T_{\lm}^{\rm DIN}$ reads
\be
\label{eq:Tdin}
T_\lm^{\rm DIN}=\dfrac{\Gamma(\ell+1-2i\hhk)}{\Gamma(\ell+1)}e^{\pi \hhk}e^{2i\hhk\log(2kr_0)} \ ,
\ee
where $r_0=2M/\sqrt{e}$~\cite{Fujita:2010xj}. As above, to match the comparable-mass results in 
radiative coordinates with the ones in the test-mass limit, in Schwarzschild coordinates, 
the tail factor we have to factor out from Eq.~\eqref{eq:hlmDIN} is obtained from the previous one
in Eq.~\eqref{eq:Tdin} via Eq.~\eqref{eq:transf_tail_factors} with $\Omega_0$ given by \eqref{eq:omega_0_value}
\be
T_\lm^{\rm DIN, radiative}=\dfrac{\Gamma(\ell+1-2i\hhk)}{\Gamma(\ell+1)}e^{\pi\hhk}e^{2i\hhk\log(2m \phi_0^{\rm DIN})} \ ,
\ee 
where the phase here is 
\be \label{eq:phase_DIN}
\phi_0^{\rm DIN}= \frac{e^{\frac{11}{12}-\gamma_E}}{4} \ ,
\ee
as pointed out in Ref.~\cite{Faye:2014fra}.
One eventually gets the following residual amplitude $\rho_{22}^{\rm DIN}$
\begin{widetext}
\begin{align}
\label{eq:rho_DIN}
\rho_{22}^{\rm DIN}(x)&=1+\left(-\dfrac{43}{42}+\dfrac{55}{84}\nu\right)x+\left(-\dfrac{20555}{10584}-\dfrac{33025}{21168}\nu+\dfrac{19583}{42336}\nu^2\right)x^2\nonumber\\
&+\bigg[\dfrac{1556919113}{122245200}-\frac{428}{105}\text{eulerlog}_2(x)
+\left(\frac{41 \pi^2}{192}-\frac{48993925}{9779616}\right)\nu-\frac{6292061}{3259872}\nu^2+\frac{10620745}{39118464}\nu^3\bigg]x^3 \nonumber\\
&+\bigg[-\frac{387216563023}{160190110080}+\frac{9202}{2205}\text{eulerlog}_2(x)+\left(-\frac{6718432743163}{145627372800}-\frac{9953 \pi ^2}{21504}+\frac{8819}{441}\text{eulerlog}_2(x)\right)\nu\nonumber\\
&+\left(\frac{10815863492353}{640760440320}-\frac{3485 \pi ^2}{5376}\right)\nu^2 -\frac{2088847783}{11650189824}\nu^3+\frac{70134663541}{512608352256} \nu ^4\bigg]x^4 \ .
\end{align}
\end{widetext}
The residual phase reads:
\begin{align}
\delta_{22}^{\rm DIN}&=\dfrac{7}{3}y^{3/2}-24\nu y^{5/2} + \dfrac{428}{105}\pi y^3 \nonumber\\
                 &+ \left(\dfrac{30995}{1134}\nu+\dfrac{962}{135}\nu^2\right)y^{7/2} -\dfrac{5536}{105}\pi\nu y^4 \ ,
\end{align}
with $y = (E_{\rm real} \Omega)^{2/3}$. 
Comparing Eq.~\eqref{eq:rho_DIN} with Eq.~\eqref{eq:rho22} one sees that, on top of the obvious
simplification  of the test-mass contributions, all $\nu$-dependent terms  are unchanged {\it except}
the coefficient of $\nu\,\text{eulerlog}_2(x)$ that from $8819/441\sim 19.997$ is reduced by less
of a factor two to $464/35\sim 13.26$. For what concerns the residual phases $\delta_{22}$ 
and $\delta_{22}^{\rm DIN}$, they differ in the contributions proportional to $\pi$ because of the 
factor $e^{i\frac{\pi}{2}(\ell-\hhl)}$ that is absent in DIN and that thus yields the absence of 
the test-mass term in Eq.~\eqref{eq:delta22}. In Appendix~\ref{sec:DIN_ILPZ}
we report the expression of $\rho_{22}^{\rm DIN}$ up to 10 PN, where from 5PN order we only have
test-mass terms.

\section{Revisiting the factorization of Ivanov et al.~\cite{Ivanov:2025ozg}, ILPZ}
\label{sec:ivanov_et_al}
Now that we have discussed our new waveform factorization proposal let us go back to the
ILPZ proposal~\cite{Ivanov:2025ozg}. As mentioned above, ILPZ has the merit of having pointed 
out, for the first time, the possibility of resumming the universal (test-mass) logs in the waveform 
amplitudes using the RG evolution as well as of introducing the concept of universal anomalous 
dimension of multipole moment, that we used above.
However, we are going to show that the factorization proposed there (with some minor 
modifications) turns out to be less powerful than the one we presented above: when 
analyzing the test-mass limit, one finds that the transcendental  numbers appear again 
starting from 5PN order, while the $\log(x)$ dependence pops up again at 8PN. 

It is thus useful to critically revisit the ILPZ proposal. The waveform is factorized as
\be
\hat{h}_{\lm}=\hat{S}_{\rm eff}^{(\epsilon)}T_\lm^{\rm ILPZ} (\rho_\lm^{\rm ILPZ})^\ell e^{i\delta_\lm^{\rm ILPZ}} \ ,
\ee
where the tail factor, expressed in  radiative coordinates, is written as
\begin{align} \label{eq:tail_parra_comp_mass}
T_\lm^{\rm ILPZ, radiative}& =\dfrac{\Gamma(\hhl+1-2i\hat{\hat{k}})}{\Gamma(\hhl+1)}e^{\pi\hat{\hat{k}}}e^{2i\hat{\hat{k}}\log(2 m \phi_0^{\rm ILPZ})} \notag \\
& \times e^{-i\frac{\pi}{2}\left(\hhl-\ell\right)}(\alpha k r_{\Omega})^{\hhl-\ell} \ ,
\end{align}
with $k=m\Omega$, $\hat{\hat{k}}=E_{\rm real}m\Omega$, $\phi_0^{\rm ILPZ} = \phi_0^{\rm DIN} = e^{11/12 - \gamma_E}/4$ 
and\footnote{In ILPZ, Ref.~\cite{Ivanov:2025ozg}, there is a typo in the argument of the logarithm at the exponent of the tail 
factor, both in their Eq.~(31), where they recall the DIN factorization, and in their Eq.~(33). In Schwarzschild coordinates 
the correct argument is $2 k r_0$, in the radiative ones it is $2 m \phi_0^{\rm DIN, ILPZ}$. 
We thank J.~Parra-Martinez for confirming this aspect.} $\alpha$ a constant  to be fixed. 
In particular, in ILPZ it was fixed to $(m \phi_0^{\rm ILPZ})^{-1}$, but the EFT procedure employed there does not allow 
to uniquely determine it. For this reason, inspired from our result in Eq.~\eqref{eq:our_tail_comp_mass}, we have restored a general constant, 
expected to be different from $2$ due to the fact that the ratio of $\Gamma$ functions between Eq.~\eqref{eq:tail_parra_comp_mass} and 
Eq.~\eqref{eq:our_tail_comp_mass} is different. 
This constant can be fixed arbitrarily, so we proceed as follows.
We start with the $(2,2)$ mode. The residual amplitude at 4PN  turns out to be
\begin{widetext}
\begin{align}
\rho_{22}^{\rm ILPZ}(x)&=1+\left(-\dfrac{43}{42}+\dfrac{55}{84}\nu\right)x+\left(-\dfrac{20555}{10584}-\dfrac{33025}{21168}\nu+\dfrac{19583}{42336}\nu^2\right)x^2\nonumber\\
&+\bigg[\dfrac{1556919113}{122245200} +\left(\frac{41 \pi^2}{192}-\frac{48993925}{9779616}\right)\nu-\frac{6292061}{3259872}\nu^2+\frac{10620745}{39118464}\nu^3 -\frac{428}{105} \left(\gamma_E + \log \left(\frac{2}{\alpha}\right) \right) \bigg]x^3\nonumber\\
&+\left[-\frac{387216563023}{160190110080}+\frac{9202 \gamma_E}{2205} + \left(\frac{10815863492353}{640760440320}-\frac{3485 \pi^2}{5376}\right) \nu^2-\frac{2088847783 \nu^3}{11650189824}+\frac{70134663541 \nu ^4}{512608352256}+ \right. \notag \\
& + \frac{9202 \log(2)}{2205}+\nu  \left(-\frac{6718432743163}{145627372800}+\frac{8819 \gamma_E}{441}-\frac{9953 \pi ^2}{21504}+\frac{73327 \log(2)}{2205}+\frac{232 \log (x)}{35}-\frac{14863 \log (\alpha)}{2205}\right) + \notag \\
& \left. -\frac{9202 \log (\alpha )}{2205}\right]x^4 \ ,
\end{align}
\end{widetext}
and we choose to fix $\alpha$ in such a way that no transcendental, nor logarithmic factors,
explicitly appear at 3PN, which yields
\be
\gamma_E + \log(2) - \log(\alpha) = 0 \Rightarrow \alpha = 2 e^{\gamma_E} \ .
\ee
Since this constant is different with respect to the one originally computed by ILPZ 
we indicate the corresponding quantities with a tilde. Thus $\tilde{\rho}_{22}^{\rm ILPZ}$ 
at 8PN accuracy (see Appendix~\ref{sec:DIN_ILPZ} for 10PN)  reads
\begin{widetext}
\begin{align}
&\tilde{\rho}_{22}^{\rm ILPZ}=1-\left(\dfrac{43}{42}+\dfrac{55}{84}\nu\right)x+\left(-\dfrac{20555}{10584}-\dfrac{33025}{21168}\nu+\dfrac{19583}{42336}\nu^2\right)x^2 + \nonumber\\
&+ \left[\dfrac{1556919113}{122245200}+\left(\dfrac{41\pi^2}{192}-\dfrac{48993925}{9779616}\right)\nu-\dfrac{6292061}{3259872}\nu^2+\dfrac{10620745}{39118464}\nu^3\right]x^3 + \nonumber\\
&+ \bigg[-\frac{387216563023}{160190110080}+\left(\frac{10815863492353}{640760440320}-\frac{3485 \pi ^2}{5376}\right) \nu ^2-\frac{2088847783 \nu ^3}{11650189824}
+\frac{70134663541 \nu ^4}{512608352256} + \nonumber\\
&+\nu \left(-\frac{6718432743163}{145627372800}-\frac{9953 \pi^2}{21504}+\frac{464}{35}{\rm eulerlog}_2(x)\right)\bigg]x^4
-\dfrac{16094530514677 }{533967033600}x^5 + \nonumber\\
&+\left(\dfrac{313425353036319023287}{1132319812111488000}-\dfrac{91592}{11025}\pi^2-\dfrac{3424}{105}\varPsi^{(2)}(3)-\dfrac{6848}{105}\zeta(3)\right)x^6 + \nonumber\\
&+\left(-\dfrac{38460677967545998977786359}{411134000579560177920000} 
+ \dfrac{1969228}{231525}\pi^2 + \dfrac{73616}{2205}\varPsi^{(2)}(3) 
+ \dfrac{147232}{2205}\zeta(3)\right)x^7 + \nonumber\\
&+\left(-\dfrac{15305094710902555724554334903}{24377827799070391726080000}+\dfrac{47066839}{2917215}\pi^2-\dfrac{128}{15} \log(2 x)
+\dfrac{1759508}{27783}\varPsi^{(2)}(3) + \dfrac{3519016 }{27783}\zeta(3)\right)x^8 \ ,
\end{align}
\end{widetext}
where the test-mass logs appear again at 8PN and $\varPsi$ indicate the polygamma function. At this stage we stress that the 
rational numbers that are independent of $\nu$, $\pi$, logarithms and other transcendental terms are the same as the DIN ones.
This is valid only up to 8 PN and we can use the fact that $\varPsi^{(2)}(3)$ and $\zeta(3)$ are related. In particular 
\be
\varPsi^{(2)}(3) = \frac{9}{4} - 2 \zeta(3) \ ,
\ee
and by doing so we arrive at
\begin{widetext}
\begin{align}
&\tilde{\rho}_{22}^{\rm ILPZ}=1-\left(\dfrac{43}{42}+\dfrac{55}{84}\nu\right)x+\left(-\dfrac{20555}{10584}-\dfrac{33025}{21168}\nu+\dfrac{19583}{42336}\nu^2\right)x^2 + \nonumber\\
&+ \left[\dfrac{1556919113}{122245200}+\left(\dfrac{41\pi^2}{192}-\dfrac{48993925}{9779616}\right)\nu-\dfrac{6292061}{3259872}\nu^2+\dfrac{10620745}{39118464}\nu^3\right]x^3 + \nonumber\\
&+ \bigg[-\frac{387216563023}{160190110080}+\left(\frac{10815863492353}{640760440320}-\frac{3485 \pi ^2}{5376}\right) \nu ^2-\frac{2088847783 \nu ^3}{11650189824}
+\frac{70134663541 \nu ^4}{512608352256} + \nonumber\\
&+\nu \left(-\frac{6718432743163}{145627372800}-\frac{9953 \pi^2}{21504}+\frac{464}{35}{\rm eulerlog}_2(x)\right)\bigg]x^4
-\dfrac{16094530514677 }{533967033600}x^5 + \nonumber\\
&+\left(\dfrac{230345430821967560887}{1132319812111488000}-\dfrac{91592}{11025}\pi^2\right)x^6 +\left(-\dfrac{7576963083194058102522359}{411134000579560177920000} + \dfrac{1969228}{231525}\pi^2 \right)x^7 + \nonumber\\
&+\left(-\dfrac{11831416136632492005314654903}{24377827799070391726080000}+\dfrac{47066839}{2917215}\pi^2-\dfrac{128}{15} \log(2 x) \right)x^8 \ .
\end{align}
\end{widetext}
From this expression one sees that: (i) up to 5PN order included the $\nu=0$ contributions are fully rational; 
(ii) the logarithms are still absent up to 7PN included, though transcendental numbers are found; 
(iii) at 8PN the $\log(x)$ appears again but not squared as it was the case of $\rho_{22}^{\rm DIN}$.
The structure of the function remains analogous for higher modes always with the choice of $\alpha = 2 e^{\gamma_E}$, with $\log(x)$ 
and transcendental numbers appearing after a certain PN order, see Appendix~\ref{sec:DIN_ILPZ}. 
Note however that the (nonuniversal) $\log(x)$-term appear at 8PN only for the $\ell=2$ modes, 
while for $\ell=3$ they are present starting from 10PN.
The use of test-mass PN knowledge allows us to complement the finding of ILPZ  and to state clearly 
that at higher PN the transcendental complexity of the residual amplitudes increases. 
This in the end shows that the factorization procedure, though being an improvement with respect to DIN, 
is actually suboptimal with respect to the one discussed in previous sections. 
For completeness, we also quote the corresponding residual phase
\begin{align}
\tilde{\delta}_{22}^{\rm ILPZ} &= \frac{7}{3} y^{3/2}-24 \nu y^{5/2} \nonumber\\
&+ \left(\frac{30995}{1134} \nu + \frac{962}{135} \nu^2 \right) y^{7/2} - \frac{4976}{105} \pi \nu y^4 \ .
\end{align}

\section{Performance: flux  and waveform for circular orbits}
\label{sec:validation}
\subsection{Test-mass}
Let us turn now to evaluate the performance of our new factorized waveform. We do so first in the 
test-mass limit, comparing the analytical expressions (either the waveform or the full flux) with
numerical data. These numerical data are obtained solving the Teukolsky equation numerically and were kindly
given to us by S.~Hughes~\cite{Hughes:2005qb}. We consider two types of data: either the modulus of the $\ell=m=2$
relativistic correction to the waveform $|\hat{h}_{22}|$ or the reduced flux function (i.e. the flux divided
by the Newtonian quadrupole) $\hat{f}$ obtained summing together all multipoles up to $\ell=8$. 
In particular, the use of $|h_{22}|$ allows us to discriminate between some choices, that a 
priori are arbitrary, concerning the values of $i_{{\rm max},a}$ and $i_{{\rm max},{\cal F}}$, that enter the NS-quantity 
$\lambda_{\rm inst}^{\rm NS}$ and the truncation of $\hl$. We recall that, since we are keeping the functions $\rho_\lm$
and $\tilde{f}_{\lm}$ at 10 PN, then $i_{\rm max}$ is always kept to 13, according to what already said in \eqref{eq:rule_imax} and below \eqref{eq:expression_galpha}.

\begin{figure}[t]
	\center
	\includegraphics[width=0.48\textwidth]{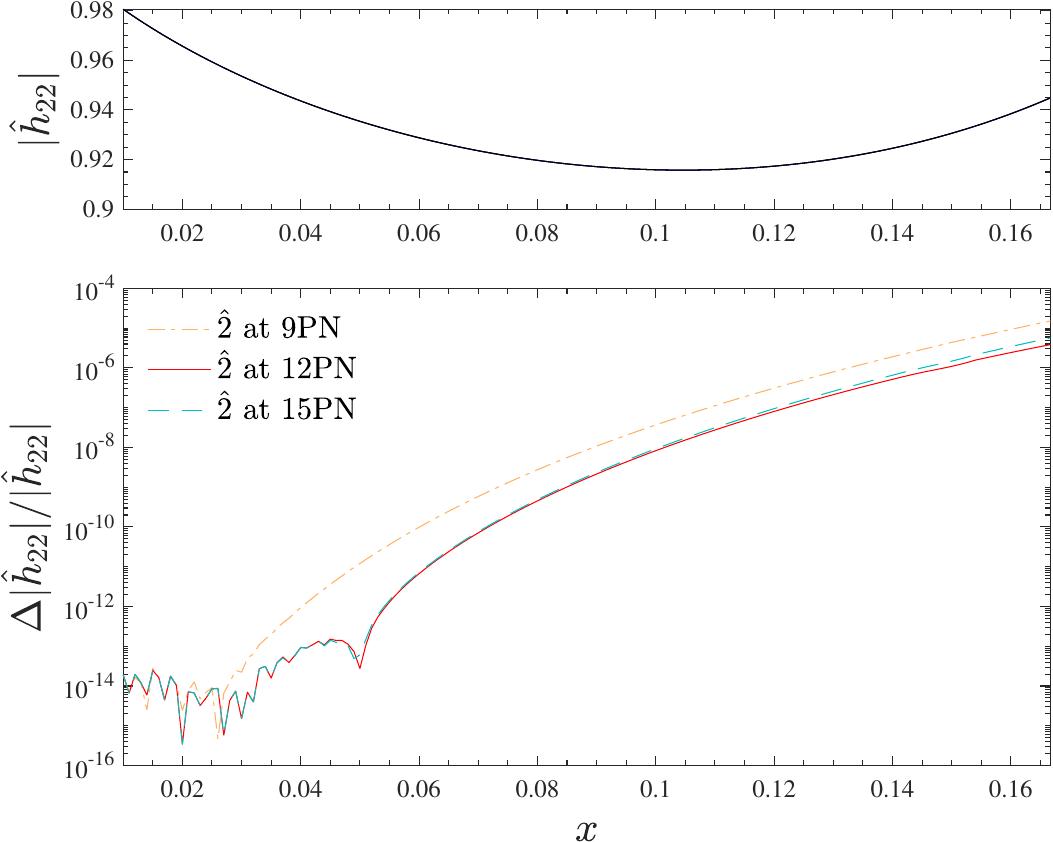}
	\caption{\label{fig:hat2}Choosing the PN accuracy of $\hat{2}$ from Eq.~\eqref{eq:hat2} in the test-mass limit.
	Top panel: $|\hat{h}_{22}|$ for a test-mass on circular orbits on a Schwarzschild black hole. 
	Bottom panel: fractional difference between the exact $|\hat{h}_{22}|$, obtained numerically, and the analytical 
	$|\hat{h}_{22}|$ obtained retaining both $(\rho_{22},\tilde{f}_{22})$ at 10PN accuracy, 
	$\lambda_{\rm inst}^{\rm NS}$ in closed form with $i_{{\rm max},{\cal F}}=10$ but truncating $\hat{2}$ from 
	Eq.~\eqref{eq:hat2} at 9PN ($i_{{\rm max},a}= 6$), 12PN ($i_{{\rm max},a}= 8$) and 15PN ($i_{{\rm max},a}= 10$). 
	The 12PN truncation is the closest one to the numerical data.}
\end{figure}

\subsubsection{Impact of $\hat{\ell}$ and $\lambda_{\rm inst}^{\rm NS}$: $\ell=m=2$ mode}
\label{sec:studyh22}
We start with $\rho_{22}$ and $\tilde{f}_{22}$ at 10PN. With this fixed, there are two residual
arbitrariness, that is: (i) the order at which $\lambda_{\rm inst}^{\rm NS}$ is computed, i.e. the
value of $i_{{\rm max},{\cal F}}$ in Eq.~\eqref{eq:gamma_inst} and (ii) the order at which $\hl$ is computed
starting from Eqs.~\eqref{eq:expression_a_general}.
As a first exploratory study, we consider $\lambda_{\rm inst}^{\rm NS}$ at the highest order available,
i.e. $i_{{\rm max},{\cal F}}=10$ and explore the effect of truncating $\hl$ at various orders. The explicit expression
of $\hl$ for $\ell=2$ up to 15PN reads
\begin{align}
\label{eq:hat2}
\hat{2} & =2 -\frac{214}{105}\w^2-\frac{3390466}{1157625}\w^4-\frac{153440219802466}{15021833990625}\w^6\nonumber\\
&-\frac{71638806585865707261481}{1520451676706008921875} \w^8 + \notag \\
& -\frac{270360664939833821554899493653643}{1099244369724415858768042968750} \w^{10} + \mathcal{O}(\w^{12}) \ .
\end{align}
Since $\rho_{22}$ and $\tilde{f}_{22}$ are truncated at 10PN, and they enter other resummed expressions,
it is a priori not evident what order of $\hat{2}$ should be retained. Pragmatically, we compare the exact $|\hat{h}_{22}|$
with three analytical expressions where $\hat{2}$ is truncated at 9PN ($i_{{\rm max},a}= 6$), 12PN ($i_{{\rm max},a}= 8$) and 15PN ($i_{{\rm max},a}= 10$).
We see in Fig.~\ref{fig:hat2} that the best agreement up to the LSO is obtained at 12PN accuracy.  
For simplicity we thus choose to keep $\hat{\ell}$ at 12PN accuracy for all modes up to $\ell=8$. 

\begin{figure}[t]
	\center
	\includegraphics[width=0.48\textwidth]{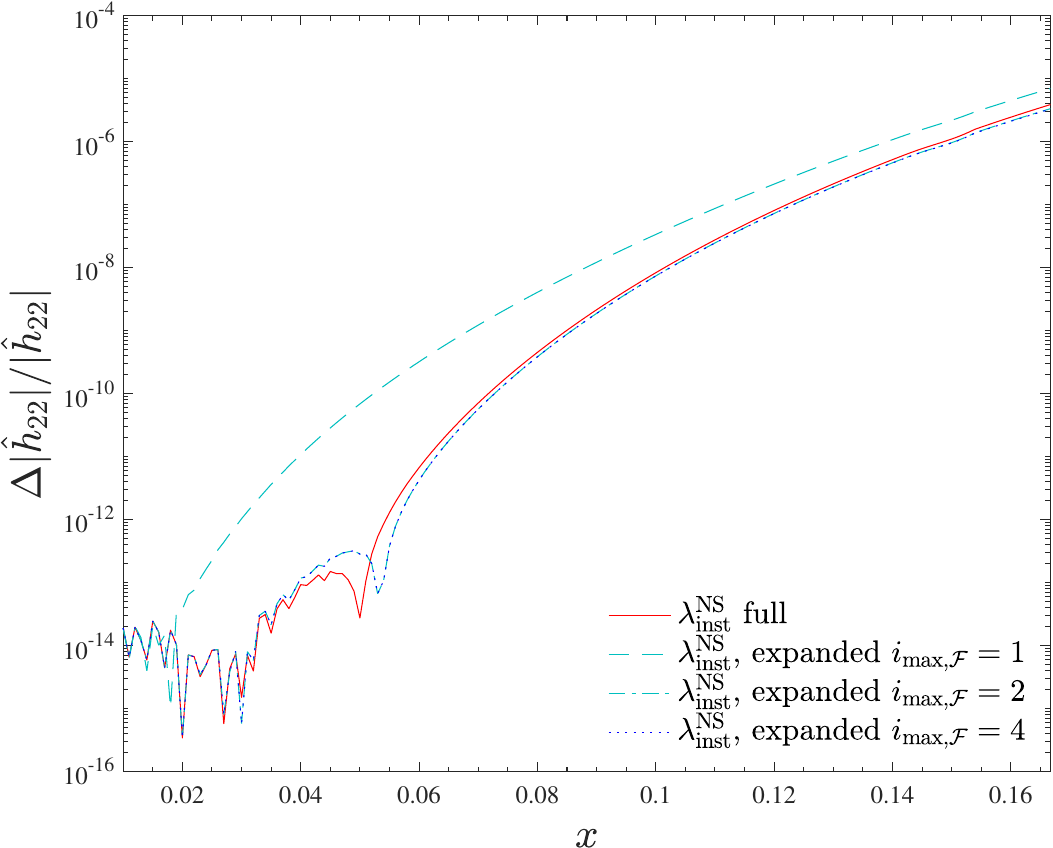}
	\caption{\label{fig:lambda_NS}Effect of various approximations to the function 
	$\lambda_{\rm inst}^{\rm NS}$ while keeping $\hat{2}$ at 12PN and both 
	$(\rho_{22},\tilde{f}_{22}e^{i\tilde{\delta}_{22}})$ at 10PN accuracy. 
	The red line is the same as  Fig.~\ref{fig:hat2} and it is very well represented by approximating
	 $\lambda_{\rm inst}^{\rm NS}$ via Eq.~\eqref{eq:expanded} with $i_{{\rm max},{\cal F}}=2$.}
\end{figure}

We move now to analyze the impact of $\lambda_{\rm inst}^{\rm NS}$ so to understand 
the importance of its complicated structure, Eq.~\eqref{eq:lambda_exp}. 
We consider the following cases:
(i) we keep the functional form with the exponential as given in 
Eq.~\eqref{eq:lambda_exp} with $i_{{\rm max},{\cal F}}=10$; 
(ii) we fix $i_{{\rm max},{\cal F}}=2$, expand the exponential Eq.~\eqref{eq:lambda_exp} 
at first order and work with 
\be
\label{eq:expanded}
\lambda_{\rm inst}^{\rm NS}\simeq 1 + \de_a u_1 X + \dfrac{1}{2} \de_a u_2 X^2 \ ,
\ee
exploring the effect of keeping either both terms ($i_{{\rm max},{\cal F}}=2$) or just the 
first one ($i_{{\rm max},{\cal F}}=1$);(iii) we also consider the same first-order 
expression with instead $i_{{\rm max},{\cal F}}=4$.

The performance of the four different choices is illustrated in Fig.~\ref{fig:lambda_NS} in terms of $\Delta|h_{22}|$.
Remarkably, one finds that the differences between using the full expression and the truncated one with 
$i_{\rm max}=2$ are practically negligible for our purposes, yielding thus an important analytical simplification.
It seems thus that the full structure of $\lambda^{\rm NS}_{\rm inst}$ is not relevant in the present context.

As a best compromise between complexity and accuracy, we thus work with: (i) $\hl$ at 12PN and (ii) $\lambda_{\rm inst}^{\rm NS}$ 
from Eq.~\eqref{eq:expanded} with $i_{{\rm max},{\cal F}}=2$. We will also see below that the accuracy is 
maintained when considering the full flux of a test-particle around circular orbits.

\begin{figure}[t]
	\center
	\includegraphics[width=0.48\textwidth]{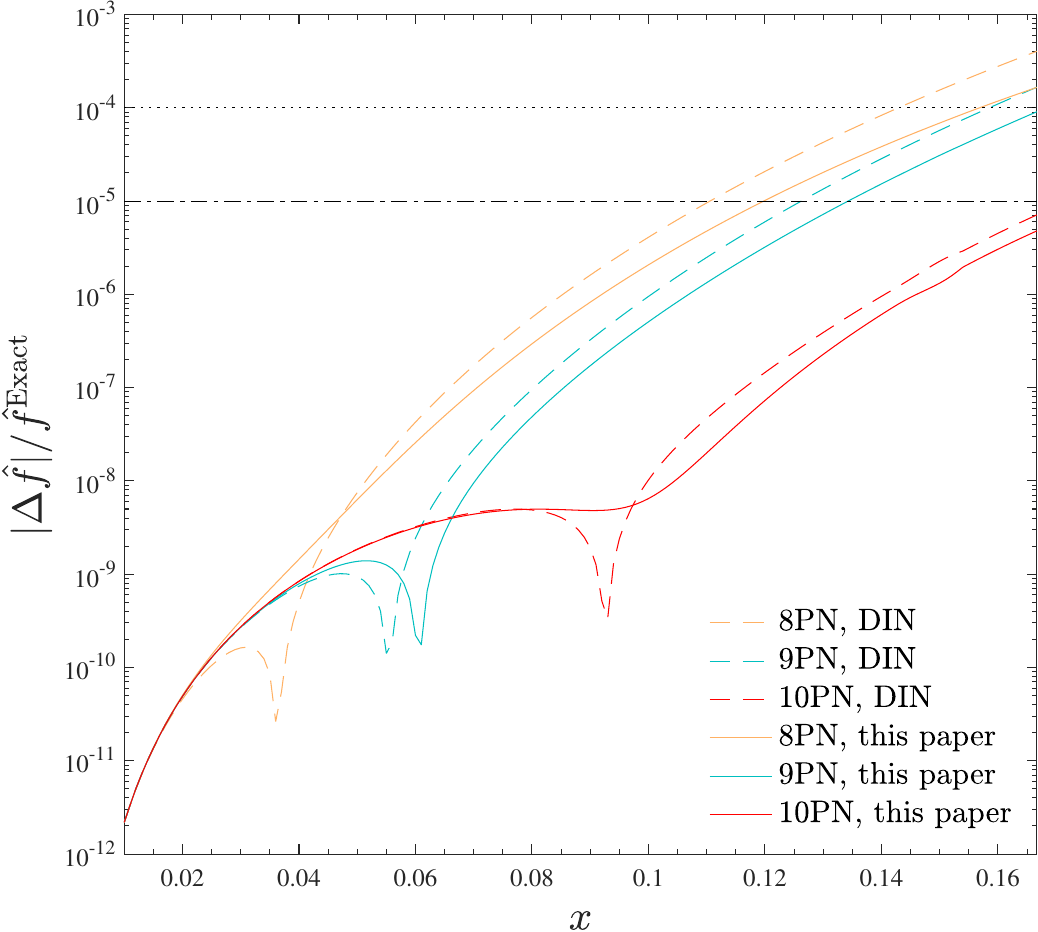}
	\caption{\label{fig:hatf}Performance of the new resummation for the full flux with all modes summed up to 
	$\ell=8$ contrasted with the DIN procedure~\cite{Damour:2008gu}. The plot shows the relative differences
	with the exact flux obtained numerically for various analytical representations. We compare the effect of 
	8PN, 9PN or 10PN truncation of the residual amplitude corrections. The relative difference at 
	$x_{\rm LSO}$ is $4.78\times 10^{-6}$ at 10PN for the new procedure versus $7.09\times 10^{-6}$ of DIN.
	Compare also with Fig.~8 of Ref.~\cite{Nagar:2022fep} that implements the DIN procedure at 22PN 
	accuracy, that yields just $4\times 10^{-6}$ at $x_{\rm LSO}$.}
\end{figure}

\subsubsection{Newton-normalized energy flux}
A more comprehensive comparison is given by studying the accuracy of the Newton-normalized
energy flux around circular orbits, $\hat{f}$ obtained by summing all multipoles up to $\ell=8$.
As target observable we consider, as usual, the reduced flux function $\hat{f}$ emitted by a test-mass
around circular orbits, considering all multipoles summed up to $\ell=8$. 
Figure~\ref{fig:hatf} evaluates the performance of the resummation in terms of the fractional difference
between the analytical and numerical Newton-normalized flux functions. We contrast the well-established
DIN resummation with our new procedure, considering three different PN orders of the residual functions,
8PN, 9PN and 10PN. Evidently, for DIN this is the PN order of the $\rho_\lm^{\rm DIN}$ functions. 
For our procedure this refers instead to the order of $(\rho_\lm,\tilde{f}_\lm e^{i\tilde{\delta}_\lm})$.
From Fig.~\eqref{fig:hatf} it is evident that the new procedure is superior with respect to the standard
DIN approach, though the actual gain seems to be reduced as the PN order is increased. 
The fractional difference at LSO with the new procedure is $4.78\times 10^{-6}$ working at 10PN. 
The DIN approach at 10PN is less accurate by almost a factor two, yielding $7.09\times 10^{-6}$.
In this respect, let us recall that Fig.~8 of Ref.~\cite{Nagar:2022fep} showed that pushing the 
$\rho_\lm^{\rm DIN}$ accuracy to 22PN allows to lower the fractional difference to 
$\sim 4\times 10^{-6}$ at the LSO, though it is larger than what we obtain here at intermediate
values of $x$, say $x\sim 0.1$. We recall that the use of the DIN 22PN-accurate resummed flux 
was probed to be sufficient to have consistency between state-of-the-art GSF 
waveforms~\cite{Pound:2019lzj,Warburton:2021kwk,Wardell:2021fyy} 
and EOB-based waveform models for large-mass-ratio 
binaries~\cite{Nagar:2022fep,Albertini:2022rfe,Albertini:2022dmc,Albertini:2023aol,Albertini:2024rrs}.
With the new resummation approach proposed here we expect to produce a similarly accurate model,
or even better, with the advantage of relying on relatively low-PN information for the residual polynomials.
Evidently, if the needs occurs, it is straightforward to push our 10PN results to even higher orders.

\begin{figure}[t]
	\center
	\includegraphics[width=0.48\textwidth]{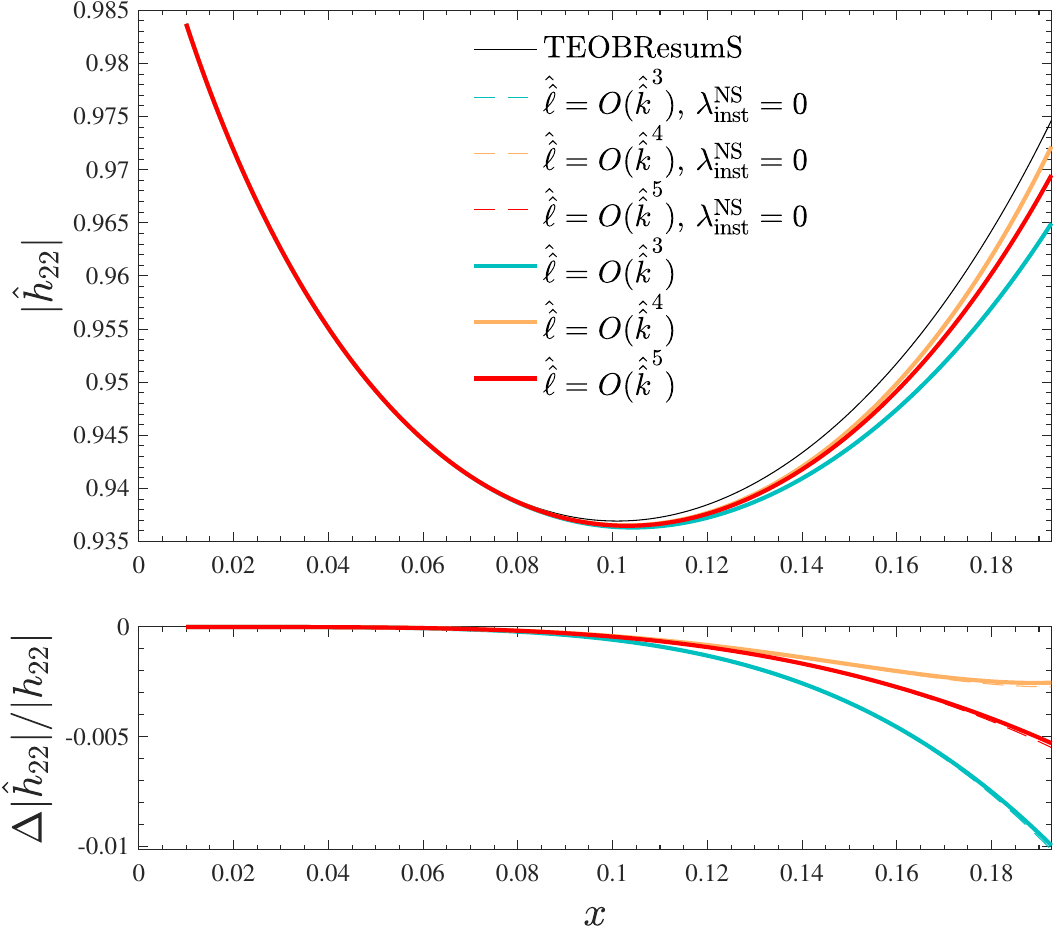}
	\caption{\label{fig:q1}Performance of the new resummation with different truncations of $\hat{\hat{2}}$, 
	taken from Eq.~\eqref{eq:gamma_2m_univ} with $\ell=2$, with respect to the DIN (resummed) function
	implemented in the state-of-the-art model \TEOBd{}. The effect of $\lambda_{\rm inst}^{\rm NS}$ is always 
	negligible.}
\end{figure}

\subsection{Equal-mass case}
\label{sec:equal_mass}
Let us move now to evaluate the performance of our new approach in the comparable mass case,
in particular focusing on the equal-mass case. Note that any conclusive statement about the performance of
the new resummed waveform would require its implementation within a waveform model, e.g. in \TEOBd{},
and assess its phasing performance  with NR simulations. This analysis requires more work that is 
postponed to the future.
Here, to get a first impression of the effects of the new resummation with respect to what implemented 
in \TEOBd{}~\cite{Nagar:2024oyk}, we consider the EOB adiabatic dynamics along circular orbits 
up to the LSO and evaluate $|\hat{h}_{22}(x)|$ on top of it.
The adiabatic EOB dynamics is defined formally in Sec.~\ref{sec:comparable_mass}.
Here we specify te EOB $A(u)$ potential to be taken from Eq.~(6) of Ref.~\cite{Nagar:2024oyk} and 
resummed as in Eq.~(8) therein, i.e. by taking separate Pad\'e approximants for the rational terms and
for the logarithmic terms. The $A(u)$ we use is analytically complete at 4PN, but depends on the 
effective 5PN function  $a_6^c(\nu)$ that is informed by NR simulations as $a_6^c(\nu)=208.19\nu^2-318.26\nu+34.85$, 
see Eq.~(36) of~\cite{Nagar:2024oyk} and related discussion. One then
specifies a grid of numerical values of $x$, from it $u$ is obtained solving numerically
Eq.~\eqref{eq:Omg_u} and then finally one gets $p_{\varphi,{\rm circ}}$ that give the effective
energy $\hat{E}^{\rm circ}_{\rm eff}=\hat{H}_{\rm eff}(u,p_{\varphi,{\rm circ}})$ and real energy 
along the adiabatic sequence of circular orbits. The adiabatic dynamics is stopped at the
the last stable orbit (LSO), defined by the condition $\de_u\hat{H}_{\rm eff}=\de_u^2\hat{H}_{\rm eff}=0$.

\begin{figure}[t]
	\center
	\includegraphics[width=0.48\textwidth]{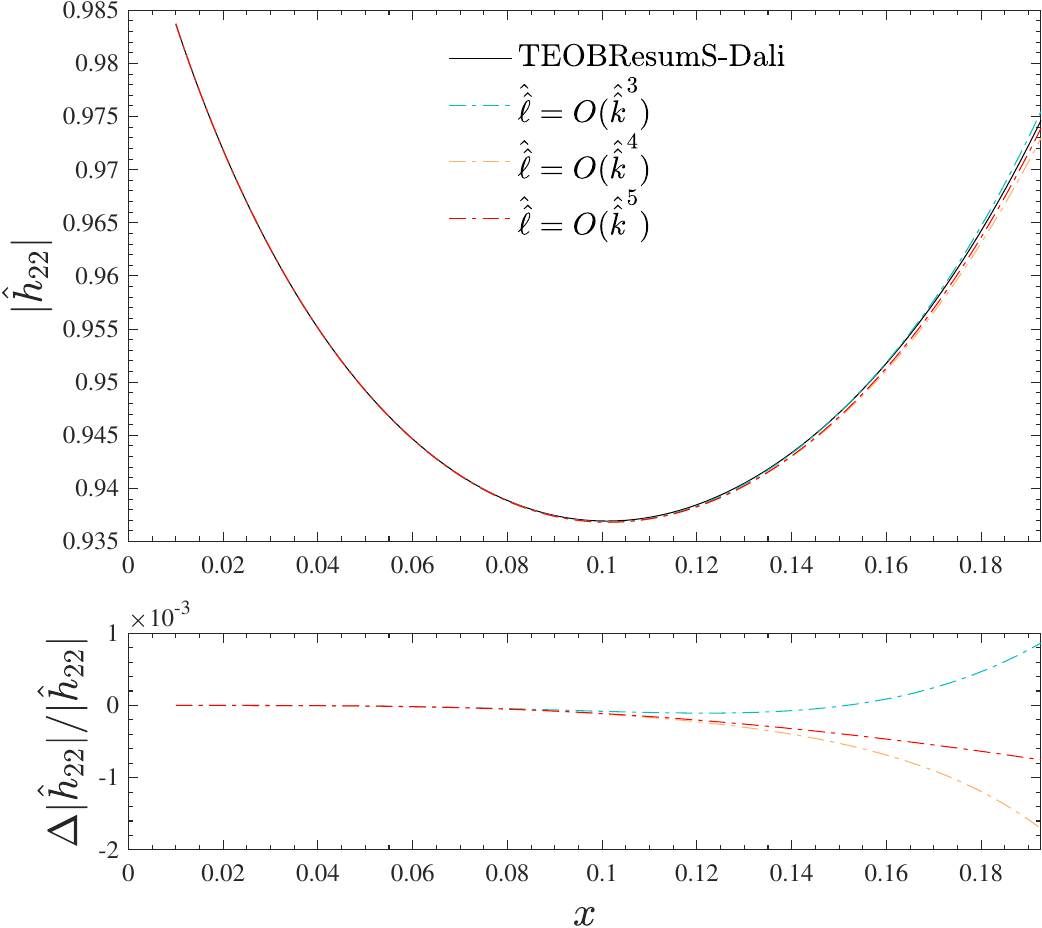}
	\caption{\label{fig:q1_ilpz}Performance of the revisited ILPZ factorization described in Sec.~\ref{sec:ivanov_et_al}: it 
	is much closer to the DIN choice of \TEOBd{}, although the effect of the truncation of $\hat{\hat{\ell}}$ with $\ell=2$ is not negligible.}
\end{figure}
On this EOB adiabatic dynamics we compute the resummed $|\hat{h}_{22}|$ in two ways at
4PN accuracy. One the one hand, we use the prescription of Ref.~\cite{Nagar:2024oyk},
i.e. the DIN tail factorization with a 4PN-accurate $\rho_{22}^{\rm DIN}$ function resummed 
as in Eq.~(23) therein. On the other hand, we use our new factorization, but including 
only the external tail for consistency with the global 4PN order we are starting from.
In doing so, we still have the arbitrariness of exploring: (i) the effect of the PN accuracy of $\hat{\hat{\ell}}$;
(i) the importance of $\lambda_{\rm inst}^{\rm NS}$.
To do so, we consider an equal-mass binary, for which we have $u_{\rm LSO}=0.1956$ 
and $p_\varphi^{\rm LSO}=10.7906$, that yield $M\Omega^{\rm LSO}=0.0845$ 
and $x_{\rm LSO}=0.1926$. In Fig.~\ref{fig:q1} we show the various $|\hat{h}_{22}|$ (top panel)
obtained truncating $\hat{\hat{\ell}}$ at $\hat{\hat{k}}^3$, $\hat{\hat{k}}^4$ or $\hat{\hat{k}}^5$ 
and either setting $\lambda_{\rm inst}^{\rm NS}=0$ or not. The bottom panel shows the fractional 
differences with respect to \TEOBd{}, where $\Delta |\hat{h}_{22}|\equiv |h_{22}|-|h_{22}|^{\TEOBd{}}$
and we omitted the superscript $\TEOBd{}$ at the denominator in the label for simplicity.
The figure conveys two messages: (i) the effect of $\lambda_{\rm inst}^{\rm NS}$ is irrelevant
in this case; (ii) whatever choice of truncation of $\hat{\hat{\ell}}$ is chosen, the new approach
always lead a waveform amplitude that is visibly smaller than the $|\hat{h}_{22}^{\rm DIN}|$
implemented in \TEOBd{}.
This fact has consequences at the level of radiation reaction, since the fractional difference
of $\sim 10^{-3}$ will gain an additional factor two. In conclusion, we can already predict that
on \TEOBd{} the straightforward change of the $\ell=m=2$ mode in radiation reaction will, by
itself, yield a longer inspiral and a delayed plunge once all the other elements entering
the EOB dynamics are kept fixed. With this understanding, the remaining open question 
is whether, once that such a new radiation reaction (and waveform) is implemented 
in the \TEOBd{}, the model still retains the current flexibility that eventually allows it to make 
it highly faithful to NR simulations via calibration as it is the case with DIN-like factorizations. 
This is a priori nontrivial given the rather large differences highlighted in the figure, but
a precise question will require more dedicated work that is postponed to the future.
For completeness, we performed a similar analysis using the (upgraded) ILPZ proposal
introduced in Sec.~\ref{sec:ivanov_et_al}, see Fig.~\ref{fig:q1_ilpz}. It is interesting to note 
that in this case the differences are much smaller than the other case and their sign depend
on the truncation order of $\hat{\hat{2}}$. Thus, we expect, a priori, that the implementation
of this kind of waveform should yield a waveform model with comparable analytical flexibility
and performance to the DIN-based one. This also will be explored in the future.

\section{Conclusions}
\label{sec:end}
The main finding of this paper is a new factorized and resummed form of the PN gravitational
waveform for circularized nonspinning binaries. The factorization is such that all logarithmic contributions 
and transcendental numbers that are present in the PN-expanded waveform, and that are independent of 
the symmetric mass ratio $\nu$, are completely resummed via exponentials and $\Gamma$-functions. 
The reminder of the $\nu$-independent waveform (written as amplitude and phase), still written in 
PN-expanded form, is thus fully rational. By contrast, $\nu$-dependent logs, representing tail of 
memory effects, and residual $\pi^2$ contributions are still present in the 3PN and 4PN residual contributions.
Our approach builds upon the findings and proposals of Refs.~\cite{Damour:2008gu,Fucito:2023afe,Cipriani:2025ikx,Ivanov:2025ozg}
but improves our current knowledge by taking ideas and procedures from each reference and blending 
them together synergically.
In particular:(i) from Ref.~\cite{Damour:2008gu} we take the idea of factoring out the source of the field, 
the leading tail contributions (in the phase) and the transcendental numbers and (ii) improve it using the 
closed form analytical structures that appear when the Teukolsky equation is solved once mapped into a 
confluent Heun equation, as  suggested in Refs.~\cite{Fucito:2023afe,Cipriani:2025ikx}. This yields 
the complete factorization of {\it all} test-mass logs (including the subleading ones) and test-mass 
transcendental contributions in both waveform amplitude and phase;(iii) the approach is then generalized
to the comparable-mass case, $\nu\neq 0$, by blending together a proposal of 
Refs.~\cite{Damour:2007xr,Damour:2008gu} (see also Ref.~\cite{Pan:2010hz}) with the new approach of 
Ref.~\cite{Ivanov:2025ozg}, that connects the renormalized angular momentum of black 
holes, $\hat{\ell}$, known from black-hole perturbation theory, with the universal part of the anomalous 
dimension of BH multipole moments. 
Our main analytical findings can be summarized as follows.
\begin{itemize}
\item[(i)] The solution of the CHE entails that the logarithmic and transcendental contributions come as two, different, 
tail-transcendental factors, that we dubbed as external and internal, with the internal one contributing starting from 
5PN order. As such, the structure of the factorized waveform is richer than the one proposed by 
Ref.~\cite{Damour:2008gu} based on the solution of the Coulomb wave equation.
\item[(ii)]We found that the factors involving $\Gamma$-functions are different and more involved than the leading 
order one introduced in Ref.~\cite{Damour:2008gu}, although the structure of the argument of the $\Gamma$-function
is similar. It is interesting to note that the renormalized angular momentum of black holes $\hat{\ell}$ appears naturally
in these functions, thus supporting a posteriori the guess of Ref.~\cite{Ivanov:2025ozg} that proposed to systematically 
replace $\ell\to\hhl$ when it appears in the argument of the $\Gamma$ functions.
\item[(iii)]Reference~\cite{Ivanov:2025ozg} used renormalization group running of the multipole moments to predict the 
existence  of a factor of type $e^{(\hhl - \ell)\log(\alpha r m \Omega)}$ in the waveform amplitude and $e^{-i\pi(\hhl-\ell)/2}$ in the 
waveform phase, where $\alpha$ is an arbitrary scale that was fixed in Ref.~\cite{Ivanov:2025ozg} according to a
reasonable prescription. The analytical structure of these factors, with $\hhl$ replaced by $\hat{\ell}$ are those 
that appear naturally in the factorized test-mass solution. An important novelty, though, is that the solution of 
CHE predicts that $\alpha=2$.

Similarly, the fact that the renormalized angular momentum of black holes $\hat{\ell}$ is present
in well precise places gives us a rationale to promote it to the quantity $\hhl$ introduced in Ref.~\cite{Ivanov:2025ozg} 
involving the anomalous universal dimension of black holes. Our analytical results allow then to put 
on a more rigorous basis some of the well motivated guesses of Ref.~\cite{Ivanov:2025ozg}.
\end{itemize}
The factorized waveform is computed up to all the $\ell=8$ modes included. In the test-mass limit, each residual 
function in the amplitude and phase is retained up to 10PN order. For the $\nu\neq 0$ case each residual 
function, $(\rho_\lm,\delta_\lm)$ is kept at a PN order compatible with the maximal analytical knowledge 
currently available, i.e. 4PN for the $\ell=m=2$ mode and {\it global} 3PN accuracy for the other modes.
In the $\nu=0$ case, the performance of the new analytical factorization is assessed comparing the 
(Newton-normalized) energy flux along circular orbits obtained analytically versus the exact one obtained
numerically. We find that at 10PN accuracy (in the sense described above) the fractional difference at the 
LSO is below $10^{-5}$, a value that is comparable to the one obtained using the factorization of 
Ref.~\cite{Damour:2008gu} but working with 22PN accurate residual amplitudes.
This indicates that the new procedure is  promising to efficiently improve the analytical description
of the waveform and radiation force implemented within state-of-the-art EOB models for computing EMRIs
waveforms~\cite{Albertini:2022dmc,Albertini:2023aol,Albertini:2024rrs} that currently rely on 22PN accurate 
results in DIN-resummed form. In this respect, the extension of the factorization and resummation procedure
discussed here to the case of a central Kerr black hole is expected to overcome the lack of accuracy of
the flux in DIN form~\cite{Nagar:2022fep}. This study is currently in progress and will be presented elsewhere.

For comparable mass binaries, in principle our factorized and resummed procedure should eventually yield 
a more accurate description of the radiation reaction force for EOB waveform models like \TEOBd{} or {\tt SEOBNRv5HM}.
The assessment of this statement would require the implementation of the new waveform and flux in
some  complete EOB model and perform the usual NR-calibration and comparison with NR data,
a work that for the moment is postponed to the future.
As a first, preliminary, analysis we focussed only on the adiabatic EOB dynamics for circular
orbits and found that, for an equal-mass binary, the amplitude of the $\ell=m=2$ relativistic correction 
$|\hat{h}_{22}|$ is always {\it smaller} than the corresponding DIN one. This entails a reduction of the corresponding
radiation reaction force and thus a delayed merger time while keeping the conservative part unchanged with
respect to the current implementation of \TEOBd{}. As a consequence, the NR-calibration (or other changes 
in the conservative pert of the model)  should be such to suitably compensate for this effect so to yield a
new version of the model as NR-faithful as the current one.

\acknowledgments
A.~C. acknowledges IPhT, IHES and the INFN Section of Turin for the kind hospitality
during the several stages of this work.
A.~C. and A.~N. are grateful to T.~Damour and J.~Parra-Martinez for 
several useful discussions. We thank T.~Damour and R.~Gamba for 
critical comments on the manuscript. This work was partly developed 
at IHES,  supported by the ``\textit{2021 Balzan Prize for 
Gravitation: Physical and Astrophysical Aspects}'', 
awarded to Thibault Damour.

\appendix

\section{Hypergeometric connection formulas and imposition of boundary conditions}
\label{app:conn_formulas}
In this Appendix we report the relevant steps concerning the application of the boundary conditions for the determination 
of the solutions of the homogeneous Teukoslky equation, as exposed in Sec.~\ref{sec:hom_solution_Teuk}. 
First of all, we have to use the connection formulas for the hypergeometric functions that allow to go to the boundary 
of the spatial interval, i.e. $r=\infty$ and $r=2M$. Let us start with the first case.

As already explained in the text, the functions $G^0_\alpha(Y)$, defined in Eq.~\eqref{eq:ansatz_G0}, are exact 
in $Y$ at the $k$-th PM order. This allows to extrapolate them to the region of large $Y$ (where the infinity is located) 
using the following hypergeometric connection formula
\begin{equation}
H^0_\alpha (Y) = \sum_{\beta=\pm} B_{\alpha \beta} \tilde{H}^0_\beta(Y) \ ,
\end{equation}
where the \textit{braiding matrix} is defined as in Eq.~\eqref{eq:braiding_matrix}, but now we write its elements 
without using the dictionary from Eqs~\eqref{eq:X}-\eqref{eq:u}
\begin{equation}
B_{\alpha\beta} = \frac{e^{\frac{i \pi}{2}(1-\beta) \left(\frac{1}{2}-\alpha a- m_3\right)} \Gamma(1-2 \alpha a)}{\Gamma \left(\frac{1}{2} - \alpha a - \beta m_3\right)} \ ,
\end{equation}
and 
\begin{align}
\tilde{H}^0_\beta(Y) & = Y^{-m_3(1+\beta)} \, e^{\frac{Y}{2} (1+\beta)} \times \notag \\
& \times  \, _2F_0 \left(\frac{1}{2}-a+ \beta m_3,\frac{1}{2}+a+ \beta m_3, \frac{\beta}{Y}\right) \ .
\end{align}
In this way it is possibile to define an equivalent basis of solutions of the same equation 
that is naturally defined very close to infinity, i.e.
\begin{equation}
\tilde{G}^0_\alpha (Y) = P_0(Y) \tilde{H}^0_\alpha (Y) + \hat{P}_0(Y) Y \tilde{H}^{0'}_{\alpha} (Y) \ ,
\end{equation}
related to the previous one by the connection formula
\begin{equation}
G^0_\alpha(Y) = \sum_{\beta=\pm} B_{\alpha \beta} \, \tilde{G}^0_\beta (Y) \ .
\end{equation}
At this point, using the leading order contribution of $\tilde{H}^0_\beta(Y)$ for $Y \to \infty$, which is simply
\begin{equation}
\tilde{H}^0_\beta (Y) \underset{Y \to \infty}{\approx} Y^{-m_3(1+\beta)} \, e^{\frac{Y}{2} \left(1+\beta\right)} \ ,
\end{equation}
and the fact that the coefficients $\hat{c}_{i,i-1}$ satisfy the relation 
\begin{equation}
1+ \sum_{i=1}^\infty \hat{c}_{i,i-1} x^i = e^{-\partial_{m_3} \mathcal{F}^{\rm NS}_{\rm inst}} \ ,
\end{equation}
it can be shown that
\begin{equation}
\tilde{G}^0_\beta(Y) \underset{Y \to \infty}{\approx} Y^{-m_3(1+\beta)} \, e^{\frac{Y}{2} \left(1+\beta\right)} e^{-\frac{1+\beta}{2}\partial_{m_3} \mathcal{F}^{\rm NS}_{\rm inst}} \ .
\end{equation}
We can move now to the asymptotic behavior at infinity of $R_{\rm in,out}(r)$. From Eqs.~\eqref{eq:R_alpha}-\eqref{eq:Rin_Rup}, 
the dictionary of Eqs.~\eqref{eq:X}-\eqref{eq:u} and the relations derived above, we obtain that
\begin{widetext}
\begin{align}
R_{\rm in,out}(r) & \underset{r \to \infty}{\approx} \sum_{\alpha} c_\alpha^{\rm in,out} g_\alpha(X) e^{-i \omega r} (2 i \omega r)^{-1-2 i \w} \sum_\beta B_{\alpha \beta} (2 i \omega r)^{\left(2+2 i \w\right)(1+\beta)} \, e^{i \omega r(1+\beta)} e^{-\frac{1+\beta}{2} \partial_{m_3} \mathcal{F}^{\rm NS}_{\rm inst}|_{m_3=-2-2 i \w}} = \notag \\
& = \sum_{\alpha \beta} c_\alpha^{\rm in,out} \, g_\alpha \, B_{\alpha \beta} \, e^{\beta i \omega r} \, (4 i \w)^{1+(2+2 i \w) \beta} \, e^{-\frac{1+\beta}{2} \partial_{m_3} \mathcal{F}^{\rm NS}_{\rm inst}|_{m_3=-2-2 i \w}}  \, \left(\frac{2M}{r}\right)^{-1-(2-2 i \w) \beta} \ .
\end{align}
\end{widetext}
{Comparing now with the asymptotic behavior at infinity of the two solutions 
in Eq.~\eqref{bcpsi}, we see that
\begin{align}
B_\beta^{\rm in,out} = & (4 i \w)^{1+ (2+2 i \w) \beta} e^{-\frac{1+\beta}{2} \partial_{m_3} \mathcal{F}^{\rm NS}_{\rm inst}|_{m_3=-2-2 i \w}} \times \notag \\
& \times \sum_\alpha c_\alpha^{\rm in,out} \,  g_\alpha \, B_{\alpha \beta} \ .
\end{align}
Imposing the outgoing boundary condition at infinity consists in taking $B^{\rm out}_-=0$, as already written in Eq.~\eqref{bout}. 
This condition yields
\begin{equation}
 \sum_\alpha c_\alpha^{\rm out} g_\alpha B_{\alpha -} = 0 \ ,
\end{equation}
and then
\begin{equation}
\frac{c_-^{\rm out}}{c_+^{\rm out}} = - \frac{g_+(X)}{g_-(X)} \ ,
\end{equation}
that precisely coincides with Eq.~\eqref{eq:ratio_out}.

We can move now to the horizon. We are going to follow the same procedure explained above, 
but using now the other basis of solutions, i.e. the $G^1_\alpha(Z)$. 
These functions, defined in Eq.~\eqref{eq:ansatz_G1}, 
are exact in $Z$ at the $k$-th PN order. This allows us to extrapolate 
them to the region of $Z$ close to 1 (where the horizon is located) using the 
following hypergeometric connection formula 
\begin{equation}
H^1_\alpha (Z) = \sum_{\beta=\pm} F_{\alpha \beta} \tilde{H}^1_\beta(Y) \ ,
\end{equation}
where the \textit{fusion matrix} is defined as in Eq.~\eqref{eq:fusion_matrix}, 
but now we write its elements without using the dictionary from Eqs.~\eqref{eq:X}-\eqref{eq:u}
\begin{equation}
F_{\alpha\beta} = \frac{\Gamma(1+2 \alpha a) \Gamma[-\beta(m_1+m_2)]}{\Gamma \left(\frac{1}{2} + \alpha a - \beta m_1\right)\Gamma \left(\frac{1}{2} + \alpha a - \beta m_2\right)} \ ,
\end{equation}
and 
\begin{align}
\tilde{H}^1_\beta(Z) = & Z^{a+m_3-\frac{1}{2}} \, (1-Z)^{\frac{1}{2}(m_1+m_2)(1+\beta)} \times \notag \\
& \times  \, _2F_1 \left(\frac{1}{2} + a + \beta m_1,\frac{1}{2} + a + \beta m_2 , \right. \notag \\
& \left. 1+\beta(m_1+m_2), 1-Z\right) \ .
\end{align}
In this way it is possibile to define an equivalent basis of solutions of the same equation and that is naturally defined very close to the horizon, i.e.
\begin{equation}
\tilde{G}^1_\alpha (Z) = P_1(Z) \tilde{H}^1_\alpha (Z) + \hat{P}_1(Z) Z \tilde{H}^{1'}_{\alpha} (Z) \ ,
\end{equation}
related to the previous one by the connection formula
\begin{equation}
G^1_\alpha(Z) = \sum_{\beta=\pm} F_{\alpha \beta} \, \tilde{G}^1_\beta (Z) \ .
\end{equation}
At this point, using the leading order contribution of $\tilde{H}^1_\beta(Z)$ for $Z \to 1$, which is simply
\begin{equation}
\tilde{H}^1_\beta (Z) \underset{Z \to 1}{\approx} (1-Z)^{\frac{1+\beta}{2}(m_1+m_2)} \ ,
\end{equation}
it can be shown that
\begin{equation}
\tilde{G}^1_\beta(Z) \underset{Z \to 1}{\approx} (1-Z)^{\frac{1+\beta}{2}(m_1+m_2)} \, h_\beta 
\end{equation}
where
\begin{equation}
h_\beta = P_1(1) + \frac{1+\beta}{2} (m_1+m_2) \tilde{P}'_1(1) \ .
\end{equation}
At this point, we can move to the asymptotic behavior at the horizon of $R_{\rm in,out}(r)$, indeed from Eqs.~\eqref{eq:Rin_Rup}-\eqref{eq:Ra_G1a}, the dictionary in Eqs~\eqref{eq:X}-\eqref{eq:u} and together with the relations derived in this Appendix, we obtain that}
\begin{widetext}
\begin{align}
R_{\rm in,out}(r) & \underset{r \to 2 M}{\approx} \sum_{\alpha} c_\alpha^{\rm in,out} e^{-i \omega r} \, \left(1-\frac{2M}{r}\right)^{2-2 i \w} (4 i \w)^{-1-2 i \w} \sum_{\beta} F_{\alpha \beta} \, \left(1-\frac{2M}{r}\right)^{(1+\beta)(2 i \w -1)} h_\beta = \notag \\
& = \sum_{\alpha \beta} c_\alpha^{\rm in,out}  \, e^{\beta i \omega r} F_{\alpha \beta} \, \left(1-\frac{2 M}{r}\right)^{(1+\beta)(2 i \w-1)} (4 i \w)^{-1-2 i \w} \, h_\beta \, e^{-(1+\beta) 2 i \w} \ .
\end{align}
\end{widetext}
Comparing now with the asymptotic behavior at the horizon of the two solutions in Eq.~\eqref{bcpsi}, we have
\begin{equation}
D_\beta^{\rm in,out} = (4 i \w)^{-1-2 i \w} \, h_\beta \, e^{-(1+\beta)2 i \w} \sum_\alpha c_\alpha^{\rm in,out} \,F_{\alpha \beta} \ .
\end{equation}
Imposing the incoming boundary condition at the horizon consists in taking $D^{\rm in}_+=0$, as already written in Eq.~\eqref{din}. 
This condition implies
\begin{align}
 \sum_\alpha c_\alpha^{\rm in} F_{\alpha +} = 0 \ ,
\end{align}
eventually yielding
\begin{align}
\frac{c_+^{\rm in}}{c_-^{\rm in}} = - \frac{F_{-+}}{F_{++}} \ ,
\end{align}
that precisely coincides with Eq.~\eqref{eq:ratio_in}.

\section{The PN-expandend multipolar waveform $\hat{h}_{\lm}$}
\label{app:PN_exp_hhat}
In this Appendix we report the PN expansions of the waveforms $\hat{h}_{\lm}$ in radiative coordinates 
with $2 \leq \ell \leq 4, 1 \leq m \leq \ell$ and for $\ell=5$ we only report the cases of $m=1,3,5$, at most at 3 PN. 
In particular, the quantity $\hat{h}_{22}$ has already been written in the text in Eq.~\eqref{eq:h224PN} up to 4PN, 
the 3PN expressions of $\hat{h}_{33}$ and $\hat{h}_{31}$ are taken from Eqs. (4.18a) and (4.18b) of 
Ref.~\cite{Faye:2014fra}, the one of $\hat{h}_{21}$ at the same order from Eq. (4.11) of Ref.~\cite{Henry:2021cek} 
and for all the other modes we consider their expansions at the order reported in Sec.~III A of Ref.~\cite{Henry:2022ccf}. 

\begin{widetext}

\begin{align}
\hat{h}_{21}^{\rm 3PN}&= 1 + \left(-\dfrac{17}{28}+\dfrac{5}{7}\nu\right)x + x^{3/2} \left[\pi + i \left(-\dfrac{1}{2} -2 \log(2) \right) \right] + \left(-\dfrac{43}{126}-\dfrac{509}{126}\nu+\dfrac{79}{168}\nu^2\right)x^2
+\left[\dfrac{17 i}{56} - \dfrac{17 \pi}{28} + \right. \nonumber \\
& \left. + \nu \left(\dfrac{3 \pi}{14} - i \left(\dfrac{353}{28} + \dfrac{3 \log(2)}{7} \right) \right) + \dfrac{17}{14} i \log(2) \right]x^{5/2} +\left[\dfrac{15223771}{1455300} - \dfrac{214 \gamma_E}{105}+ \dfrac{\pi^2}{6} + \left(-\dfrac{102119}{2376} + \frac{205 \pi^2}{128}\right) \nu + \right. \nonumber \\
& \left. - \dfrac{4211}{8316} \nu^2 + \dfrac{2263}{8316} \nu^3 + i \pi \left(\dfrac{109}{210} - 2 \log(2) \right) - \log(2) \left(1+2 \log(2)\right) - \dfrac{107}{105} \left(2 \log(2) + \log(x) \right) \right]x^3  \ , \\
\nonumber\\
\hat{h}_{33}^{\rm 3PN}&= 1 + \left(-4+2 \nu \right)x + x^{3/2} \left[3 \pi + i \left(-\dfrac{21}{5} + 6 \log \left( \dfrac{3}{2} \right) \right) \right] + \left(\dfrac{123}{110}-\dfrac{1838}{165}\nu+\dfrac{887}{330}\nu^2\right)x^2
+\left[-12 \pi + \dfrac{9 \pi}{2} \nu + \right. \nonumber \\
& \left. + i \left(\dfrac{84}{5} - 24 \log \left( \dfrac{3}{2} \right) + \nu \left(-\dfrac{48103}{1215} + 9 \log \left(\dfrac{3}{2}\right) \right) \right) \right]x^{5/2} +\left[\dfrac{19388147}{280280} + \dfrac{492}{35} \log \left(\dfrac{3}{2}\right)- 18 \log^2 \left(\dfrac{3}{2}\right) - \dfrac{78}{7} \gamma_E  + \right. \nonumber \\
& \left. +  \dfrac{3}{2} \pi^2 + 6 i \pi \left(-\dfrac{41}{35} + 3 \log \left(\dfrac{3}{2}\right)\right) + \left(-\dfrac{7055}{429} + \frac{41 \pi^2}{8}\right) \dfrac{\nu}{8} - \dfrac{318841}{17160} \nu^2 + \dfrac{8237}{2860} \nu^3 - \dfrac{39}{7} \log(16 x)\right]x^3  \ ,\\
\nonumber\\
\hat{h}_{31}^{\rm 3PN}&= 1 + \left(-\dfrac{8}{3} - \dfrac{2}{3}\nu \right)x + x^{3/2} \left[\pi + i \left(-\dfrac{7}{5} - 2 \log (2) \right) \right] + \left(\dfrac{607}{198}-\dfrac{136}{99}\nu-\dfrac{247}{198}\nu^2\right)x^2
+\left[-\dfrac{8}{3} \pi - \dfrac{7 \pi}{6} \nu + \right. \nonumber \\
& \left. + i \left(\dfrac{56}{15} + \dfrac{16}{3} \log (2) + \nu \left(-\dfrac{1}{15} + \dfrac{7}{3} \log (2) \right) \right) \right]x^{5/2} +\left[\dfrac{10753397}{1513512} - 2 \log(2) \left(\dfrac{212}{105} + \log(2) \right) - \dfrac{26}{21} \gamma_E  + \right. \nonumber \\
& \left. +  \dfrac{\pi^2}{6} - 2 i \pi \left(\dfrac{41}{105} + \log (2)\right) + \left(-\dfrac{1738843}{19305} + \frac{41 \pi^2}{8}\right) \dfrac{\nu}{8} + \dfrac{327059}{30888} \nu^2 - \dfrac{17525}{15444} \nu^3 - \dfrac{13}{21} \log(x)\right]x^3  \ ,\\
\nonumber\\
\hat{h}_{32}^{\rm 2.5PN}&= 1 + \left(-\dfrac{193}{90} + \dfrac{145}{18}\nu - \dfrac{73}{18} \nu^2 \right) \dfrac{x}{1-3 \nu} + \dfrac{x^{3/2}}{1-3 \nu} \left[2 \pi - 6 \pi \nu + i \left(-3 + \dfrac{66}{5} \nu \right) \right] + \nonumber \\
& + \left(- \dfrac{1451}{3960}-\dfrac{17387}{3960}\nu + \dfrac{5557}{220}\nu^2 - \dfrac{5341}{1320} \nu^3 \right) \dfrac{x^2}{1-3\nu} + \nonumber \\
& + \left[\dfrac{193}{30}i - \dfrac{193}{45} \pi + \nu \left(-\dfrac{258929}{5400} i + \dfrac{136}{9} \pi \right) + \nu^2 \left(\dfrac{33751}{450} i - \dfrac{46}{9} \pi \right) \right] \dfrac{x^{5/2}}{1-3\nu}\ ,
\end{align}

\begin{align}
\hat{h}_{44}^{\rm 2.5PN}&= 1 + \left(-\dfrac{593}{110} + \dfrac{1273}{66}\nu - \dfrac{175}{22} \nu^2 \right) \dfrac{x}{1-3 \nu} + \dfrac{x^{3/2}}{1-3 \nu} \left[4 \pi - 12 \pi \nu + i \left(-\dfrac{42}{5} + \left(\dfrac{1193}{40}- 24 \log(2) \right) \nu  + 8 \log(2) \right) \right] + \nonumber \\
& + \left(\dfrac{1068671}{200200}-\dfrac{1088119}{28600}\nu + \dfrac{146879}{2340}\nu^2 - \dfrac{226097}{17160} \nu^3 \right) \dfrac{x^2}{1-3\nu} + \nonumber \\
& + \left[\dfrac{12453}{275} i - \dfrac{1186}{55} \pi - \dfrac{2372}{55} i \log(2) + \nu \left(-\dfrac{31525499}{140800} i + \dfrac{2480}{33} \pi + \dfrac{4960}{33} i \log(2) \right) + \right. \nonumber \\
& \left.+ \nu^2 \left(\dfrac{4096237}{21120} i - \dfrac{284}{11} \pi - \dfrac{568}{11} i \log(2) \right) \right] \dfrac{x^{5/2}}{1-3\nu}\ , \\
\nonumber \\
\hat{h}_{43}^{\rm 2PN}&= 1 + \left(-\dfrac{39}{11} + \dfrac{1267}{132}\nu - \dfrac{131}{33} \nu^2 \right) \dfrac{x}{1-2 \nu} + \dfrac{x^{3/2}}{1-2 \nu} \left[3 \pi - 6 \pi \nu + i \left(-\dfrac{32}{5} + \left(\dfrac{16301}{810}- 12 \log \left(\dfrac{3}{2} \right) \right) \nu  + \right. \right. \nonumber \\
& \left. \left. +6 \log \left(\dfrac{3}{2} \right) \right) \right] + \dfrac{x^2}{1-2\nu} \left(\dfrac{7206}{5005} - \dfrac{82869}{5720} \nu + \dfrac{104839}{3432} \nu^2 - \dfrac{2987}{572} \nu^3 \right) \ ,\\
\nonumber\\
\hat{h}_{42}^{\rm 2.5PN}&= 1 + \left(-\dfrac{437}{110} + \dfrac{805}{66}\nu - \dfrac{19}{22} \nu^2 \right) \dfrac{x}{1-3 \nu} + \dfrac{x^{3/2}}{1-3 \nu} \left[2 \pi - 6 \pi \nu + i \left(-\dfrac{21}{5} + \dfrac{84}{5} \nu \right) \right] + \nonumber \\
& + \dfrac{x^2}{1-3 \nu} \left(\dfrac{1038039}{200200}-\dfrac{606751}{28600} \nu + \dfrac{400453}{25740} \nu^2 + \dfrac{25783}{17160} \nu^3 \right) + \nonumber \\
& + \dfrac{x^{5/2}}{1-3 \nu} \left[\dfrac{9177}{550} i - \dfrac{437}{55} \pi + \nu \left(-\dfrac{83029}{880} i + \dfrac{772}{33} \pi \right) + \nu^2 \left(\dfrac{93081}{1100} i + \dfrac{14}{11} \pi \right) \right]\ ,\\
\nonumber\\
\hat{h}_{41}^{\rm 2PN}&= 1 + \left(-\dfrac{101}{33} + \dfrac{337}{44}\nu - \dfrac{83}{33} \nu^2 \right) \dfrac{x}{1-2 \nu} + \dfrac{x^{3/2}}{1-2 \nu} \left[\pi - 2 \pi \nu + i \left(-\dfrac{32}{15} + \left(\dfrac{1661}{30} + 4 \log (2) \right) \nu  -2 \log (2) \right) \right] + \nonumber \\
& + \dfrac{x^2}{1-2 \nu} \left(\dfrac{42982}{15015} - \dfrac{513989}{51480} \nu + \dfrac{196957}{10296} \nu^2 - \dfrac{1195}{572} \nu^3 \right) \ ,\\
\nonumber\\
\hat{h}_{55}^{\rm 2PN}&= 1 + \left(-\dfrac{263}{39} + \dfrac{688}{39}\nu - \dfrac{256}{39} \nu^2 \right) \dfrac{x}{1-2 \nu} + \dfrac{x^{3/2}}{1-2 \nu} \left[5 \pi - 10 \pi \nu + i \left(-\dfrac{181}{14} + \left(\dfrac{105834}{3125} - 20 \log \left(\dfrac{5}{2}\right) \right) \nu + \right. \right. \nonumber \\
&\left. \left. +10 \log \left(\dfrac{5}{2}\right) \right) \right] + \dfrac{x^2}{1-2 \nu} \left(\dfrac{9185}{819} - \dfrac{188765}{3276} \nu + \dfrac{54428}{819} \nu^2 - \dfrac{10567}{819} \nu^3 \right) \ ,\\
\nonumber\\
\hat{h}_{53}^{\rm 2PN}&= 1 + \left(-\dfrac{69}{13} + \dfrac{464}{39}\nu - \dfrac{88}{39} \nu^2 \right) \dfrac{x}{1-2 \nu}+ \dfrac{x^{3/2}}{1-2 \nu} \left[3 \pi - 6 \pi \nu + i \left(-\dfrac{543}{70} + \left(\dfrac{83702}{3645} - 12 \log \left(\dfrac{3}{2}\right) \right) \nu +  \right. \right. \nonumber \\
& \left. \left. +6 \log \left(\dfrac{3}{2}\right) \right) \right] + \dfrac{x^2}{1-2 \nu} \left(\dfrac{12463}{1365} - \dfrac{56969}{1820} \nu + \dfrac{2172}{91} \nu^2 - \dfrac{365}{273} \nu^3 \right) \ ,\\
\nonumber\\
\hat{h}_{51}^{\rm 2PN}&= 1 + \left(-\dfrac{179}{39} + \dfrac{352}{39}\nu - \dfrac{4}{39} \nu^2 \right) \dfrac{x}{1-2 \nu} + \dfrac{x^{3/2}}{1-2 \nu} \left[\pi - 2 \pi \nu + i \left(-\dfrac{181}{70} + \left(\dfrac{626}{5} + 4 \log (2) \right) \nu  - 2 \log (2) \right) \right] + \nonumber \\
& + \dfrac{x^2}{1-2\nu} \left(\dfrac{5023}{585} - \dfrac{49447}{2340} \nu + \dfrac{68}{9} \nu^2 + \dfrac{287}{117} \nu^3 \right) \ .
\end{align}
\end{widetext}

\section{Simplified rewriting of tail factors}
\label{sec:AppTails}
The evaluation of the $\Gamma$ functions could be computationally expensive. We thus 
note that $T_{\ell m}^{2,\text{trnsc}}$ and  $\tilde{T}_{\ell m}^{\text{trnsc}}$ can be conveniently 
rewritten with a reduced number of $\Gamma$ function using the relation
$\Gamma(z+1) = z \Gamma(z)$ and $\Gamma(z) \Gamma(1-z) = \pi/\sin(\pi z)$ so that
\begin{widetext}
\begin{align}
\textcolor{black}{T_{\ell m}^{2,\text{trnsc}}} & = \textcolor{black}{\frac{e^{-i \left(\hl + \frac{1}{2}\right) \pi}  \pi^2 \Gamma \left(\hl - 1 - 2 i \w \right)^3 \left(\hl - 2 i \w \right)^2 \left(\hl - 1 -2 i \w \right)^2 \left( \hl + 1 - 2 i \w \right) \left(\hl + 2 - 2 i \w \right)}{\left(2 \hl + 1\right)^2 \sin(2 \pi \hl)^2 \Gamma(2 \hl + 1)^4 \Gamma \left(-2 -\hl  - 2 i \w \right)^3 \left(-1 -\hl - 2 i \w \right)^2 \left(-2 -\hl - 2 i \w \right)^2 \left(-\hl -2 i \w\right) \left(1-\hl - 2 i \w \right)}} \ , \label{eq:T2_trscn_easy}\\
\textcolor{black}{\tilde{T}_{\ell m}^{\text{trnsc}}} & = \textcolor{black}{\frac{\pi \, \Gamma \left(\hl + 1 - 2 i \w \right)^2 \left(\hl + 1 -2 i \w \right) \left(\hl + 2 - 2 i \w \right)}{\left(2 \hl + 1 \right) (-\sin(2 \pi \hl)) \Gamma(2\hl + 1)^2 \Gamma \left(-\hl - 2 i \w \right)^2 \left(1 - \hl -2 i \w \right) \left(-\hl -2 i \w \right)}} \ . \label{eq:Tt_trscn_easy}
\end{align}
\end{widetext}
When we expand the $\Gamma$-functions that contain $\hl$ in order to obtain the PN expansions of the waveform, using the 
formulas reported above and in the text is not very efficient, especially when we want to go to high order. A more convenient 
approach is to exploit again the properties of the $\Gamma$-function, together with the fact that its derivatives always yield 
polygamma functions. Hence we use the following representation for the function $\Gamma(x)$, for $x$ close to 0
\be
\Gamma(x) = \frac{1}{x} \text{Exp} \left(\sum_{m=0}^\infty \varPsi^{(m-1)}(1) \frac{x^m}{m!}\right) \ ,
\ee
where $\psi^{(m-1)}(1)$ is the $(m-1)^{\rm th}$ derivative of the digamma function evaluated in 1 and it is related to the Riemann
$\zeta$ function by the relation
\be
\varPsi^{m-1}(1) = (-1)^{m+1} \, m! \, \zeta(m+1) \ ,
\ee
Obviously, we have to truncate the series to a given order $n_{\Gamma}$ and, for getting the $h_{22}$ waveform at 10 PN, we
took $n_{\Gamma}=16$. We call the truncated summation as $\tilde{\Gamma}(x)$
\be
\tilde{\Gamma}(x) = \frac{1}{x} \text{Exp} \left(\sum_{m=0}^{n_\Gamma} \varPsi^{(m-1)}(1) \frac{x^m}{m!}\right)\ .
\ee
At this point, using the recursive relation we already mentioned above, we can define the function that is obtained from the
original $\Gamma(x)$ through a shift by an integer $n$. In particular
\begin{widetext}
\begin{align}
& \tilde{\Gamma}_n (x+n) = \left(\prod_{k=1}^n (x+n-k) \right) \tilde{\Gamma}(x) \quad \text{for} \, \, n >0 \ ,\\
& \notag \\
& \tilde{\Gamma}_n (x+n) = \left(\prod_{k=0}^{|n|-1} (x+n+k) \right)^{-1} \tilde{\Gamma}(x) \quad \text{for} \, \, n <0 \ ,
\end{align}
In this way we rewrite $T^{1,\rm trscn}$ in \eqref{eq:T1_trscn} and \eqref{eq:T2_trscn_easy}-\eqref{eq:Tt_trscn_easy} in the following way
\begin{align}
T_{\ell m}^{1,\text{trnsc}} & = \frac{\tilde{\Gamma}_{\ell-1} \left(\hl - 1 - 2 i \w \right)}{\tilde{\Gamma}_{2 \ell + 2} \left(2 \hl+ 2\right)} e^{\pi \w} e^{-i\frac{\pi}{2} \left(\hl- \ell \right)}\ ,\\
\textcolor{black}{T_{\ell m}^{2,\text{trnsc}}} & = \textcolor{black}{\frac{e^{-i \left(\hl + \frac{1}{2}\right) \pi}  \pi^2 \, \tilde{\Gamma}_{\ell-1} \left(\hl - 1 - 2 i \w \right)^3 \left(\hl - 2 i \w \right)^2 \left(\hl - 1 -2 i \w \right)^2 \left( \hl + 1 - 2 i \w \right) \left(\hl + 2 - 2 i \w \right)}{(2 \hl + 1)^2 \sin(2 \pi \hl)^2 \tilde{\Gamma}_{2\ell+1}(2 \hl + 1)^4 \tilde{\Gamma}_{-2-\ell} \left(-2 -\hl  - 2 i \w \right)^3 (-1 -\hl - 2 i \w)^2 (-2 -\hl - 2 i \w)^2 (-\hl -2 i \w) (1-\hl - 2 i \w)}} \ ,\\
\textcolor{black}{\tilde{T}_{\ell m}^{\text{trnsc}}} & = \textcolor{black}{\frac{\pi \, \tilde{\Gamma}_{\ell+1} \left(\hl + 1 - 2 i \w \right)^2 \left(\hl + 1 -2 i \w \right) \left(\hl + 2 - 2 i \w \right)}{\left(2 \hl + 1 \right) (-\sin(2 \pi \hl)) \tilde{\Gamma}_{2 \ell+1}(2\hl + 1)^2 \tilde{\Gamma}_{-\ell} \left(-\hl - 2 i \w \right)^2 \left(1 - \hl -2 i \w \right) \left(-\hl -2 i \w \right)}} \ .
\end{align}
\end{widetext}

\section{Explicit evaluation of $\hat{\ell}$ and $\hhl$ up to $\ell=8$.} 
\label{app:a_expressions}
Let us list explicitly the PN-expansion of $\hat{\ell}=a-1/2$ up to $\ell=8$ as a function of $\w\equiv M\omega$,
where $\omega$ is the gravitational wave frequency. For each multipole, we retain terms up 
to $\w^{10}$, that  corresponds to 15PN. For circular orbits $\w=m\Omega=m x^{3/2}$.
\begin{widetext}
 \begin{align}
\hat{2} & =2 -\frac{214 \w^2}{105}-\frac{3390466 \w^4}{1157625}-\frac{153440219802466 \w^6}{15021833990625}-\frac{71638806585865707261481 \w^8}{1520451676706008921875} + \notag \\
& -\frac{270360664939833821554899493653643 \w^{10}}{1099244369724415858768042968750} + \mathcal{O}(\w^{12}) \ , \\
\nonumber\\
\hat{3} & =3 -\frac{26 \w^2}{21}-\frac{21842 \w^4}{33957}-\frac{381415329076 \w^6}{481821815475}-\frac{47254211021655226 \w^8}{35059764403038375}
-\frac{225004388212297377065114 \w^{10}}{80542621464278043695625} + \mathcal{O}(\w^{12}) \ , \\
\hat{4} & = 4-\frac{3142 \w^2}{3465}-\frac{136964836738 \w^4}{540820405125}-\frac{13932003344124287414 \w^6}{84411749090784740625}-\frac{6515321108662855725628955741 \w^8}{44795217188455810394959265625}+\notag \\
&-\frac{769810485256571907180627931975005951024593 \w^{10}}{5068298023848664664410731441360435855468750}+ \mathcal{O}(\w^{12})\ , \\
\nonumber\\
\hat{5} & = 5-\frac{1546 \w^2}{2145}-\frac{2934884558 \w^4}{23028130125}-\frac{5931370713515016362 \w^6}{113475665579372971875}-\frac{20361021135584219709260016691 \w^8}{708286794070019020577861765625}+\notag \\
&-\frac{585354059401551019020912248530563931279 \w^{10}}{31578210440387429769654744808490378906250} + \mathcal{O}(\w^{12}) \ ,\\
\nonumber\\
\hat{6} & = 6-\frac{1802 \w^2}{3003}-\frac{169547733896 \w^4}{2301891887295}-\frac{82446690884152932671 \w^6}{3944111533764126174450}-\frac{18568771026973098936173491476839 \w^8}{2343648850762937006487915050934600}+\notag \\
&-\frac{1387078951025544741383806648117966366402577 \w^{10}}{395829226692071252556053586517985515259220000} + \mathcal{O}(\w^{12}) \ ,\\
\nonumber\\
\hat{7} & = 7-\frac{11948 \w^2}{23205}-\frac{486532547934943 \w^4}{10446023432344500}-\frac{7620497202213615148920871 \w^6}{783734867119804415490675000} + \notag \\
& -\frac{316031621931666007014822597904807050533 \w^8}{116850054511874382038316120275255124000000}+ \notag \\
& -\frac{59734177894542379514105203127666538672120603497653 \w^{10}}{68223783535780518923889918622775685235766757000000000} + \mathcal{O}(\w^{12}) \ , \\
\nonumber\\
\hat{8} & = 8-\frac{1312 \w^2}{2907}-\frac{1403375484947 \w^4}{44710186690260}-\frac{109429312446368161130707 \w^6}{21746955660952413168827550} + \notag \\
& -\frac{2906196929563736466118451550378731629 \w^8}{2707886268776170104684184599532920864000} + \notag \\
& -\frac{12511597984300589057255659344921316318815343747 \w^{10}}{47039681702851577180201823360757423055889737940000} + \mathcal{O}(\w^{12}) \ .
\end{align}
\end{widetext}
For the expression of $\hat{\hat{\ell}}$ we rely on the expression of $\hat{\ell}$ for a Kerr black hole,
since, following Ref.~\cite{Ivanov:2025ozg} (see also our Sec.~\ref{sec:comparable_mass}), we need 
explicit the dependence on the black hole spin. These quantities can be (partly) found explicitly in Ref.~\cite{Ivanov:2025ozg}, 
though we recomputed explicitly from our method and the quantum period $a$ from a Kerr background, 
following a procedure that will be detailed elsewhere. 
Beyond Eq.~\eqref{eq:gamma_2m_univ}, we have
\begin{widetext}
\begin{align}
& \hat{\hat{3}} = 3 + \gamma_{3 m}^{\rm univ} = 3 - \frac{26}{21} (\hhk)^2 + \frac{7 m \mathcal{J}}{45} (\hhk)^3 - \frac{21842}{33957} (\hhk)^4 + \frac{286631 m \mathcal{J}}{935550} (\hhk)^5 \ , \\
& \notag \\
& \hat{\hat{4}} = 4 + \gamma_{4 m}^{\rm univ} = 4 - \frac{3142}{3465} (\hhk)^2 + \frac{53 m \mathcal{J}}{825} (\hhk)^3 - \frac{136964836738}{540820405125} (\hhk)^4 + \frac{17165882093 m \mathcal{J}}{257533526250} (\hhk)^5 \ , \\
& \notag \\
& \hat{\hat{5}} = 5 + \gamma_{5 m}^{\rm univ} = 5 - \frac{1546}{2145} (\hhk)^2 + \frac{7486 m \mathcal{J}}{225225} (\hhk)^3 - \frac{2934884558}{23028130125} (\hhk)^4 + \frac{470191444829 m \mathcal{J}}{21761582968125} (\hhk)^5 \ , \\
& \notag \\
& \hat{\hat{6}} = 6 + \gamma_{6 m}^{\rm univ} = 6 - \frac{1802}{3003} (\hhk)^2 + \frac{2053 m \mathcal{J}}{105105} (\hhk)^3 - \frac{169547733896}{2301891887295} (\hhk)^4 + \frac{8470028811857 m \mathcal{J}}{966794592663900} (\hhk)^5 \ , \\
& \notag \\
& \hat{\hat{7}} = 7 + \gamma_{7 m}^{\rm univ} = 7 - \frac{11948}{23205} (\hhk)^2 + \frac{4867 m \mathcal{J}}{389844} (\hhk)^3 - \frac{486532547934943}{10446023432344500} (\hhk)^4 + \frac{1082205751733021 m \mathcal{J}}{263239790495081400} (\hhk)^5 \ , \\
& \notag \\
& \hat{\hat{8}} = 8 + \gamma_{8 m}^{\rm univ} = 8 - \frac{1312}{2907} (\hhk)^2 + \frac{3449 m \mathcal{J}}{406980} (\hhk)^3 - \frac{1403375484947}{44710186690260} (\hhk)^4 + \frac{3349089392933 m \mathcal{J}}{1564856534159100} (\hhk)^5 \ .
\end{align}

\end{widetext}

\section{The residual waveform amplitudes $\rho_{\lm}$ and $\tilde{f}_{\lm}$} 
\label{app:rholm_flm_PN}
In this Appendix we list the 10PN accurate functions $\rho_{\lm}$ and $\tilde{f}_{\lm}$ up to $\ell=8$. 
Note that we also include the terms dependent on the symmetric mass ratio $\nu$, up to the corresponding 
order known in the literature, that is obtained as explained in the main text factorizing the $\hat{h}_\lm$'s modes 
listed in Appendix~\ref{app:PN_exp_hhat}. 

\begin{widetext}
\begin{align}
\rho_{22} & = 1+x \left(-\frac{43}{42}+\frac{55 \nu }{84}\right)+x^2 \left(-\frac{20555}{10584}-\frac{33025 \nu }{21168}+\frac{19583 \nu^2}{42336}\right)+x^3 \left[-\frac{4296031}{4889808}+\left(-\frac{48993925}{9779616}+\frac{41 \pi^2}{192}\right) \nu + \right. \notag \\
& \left.-\frac{6292061 \nu^2}{3259872}+\frac{10620745 \nu^3}{39118464}\right]+x^4 \left[\frac{9228174993589}{800950550400}+\left(\frac{10815863492353}{640760440320}-\frac{3485 \pi ^2}{5376}\right) \nu^2-\frac{2088847783 \nu^3}{11650189824} + \right. \notag \\
& \left. +\frac{70134663541 \nu ^4}{512608352256}+\nu \left(-\frac{2487107795131}{145627372800}+\frac{464 \gamma }{35}-\frac{9953\pi ^2}{21504}+\frac{928 \log (2)}{35}+\frac{232 \log (x)}{35}\right)\right]-\frac{8938613036677 x^5}{2116091577600}+ \notag \\
& -\frac{1060700697798333909671 x^6}{24231643979185843200} +\frac{3567168919606240724303840051 x^7}{43991338062012939037440000}+\frac{8339316227220569285816625738049 x^8}{279101750471556914871889920000}+ \notag \\
& -\frac{522338057689474511990262498143822507399 x^9}{857097472947610731676894961786880000}+\frac{1523513000214555169284583871085138536795675131x^{10}}{1333729377653777059562416250036563968000000} + \notag \\
& + \mathcal{O}(x^{11})
\end{align}

\begin{align}
\rho_{21} & = 1+x \left(-\frac{59}{56}+\frac{23 \nu }{84}\right)+x^2 \left(-\frac{47009}{56448}-\frac{10993 \nu }{14112}+\frac{617 \nu^2}{4704}\right)+x^3 \left[-\frac{252637559}{521579520}+\left(\frac{1024181}{17385984}-\frac{41 \pi ^2}{768}\right) \nu + \right. \notag \\
&\left. + \frac{622373 \nu ^2}{8692992}+\frac{2266171 \nu^3}{39118464}\right]+\frac{52482698065069 x^4}{22782593433600}-\frac{402632667232813 x^5}{1772899998105600}-\frac{55196138528767137659521 x^6}{18380150692360224768000} \notag \\
& +\frac{101595027514125796087705740847 x^7}{26694595659556296044052480000}+\frac{11469905672332070115751134880510891 x^8}{1279632137536490607167699681280000}+ \notag \\
& +\frac{43091271838801491480129834661226441842151 x^9}{13869318515332004175599256349546905600000}+ \frac{58845426920265862565878145589551391900988791691 x^{10}}{1918404137040722817568889138269327982592000000} + \notag \\
& + \mathcal{O}(x^{11})
\end{align}

\begin{align}
\rho_{33} & = 1+x \left(-\frac{7}{6}+\frac{2 \nu }{3}\right)+x^2 \left(-\frac{6719}{3960}-\frac{1861 \nu }{990}+\frac{149 \nu^2}{330}\right)+x^3 \left[\frac{3375943}{4633200}+\left(-\frac{129509}{25740}+\frac{41 \pi^2}{192}\right) \nu + \right. \notag \\
&\left. -\frac{274621 \nu ^2}{154440} +\frac{12011 \nu^3}{46332}\right]+\frac{76091363683 x^4}{8562153600}-\frac{209877220896767 x^5}{30566888352000}-\frac{10477588355364136967 x^6}{427115497640832000} +\notag \\
& +\frac{37752180939909783102121x^7}{627859781532023040000}+\frac{60686120786493671963280239 x^8}{13959119588370179198976000}-\frac{478820074436573768980537201143457 x^9}{1728876844218470608590888960000}+ \notag \\
& +\frac{3481966952168645181645222016443568249x^{10}}{5082897922002303589257213542400000}+ \mathcal{O}(x^{11})
\end{align}

\begin{align}
\rho_{32} & = 1+\frac{x \left(328-1115 \nu +320 \nu^2\right)}{270 (-1+3 \nu )}+\frac{x^2 \left(-1444528+8050045 \nu -4725605 \nu ^2-20338960 \nu ^3+3085640 \nu ^4\right)}{1603800 (1-3 \nu )^2} +\frac{36836452 x^3}{134336475}+ \notag \\
&+\frac{1655232970982 x^4}{438877263825}-\frac{368441088146531152 x^5}{176264081083715625}-\frac{7823394309761517497643574 x^6}{1357708456043130926109375}+\frac{145905104121530562581027648 x^7}{14099280120447898078828125}+ \notag \\
& -\frac{4486340035923020870882627246806 x^8}{29798989266360005695641340828125}-\frac{10151083533213254195947791183849086813144 x^9}{312998898582333932825165560622855859375}+ \notag \\ 
& + \frac{8331545419888807518348040643491314189335456 x^{10}}{211274256543075404656986753420427705078125} + \mathcal{O}(x^{11})
\end{align}

\begin{align}
\rho_{31} & = 1+x \left(-\frac{13}{18}-\frac{2}{9} \nu \right)+ x^2 \left(\frac{101}{7128}-\frac{1685 \nu}{1782} - \frac{829 \nu^2}{1782}\right) + x^3 \left[\frac{52932229}{125096400} + \left(-\frac{9688441}{2084940}+\frac{41 \pi^2}{192}\right) \nu + \right. \notag \\
&\left.+ \frac{174535 \nu^2}{75816} - \frac{727247 \nu^3}{1250964}\right]+\frac{1116965355763 x^4}{693534441600} +\frac{84898244211959 x^5}{59422030956288}+\frac{1193868859143706092923 x^6}{435914076892233139200}+ \notag \\
& +\frac{8920956949869915872083867 x^7}{1373129342210534388480000}+\frac{112201791409697214826977244735 x^8}{9524921496406861555346079744} + \notag \\
& +\frac{26378952364648431230144036956143416383 x^9}{1191031902366325494611306358988800000}+\frac{765331488484898455876880257974326109761 x^{10}}{17655296435077295567179364850892800000}+ \mathcal{O}(x^{11})
\end{align}

\begin{align}
\rho_{44} & = 1+\frac{x \left(1614-5870 \nu +2625 \nu ^2\right)}{1320 (-1+3 \nu )}+\frac{x^2 \left(-511573572+2338945704 \nu -313857376 \nu ^2-6733146000 \nu^3+1252563795 \nu ^4\right)}{317116800 (1-3 \nu )^2}+ \notag \\
& +\frac{24690609991 x^3}{17441424000}+\frac{17020275511082521 x^4}{2158550634240000}-\frac{510946845905859837493 x^5}{63159191557862400000} -\frac{28535698529573645952120671 x^6}{1540153506978358272000000} + \notag \\
& +\frac{978273446431624777735697604185239 x^7}{17620372621147024348139520000000} -\frac{15680384351704357459403788529334974083 x^8}{1191544415579005354941388554240000000}+ \notag \\
& -\frac{17487670055656300353621437544938173920755633309 x^9}{79004719075039773851279036087599104000000000} + \notag \\
&+ \frac{382857189755878843975847634249673771736872582848516356683 x^{10}}{681515061517115582488468008822487480137154560000000000}+ \mathcal{O}(x^{11})
\end{align}

\begin{align}
\rho_{43} & = 1+\frac{x \left(222-547 \nu +160 \nu ^2\right)}{176 (-1+2 \nu )}-\frac{6894273 x^2}{7047040}+\frac{506924181 x^3}{620139520}+\frac{26781110777273 x^4}{6139877359616}-\frac{28768376380442572667 x^5}{8982640577118208000} + \notag \\
& -\frac{5669316100034808751407049 x^6}{924852673820090695680000} +\frac{299611836851443183912839431461 x^7}{17135101200431674199900160000} + \notag \\
& -\frac{21268794191597811953895555624437143 x^8}{271142551456200329657773751009280000}-\frac{336865316122289780748845499803912441189725229 x^9}{8434105977086276214268778744144461824000000}+ \notag \\
& + \frac{28686781774593283181871565750327261318744729306834027 x^{10}}{306335878405937495294679282758508501359866675200000} + \mathcal{O}(x^{11})
\end{align}

\begin{align}
\rho_{42} & = 1+\frac{x \left(1146-3530 \nu +285 \nu ^2\right)}{1320 (-1+3 \nu )}+\frac{x^2 \left(-114859044+295834536 \nu +1204388696 \nu ^2-3047981160 \nu ^3-379526805 \nu ^4\right)}{317116800 (1-3 \nu )^2} + \notag \\
& +\frac{1523288561 x^3}{3488284800}+\frac{4985601363054409 x^4}{2158550634240000}+\frac{4641602014445914289 x^5}{7017687950873600000} +\frac{38419606505272976130955243 x^6}{29262916632588807168000000}  + \notag \\
& +\frac{145058623694284996827653984925973 x^7}{17620372621147024348139520000000}+\frac{28631337402878870733335998802795474191 x^8}{2331282552219793085754890649600000000}+ \notag \\
& +\frac{31788515917506624732022490108579280412370671459 x^9}{1501089662425755703174301685664382976000000000} + \notag \\
& + \frac{39730267245587636539364034402243138450376132405595215123 x^{10}}{681515061517115582488468008822487480137154560000000000}+ \mathcal{O}(x^{11})
\end{align}

\begin{align}
\rho_{41} & = 1+\frac{x \left(602-1385 \nu +288 \nu ^2\right)}{528 (-1+2 \nu )}-\frac{7775491 x^2}{21141120}-\frac{2639236841 x^3}{16743767040}+\frac{756631624067659 x^4}{4144417217740800}-\frac{9886295843201751293 x^5}{18656253506322432000} + \notag \\
& -\frac{60281579848409508278066009 x^6}{74913066579427346350080000} -\frac{31070286422874133517320026060091 x^7}{32477870815298195278490763264000}+ \notag \\
& -\frac{122314296664176349124913244342128750421 x^8}{65887640003856680106839021495255040000}-\frac{20774409877077429802225089670680044495266041349 x^9}{6148463257295895360201939704481312669696000000}+ \notag \\
& - \frac{58553041735688248975694376316532124661587569538152493001 x^{10}}{10049348491106779533141953870892871387110426279936000000} + \mathcal{O}(x^{11})
\end{align}

\begin{align}
\rho_{55} & = 1+\frac{x \left(487-1298 \nu +512 \nu ^2\right)}{390 (-1+2 \nu )}-\frac{3353747 x^2}{2129400} +\frac{178480932157 x^3}{98825454000}+\frac{21607314190081609 x^4}{2929186456560000}-\frac{3853382562691891616251 x^5}{439817346452484000000} + \notag \\
& -\frac{14351664402623276428715235427 x^6}{938948460247549992240000000}+\frac{592982855291865037119711733345769 x^7}{10933384142111114266783200000000} + \notag \\
& -\frac{408020694009589851775646530433754459227 x^8}{18386453444105418640163971776000000000}-\frac{8565825351697101879212948148444655068785476699 x^9}{43781639076569381811772047953293440000000000}+ \notag \\
& + \frac{582716042044653399715960479647006675247244582205014897 x^{10}}{1106278834350662796558037284888613971264000000000000}+ \mathcal{O}(x^{11})
\end{align}

\begin{align}
\rho_{54} & = 1+\frac{x \left(-17448+96019 \nu -127610 \nu ^2+33320 \nu ^3\right)}{13650\left(1-5 \nu +5 \nu ^2\right)}-\frac{16213384 x^2}{15526875}+\frac{23907047147974 x^3}{19816562390625}+\frac{3760125034976608 x^4}{799786096484375}+ \notag \\
& -\frac{51404108646315165422368 x^5}{12527349549020947265625}-\frac{74089274875278791686646685725474 x^6}{12135152371668622488819580078125} + \notag \\
& +\frac{84717988285768787787337675831275187208 x^7}{3671793728857633449554584442138671875}-\frac{38954287305412322519905252750338181095184 x^8}{25059992199453348293210038817596435546875}+ \notag \\
& -\frac{2488682065054446388634785298772309212618963458881404 x^9}{54152887064065308968245247191355754146575927734375} +\notag \\
& + \frac{53389045824413257119522042108423331533938716978843738949344 x^{10}}{362817574218354561923622125526184633312790393829345703125} + \mathcal{O}(x^{11})
\end{align}

\begin{align}
\rho_{53} & = 1+\frac{x \left(375-850 \nu +176 \nu ^2\right)}{390 (-1+2 \nu )}-\frac{410833 x^2}{709800} +\frac{2363770999 x^3}{3660202000}+\frac{10508152047048587 x^4}{3580116780240000}-\frac{401450585712686173 x^5}{3257906270018400000} + \notag \\
& +\frac{441984540599627084261833 x^6}{2318391259870493808000000}+\frac{23594980327555218075118994943313 x^7}{2078692787512483453190880000000} + \notag \\
& +\frac{190984192472829048613111096320795163 x^8}{15132883493090879539229606400000000}+\frac{209195276474917254773516793817406424031097 x^9}{12011423614971023816672715487872000000000}+ \notag \\
& + \frac{260252264560418527510988831422707763777911698582999 x^{10}}{3338564383499942595922746264410083315200000000000}+ \mathcal{O}(x^{11})
\end{align}

\begin{align}
\rho_{52} & = 1+\frac{x \left(-15828+84679 \nu -104930 \nu ^2+21980 \nu ^3\right)}{13650 \left(1-5 \nu +5 \nu ^2\right)}-\frac{7187914 x^2}{15526875}+\frac{3687844215413 x^3}{39633124781250}+\frac{1839059977566593 x^4}{1942337662890625}+ \notag \\
& -\frac{27080815463058154841 x^5}{38784363928857421875}-\frac{94646988016901552664228745857037 x^6}{97081218973348979910556640625000} +\frac{7563780090736269853904808183789887747 x^7}{14687174915430533798218337768554687500}+ \notag \\
& -\frac{83798029998498621460241150434226649988291 x^8}{100239968797813393172840155270385742187500}-\frac{2243158390827967605967018890175160986344596269408349 x^9}{866446193025044943491923955061692066345214843750000} + \notag \\
& + \frac{4099599609326793470635329972775224204805740335875194721423 x^{10}}{2902540593746836495388977004209477066502323150634765625000} + \mathcal{O}(x^{10})
\end{align}

\begin{align}
\rho_{51} & = 1+\frac{x \left(319-626 \nu +8 \nu ^2\right)}{390 (-1+2 \nu )}-\frac{31877 x^2}{304200} +\frac{7644117109 x^3}{98825454000}+\frac{12434529945101507 x^4}{32221051022160000}+\frac{15788148943867940459 x^5}{62831049493212000000} + \notag \\
& +\frac{29825384176217937274318621 x^6}{55232262367502940720000000}+\frac{1011659203961901985836599087085037 x^7}{841870578942555798542306400000000} + \notag \\
& +\frac{5981016583096207127479332693999504691 x^8}{2626636206300774091451995968000000000}+\frac{194694053752885604371116858964284344972136941 x^9}{43781639076569381811772047953293440000000000}+ \notag \\
& + \frac{107227341527224178345273875870653477415995987655216043 x^{10}}{12169067177857290762138410133774753683904000000000000}+ \mathcal{O}(x^{11})
\end{align}

\begin{align}
\rho_{66} & = 1+\frac{x \left(-106+602 \nu -861 \nu ^2+273 \nu ^3\right)}{84 \left(1-5 \nu +5 \nu ^2\right)}-\frac{1025435 x^2}{659736}+\frac{1627273547 x^3}{789703992}+\frac{114168486108508411 x^4}{16126071398236800} + \notag \\
& -\frac{1574036212967171426017 x^5}{171355634677664236800}-\frac{919173135098090560682410699 x^6}{69738316201115791092864000}+\frac{146953358169336276155911735316717 x^7}{2722514126175359368474317696000}+ \notag \\
& -\frac{1864716560214036044086534607425523950127 x^8}{67116289442995335266626791623454720000}-\frac{13434996648252344775338680008002415754324959049 x^9}{74714524570836837172161610703146028851200000}+ \notag \\
& + \frac{102719304818904051111464796394418007988761044900866333 x^{10}}{198341062081021151472753164176328011781895782400000}+ \mathcal{O}(x^{11})
\end{align}

\begin{align}
\rho_{65} & = 1+\frac{x \left(-185+838 \nu -910 \nu ^2+220 \nu ^3\right)}{144 \left(1-4 \nu +3 \nu ^2\right)}-\frac{59574065 x^2}{54286848}+\frac{8657995830811 x^3}{5792623828992}+\frac{21530564223492384535 x^4}{4367545685078704128}+ \notag \\
& -\frac{5375486196497271710568217 x^5}{1113828970791511441539072}-\frac{2176198626047627791581107566390985 x^6}{367721920108950029638606019100672}+ \notag \\
& +\frac{16019848709450202549793808488474808585 x^7}{582471521452576846947551934255464448}-\frac{126005227243759942526206577870950649250917177 x^8}{34462258410565894213300976399723211169726464} + \notag \\
& -\frac{1575111162607453652075575640156430324736648915265053229 x^9}{31120658331855379139639538285653582535150970997833728}+ \notag \\
&+ \frac{324720281805042623060753243420680740358292464587253339808825 x^{10}}{1676034175120403298944426973912159341013090694059333255168} + \mathcal{O}(x^{11})
\end{align}

\begin{align}
\rho_{64} & = 1+\frac{x \left(-86+462 \nu -581 \nu ^2+133 \nu ^3\right)}{84 \left(1-5 \nu +5 \nu ^2\right)}-\frac{476887 x^2}{659736}+\frac{1385165879 x^3}{1579407984}+\frac{4229377153108507 x^4}{1240467030633600} + \notag \\
& -\frac{22359580333328900731 x^5}{24479376382523462400}-\frac{45344566934390941710358429 x^6}{69738316201115791092864000}+\frac{6176785620187710739430421252263 x^7}{418848327103901441303741184000}+ \notag \\
& +\frac{816994182204267268625341809021785062193 x^8}{67116289442995335266626791623454720000}+\frac{48569278118055972634976041398186147184113319 x^9}{3932343398465096693271663721218212044800000}+ \notag \\
& + \frac{2841014621135065712399612734522969959847335715991 x^{10}}{27689663839316089832856786845780819737804800000}+ \mathcal{O}(x^{11})
\end{align}

\begin{align}
\rho_{63} & = 1+\frac{x \left(-169+742 \nu -750 \nu ^2+156 \nu ^3\right)}{144 \left(1-4 \nu+3 \nu ^2\right)}-\frac{152153941 x^2}{271434240}+\frac{51586722827411 x^3}{144815595724800}+\frac{175122971512190296447 x^4}{109188642126967603200}+ \notag \\
& -\frac{10829698501758916984877689 x^5}{10709893949918379245568000}-\frac{283718550047151325867884589375929689 x^6}{229826200068093768524128761937920000}+ \notag \\
& +\frac{1045200261351012276786959851724092331401 x^7}{364044700907860529342219958909665280000}+\frac{2540982735596923171164018613039703020522538687 x^8}{8284196733309109185889657788395002685030400000}+ \notag \\
& -\frac{1148374696461470846979928631877103217672448735046307450437 x^9}{486260286435240299056867785713337227111733921841152000000}+ \notag \\
& + \frac{156361119625394281735446698257853981792745918687338825116453 x^{10}}{18429299075479452178752055923559106054419100699983872000000} + \mathcal{O}(x^{11})
\end{align}

\begin{align}
\rho_{62} & = 1+\frac{x \left(-74+378 \nu -413 \nu ^2+49 \nu ^3\right)}{84 \left(1-5 \nu +5 \nu ^2\right)}-\frac{817991 x^2}{3298680}+\frac{685764011 x^3}{3948519960}+\frac{13781815719703963 x^4}{16126071398236800}+ \notag \\
& +\frac{32029128714544374911 x^5}{171355634677664236800} +\frac{36812330527951150296634997 x^6}{69738316201115791092864000}+\frac{1119149788217510433763578329381 x^7}{544502825235071873694863539200}+ \notag \\
& +\frac{6341786996944807503294843179503298491 x^8}{1917608269799866721903622617812992000}+\frac{27169468775463900930633544102161844094878007 x^9}{4394972033578637480715388864890942873600000}+ \notag \\
& + \frac{2647626554883558575890732334755753676979804972645853 x^{10}}{198341062081021151472753164176328011781895782400000}+ \mathcal{O}(x^{11})
\end{align}

\begin{align}
\rho_{61} & = 1+\frac{x \left(-161+694 \nu -670 \nu ^2+124 \nu ^3\right)}{144 \left(1-4 \nu+3 \nu ^2\right)}-\frac{79192261 x^2}{271434240}-\frac{27314166599861 x^3}{144815595724800}-\frac{181709504508887713 x^4}{1679825263491809280}+ \notag \\
& -\frac{64866662002972297454068253 x^5}{139228621348938930192384000} -\frac{157010360117387018183797863431034089 x^6}{229826200068093768524128761937920000}+ \notag \\
& -\frac{300369765056177772932142289748156789 x^7}{294773037172356703920825877659648000}-\frac{184665498019232228264348136106613385088306822669 x^8}{107694557533018419416565551249135034905395200000}+ \notag \\
& -\frac{1414165612424505478412395849261428695007018508826886065693 x^9}{486260286435240299056867785713337227111733921841152000000}+ \notag \\
& - \frac{347140144927021252743790052222963714998721568561459034612199 x^{10}}{69464281130653319750680826173415092051271994946093056000000} + \mathcal{O}(x^{11})
\end{align}

\begin{align}
\rho_{77} & = 1+\frac{x \left(-906+4246 \nu -4963 \nu ^2+1380 \nu ^3\right)}{714 \left(1-4 \nu +3 \nu ^2\right)}-\frac{32358125 x^2}{20986602}+\frac{2417601440573 x^3}{1078879235616}+\frac{5528558650842721713007 x^4}{802300883523265526400}+ \notag \\
& -\frac{88210686912309264592171853 x^5}{9308696001078688270056000}-\frac{12486670741638741400197571136971 x^6}{1069680874875954226488675072000}+ \notag \\
& +\frac{45989028721489063619400467219656391933087 x^7}{849886838616106439339046920963458560000}-\frac{14673195200298480999117846123477461247588080684783 x^8}{464641463562443828229762474350548236645888000000} + \notag \\
& -\frac{167793384822059840577494615632956060087271939153385261 x^9}{995262014950754680068151220058874322895492096000000}+ \notag \\
& + \frac{15352121877483004480583063788584347936801974932787398932681693 x^{10}}{29604781242312902698978086836924778888541605567860736000000}+ \mathcal{O}(x^{11})
\end{align}

\begin{align}
\rho_{76} & = 1+\frac{x \left(2144-16185 \nu +37828 \nu ^2-29351 \nu ^3+6104 \nu^4\right)}{1666 \left(-1+7 \nu -14 \nu ^2+7 \nu ^3\right)}-\frac{195441224 x^2}{171390583}+\frac{3674137194436 x^3}{2141525334585}+ \notag \\
&+\frac{946986560439949477346 x^4}{185795179527330297925}-\frac{54457189205362162598715837 x^5}{10059879995507298981149125} -\frac{1275209335896771690176000049467 x^6}{224777958619615088434796049000}+ \notag \\
& +\frac{2429258085241577361478337724665141517218 x^7}{78133747170605976162028633155324961875}-\frac{24497758098484671252388796475018499500416987087 x^8}{4152991625308998802006001269254827109608640625}+ \notag \\
& -\frac{3675574239242162169459751653055328928964879902440371 x^9}{67930861559872503313394163306771139288878499125000}+ \notag \\
& + \frac{4557212424377173046821902060655574739663903857970770184256249 x^{10}}{19448776612604613117779337111140538667977934762230548062500} +\mathcal{O}(x^{11})
\end{align}

\begin{align}
\rho_{75} & = 1+\frac{x \left(-762+3382 \nu -3523 \nu ^2+804 \nu ^3\right)}{714 \left(1-4 \nu+3 \nu ^2\right)}-\frac{17354227 x^2}{20986602}+\frac{1175826860213 x^3}{1078879235616}+\frac{120944989240326271063 x^4}{32092035340930621056}+ \notag \\
& -\frac{123353734363929325730879 x^5}{74469568008629506160448}-\frac{10789670516660767939133815751 x^6}{8557446999007633811909400576}+ \notag \\
& +\frac{24483664498826454932668376595898634183 x^7}{1359818941785770302942475073541533696}+\frac{327578252367522212094173267481043399474615865 x^8}{29737053667996405006704798358435087145336832} + \notag \\
& +\frac{432284730469099180791217385102239805437733135683 x^9}{63696768956848299524361678083767956665311494144}+ \notag \\
& + \frac{244516357553978806118185063657119714687583808863864382093 x^{10}}{1894705999508025772734597557563185848866662756343087104}+ \mathcal{O}(x^{11})
\end{align}

\begin{align}
\rho_{74} & = 1+\frac{x \left(17756-131805 \nu +298872 \nu ^2-217959 \nu ^3+41076 \nu^4\right)}{14994 \left(-1+7 \nu -14 \nu ^2+7 \nu ^3\right)}-\frac{2995755988 x^2}{4627545741}+\frac{16442111460952 x^3}{27388981910745}+ \notag \\
& +\frac{2606604306485596127760496 x^4}{1219002172878814084685925}-\frac{8398739013306566134647316112 x^5}{6000261150047580783210855375} -\frac{35724718138029692768043706630094 x^6}{24045133475597194487293970828625}+ \notag \\
& +\frac{78160311142026706323920266626470169518986916 x^7}{13841158910031336859174886277566351021270625} +\frac{704931738675914666790960014196215100308398799230904 x^8}{601928187912501717910055814691105792897506977015625}+ \notag \\
& -\frac{1317041709208439955364910434316964781737252922696099210328 x^9}{446752906853222512537997155831182792805908370861927921875}+ \notag \\
& + \frac{4848689906147544773286561230100462086457401422188295682509207951856 x^{10}}{209301515158528966048496816047053189957539447966484935030882015625} + \mathcal{O}(x^{11})
\end{align}

\begin{align}
\rho_{73} & = 1+\frac{x \left(-666+2806 \nu -2563 \nu ^2+420 \nu ^3\right)}{714 \left(1-4 \nu+3 \nu ^2\right)}-\frac{7804375 x^2}{20986602}+\frac{39039452957 x^3}{119875470624}+\frac{40071461417809707661 x^4}{29714847537898723200}+ \notag \\
& +\frac{33993287863873347431633 x^5}{1034299555675409807784000}+\frac{5641700867271007980534850949 x^6}{13205936726863632425786112000}+ \notag \\
& +\frac{12287674768596443789121822439479700349 x^7}{3497476702123894812094843296145920000}+\frac{26496306450213349105417381814984837836294720337 x^8}{5736314364968442323824228078401830082048000000} + \notag \\
& +\frac{10937666510038480768456575866186616629459869081091 x^9}{1365242818862489273070166282659635559527424000000}+ \notag \\
& + \frac{284061581330479993100185966614224071721365694068940081839 x^{10}}{13536708386974349656597204772256414672401282838528000000}+ \mathcal{O}(x^{11})
\end{align}

\begin{align}
\rho_{72} & = 1+\frac{x \left(16832-123489 \nu +273924 \nu ^2-190239 \nu ^3+32760 \nu^4\right)}{14994 \left(-1+7 \nu -14 \nu ^2+7 \nu ^3\right)}-\frac{1625746984 x^2}{4627545741}-\frac{1714105203716 x^3}{40030050484935}+ \notag \\
& +\frac{367401703087127453011714 x^4}{1219002172878814084685925}-\frac{33272026750411719803604199213 x^5}{66002872650523388615319409125}+ \notag \\
& -\frac{227576724781159396029375218244293 x^6}{340331119962298752743237740959000}-\frac{7418751670492904875699767659468040614961184 x^7}{13841158910031336859174886277566351021270625}+ \notag \\
&  -\frac{8292519424967403642246238490122506180896385513154811 x^8}{6621210067037518897010613961602163721872576747171875} + \notag \\
&-\frac{619340400734377998141895637562403591436428242347703170101 x^9}{274924865755829238484921326665343257111328228222724875000} + \notag \\
& - \frac{2885680423472576962277776444734794704139221393744810506249700270207 x^{10}}{837206060634115864193987264188212759830157791865939740123528062500} +\mathcal{O}(x^{11})
\end{align}

\begin{align}
\rho_{71} & = 1+\frac{x \left(-618+2518 \nu -2083 \nu ^2+228 \nu ^3\right)}{714 \left(1-4 \nu+3 \nu ^2\right)}-\frac{1055091 x^2}{6995534}-\frac{52111064275 x^3}{1078879235616}+\frac{55948333649900034703 x^4}{802300883523265526400}+ \notag \\
& -\frac{50319685880557431724273 x^5}{1034299555675409807784000}+\frac{3692567644786828116118138613 x^6}{1069680874875954226488675072000}+ \notag \\
& +\frac{118739450850098509971130520632900320527 x^7}{849886838616106439339046920963458560000}+\frac{19711898690463763439818791170617786170840147113 x^8}{51626829284715980914418052705616470738432000000}+ \notag \\
& +\frac{83769722740149343977138430254415666195156596720137 x^9}{90478364995523152733468292732624938445044736000000}+ \notag \\
& + \frac{61972581228530138869408209034899874927799293636354218421517 x^{10}}{29604781242312902698978086836924778888541605567860736000000}+ \mathcal{O}(x^{11})
\end{align}

\begin{align}
\rho_{88} & = 1+\frac{x \left(3482-26778 \nu +64659 \nu ^2-53445 \nu ^3+12243 \nu^4\right)}{2736 \left(-1+7 \nu -14 \nu ^2+7 \nu ^3\right)}-\frac{9567401 x^2}{6238080}+\frac{6994167352063 x^3}{2944124236800} + \notag \\
&+\frac{1324395872659094084407 x^4}{195820062297919488000}-\frac{42377713606135758737056277 x^5}{4375403471984713039872000}-\frac{19396231408830185467690920493889 x^6}{1842629146353823071465897984000}+ \notag \\
& +\frac{57107372272416480500837297460787595577751351 x^7}{1049549603468503614466006292301663436800000}-\frac{139277139147775794323417896196237845334550465283 x^8}{4060802829419955802879384345487163260928000000}+ \notag \\
& - \frac{3911545725831905541593420613528417792905432999087943983 x^9}{24455005783040020267675936047482511266727906508800000}+ \notag \\
& + \frac{290409219389711309120651089076609909167725765796536603595767691246517 x^{10}}{555033516885242119350503403449734748241752085352462795407360000000}+ \mathcal{O}(x^{11})
\end{align}

\begin{align}
\rho_{87} & = 1+\frac{x \left(23478-154099 \nu +309498 \nu ^2-207550 \nu ^3+38920 \nu^4\right)}{18240 \left(-1+6 \nu -10 \nu ^2+4 \nu ^3\right)}-\frac{195527087 x^2}{166348800}+\frac{44847192774761 x^3}{23727460515840}+ \notag \\
& +\frac{3234485022267774075143 x^4}{618888098126757888000}-\frac{499312774703090880001192471 x^5}{84663891823740479078400000}-\frac{4836741975445393086759059408550804473 x^6}{893033999460789216201223962624000000}+ \notag \\
& +\frac{45532622927086057554091771713152836736628767 x^7}{1337201327512602547786319834665451520000000} -\frac{4260729361268659008086620487617542426942676064627 x^8}{526835927818725202187045433740833251655680000000}+ \notag \\
& -\frac{2425342339212821992617000177896359858765013069248653581687 x^9}{42746621652073054452947756454044320135883217960960000000}+ \notag \\
& + \frac{3736895433245882001297548100800648014495989287023347559830675991507811 x^{10}}{13854415012869476741707641267625235583623217755313049436160000000000} + \mathcal{O}(x^{11})
\end{align}

\begin{align}
\rho_{86} & = 1+\frac{x \left(1002-7498 \nu +17269 \nu ^2-13055 \nu ^3+2653 \nu^4\right)}{912 \left(-1+7 \nu -14 \nu ^2+7 \nu ^3\right)}-\frac{376847 x^2}{415872}+\frac{46405282477 x^3}{36347212800}+\frac{9783772540250729687 x^4}{2417531633307648000}+ \notag \\
& -\frac{41977425301170767125213 x^5}{18005775604875362304000}-\frac{55899135842021735882399182429 x^6}{32858955970644307172917248000}+\frac{124359369957766462165284766751401921007 x^7}{5924738231347432440097807427174400000}+ \notag \\
& +\frac{575397439221284455013877513167229873295153943 x^8}{61274116767653654090086732236431818752000000}+\frac{49748319315014356704012915119583863163630052871 x^9}{40078806802398697201737762974570345240985600000}+ \notag \\
& + \frac{190613877004247835365013506223805980660471211874481388582748521 x^{10}}{1226027131966102108309097904944975134670952168727511040000000}+ \mathcal{O}(x^{11})
\end{align}

\begin{align}
\rho_{85} & = 1+\frac{x \left(4350-28055 \nu +54642 \nu ^2-34598 \nu ^3+6056 \nu^4\right)}{3648 \left(-1+6 \nu -10 \nu ^2+4 \nu ^3\right)}-\frac{4804679 x^2}{6653952}+\frac{190097528771365 x^3}{232529113055232}+ \notag \\
& +\frac{624110733686390081251 x^4}{242604134465689092096}-\frac{185444148840350898737207 x^5}{102117678753557747073024}-\frac{232581574768768666209006265126226705 x^6}{137227176493142714118344878992654336}+ \notag \\
& +\frac{1051939072286046472493406110588438081806705 x^7}{123287822874537934741822230644419165421568} +\frac{26482785694332631195469340189534169794047002433 x^8}{16191143202467317893774029906070280156483682304}+ \notag \\
& -\frac{3501401713126568590234990336138850998100374600126949360167 x^9}{836841789179925199146919139683893540552509711725526777856}+ \notag \\
& + \frac{6879718426573587912342032163873937876821540132737064373286569142699 x^{10}}{166907706557345901484560618239471753336595245639150075198954274816} + \mathcal{O}(x^{11})
\end{align}

\begin{align}
\rho_{84} & = 1+\frac{x \left(2666-19434 \nu +42627 \nu ^2-28965 \nu ^3+4899 \nu^4\right)}{2736 \left(-1+7 \nu -14 \nu ^2+7 \nu ^3\right)}-\frac{1387201 x^2}{2911104}+\frac{10153126272529 x^3}{20608869657600} + \notag \\
& +\frac{2462334195606104843809 x^4}{1370740436085436416000}-\frac{103538033652943893682741 x^5}{486155941331634782208000}+\frac{72269516570650762726364447863 x^6}{281813869442349410930078515200}+ \notag \\
& +\frac{39241168264288473786524541023384623623906553 x^7}{7346847224279525301262044046111644057600000}+\frac{258752365309146016064951498983149164278404045727 x^8}{44668831123619513831673227800358795870208000000}+ \notag \\
& +\frac{60037037845449314773882636807628632905106425837853 x^9}{6509184397934527619823246219718528418080358400000}+ \notag \\
& + \frac{2477177039690846680278395700132978633355193239841215886162103179027 x^{10}}{79290502412177445621500486207104964034536012193208970772480000000} + \mathcal{O}(x^{11})
\end{align}

\begin{align}
\rho_{83} & = 1+\frac{x \left(20598-131059 \nu +249018 \nu ^2-149950 \nu ^3+24520 \nu^4\right)}{18240 \left(-1+6 \nu -10 \nu ^2+4 \nu ^3\right)}-\frac{7756983 x^2}{18483200}+\frac{5499964522491 x^3}{43060946862080}+ \notag \\
& +\frac{275767970968712359847 x^4}{374389096397668352000}-\frac{31057368287764656198571159 x^5}{51216428387201030553600000}-\frac{1421966033677346414273632820025134381 x^6}{2036253212445731156789305540608000000}+ \notag \\
& +\frac{1148836240779786307640988945823184787635743 x^7}{2948228087564516728406752913711431680000000} -\frac{984888507583447229640584799155855288299270316363 x^8}{1735161951567570933403423986847380846673920000000}+ \notag \\
& -\frac{42990430864192773582595886014178221278582939125186886660913 x^9}{29894039679781278547486537625882113788598455065640960000000}+ \notag \\
& - \frac{15135797809549718648951142381952977652240740910671738399342630434773 x^{10}}{67754003901113233952078907906008286202672317649533397893120000000000} + \mathcal{O}(x^{11})
\end{align}

\begin{align}
\rho_{82} & = 1+\frac{x \left(2462-17598 \nu +37119 \nu ^2-22845 \nu ^3+3063 \nu^4\right)}{2736 \left(-1+7 \nu -14 \nu ^2+7 \nu ^3\right)}-\frac{9876487 x^2}{43666560}+\frac{634049806651 x^3}{20608869657600}+\frac{70401540746476908487 x^4}{195820062297919488000}+ \notag \\
& -\frac{16966826976534286320197 x^5}{397763951998610276352000}+\frac{68381401748054600733128608771 x^6}{958167156103987997162266951680}+ \notag \\
& +\frac{32700325002382756840771843212439482531067 x^7}{60717745655202688440182182199269785600000}+\frac{41928264097308069280209974408841883351995035647 x^8}{44668831123619513831673227800358795870208000000}+ \notag \\
& +\frac{46860305034538364463168079558048840018205529936064411 x^9}{24455005783040020267675936047482511266727906508800000}+ \notag \\
& + \frac{210890927187523979065576537521985537193351515027078418864144948167 x^{10}}{50457592444112919940954854859066795294704735032042072309760000000}+ \mathcal{O}(x^{11})
\end{align}

\begin{align}
\rho_{81} & = 1+\frac{x \left(20022-126451 \nu +236922 \nu ^2-138430 \nu ^3+21640 \nu^4\right)}{18240 \left(-1+6 \nu -10 \nu ^2+4 \nu ^3\right)}-\frac{44651567 x^2}{166348800}-\frac{1217603285743427 x^3}{5813227826380800}+ \notag \\
& -\frac{6078495033445038459049 x^4}{30325516808211136512000}-\frac{1865859202236857382751848871 x^5}{4148530699363283474841600000}-\frac{9977384137433732209377556805840281111 x^6}{14994228200736747609084886253568000000}+ \notag \\
& -\frac{9927997941551779487113652607478804451599119491 x^7}{9631861162073276151704861769095247298560000000} + \notag \\
& -\frac{429330521421570328318485889671010845649516543469239 x^8}{252986612538551842090219217282348127445057536000000}+ \notag \\
& -\frac{931515065686357247631661312349476534275650528984976113993082727 x^9}{326891323898408280916765288939020914278324106142783897600000000}+ \notag \\
& - \frac{7942760153876491020500405029001238710864020671889639177696341219373316\
461 x^{10}}{1629958071849081069185162287494841341177687945694824953114787840000000\
000} + \mathcal{O}(x^{11})
\end{align}

\end{widetext}
For what concerns the functions $\tilde{f}_{\lm}$'s, we only consider the terms deriving from the test-mass regime
since the $\nu$-dependent PN knowledge stops at 4PN. Hence we always report the expressions at 10 PN.
\begin{widetext}

\begin{align}
\tilde{f}_{2,2} & = 1+\frac{4391 x}{2247}+\frac{53185 x^2}{2646}+\frac{17096210 x^3}{305613}+\frac{4747421406107252 x^4}{71641272277575}+\frac{8197825650198689 x^5}{18747248820300}+ \notag \\
& + \frac{93413981315288045717 x^6}{265033606022345160}-\frac{84886593520942215307406177173 x^7}{115729970007349756213155000}+\frac{12091990099120207716578842317287 x^8}{2316762577156478297276430000} + \notag \\
& -\frac{18745458158059179828839098304527937 x^9}{5506639808719744112850116685000}-\frac{12907954629421965590241825710607690689624960837 x^{10}}{465999378807314866788638416951557033375000} + \mathcal{O}(x^{11})
\end{align}

\begin{align}
\tilde{f}_{2,1} & = 1+\frac{30209 x}{8988}+\frac{67505 x^2}{7056}+\frac{731322941 x^3}{32598720}+\frac{2569593492526159 x^4}{61751399673600}+\frac{102953731748008727 x^5}{1599765232665600} + \notag \\
& +\frac{3503605123952056862449 x^6}{41881853791185408000}+\frac{8488851720112108422701257701971 x^7}{149231653176884781937674240000} + \notag \\
& -\frac{13371545762185209389010195172133 x^8}{117153821185591791427706880000}-\frac{123525354334206981063755635013970833759 x^9}{211629005666076723870838262169600000} + \notag \\
& -\frac{3049386805706348159208619032193343388549694519 x^{10}}{1642848666694434337118276838717407232000000} + \mathcal{O}(x^{11})
\end{align}

\begin{align}
\tilde{f}_{3,3} & = 1+7 x+\frac{154109 x^2}{4290}+\frac{4448633 x^3}{28600}+\frac{1094520121 x^4}{2202200}+\frac{351681057505831 x^5}{306612306000} +\frac{378170389349492431 x^6}{122060898960000} + \notag \\
& +\frac{2104918379257053487 x^7}{543725822640000}-\frac{165316579227648026308503643 x^8}{42136259902880663520000} -\frac{301151351979811613862416479 x^9}{254155218461819875200000} + \notag \\
& -\frac{198078160021185703918032455281171 x^{10}}{2054844941263813690992000000}+ \mathcal{O}(x^{11})
\end{align}

\begin{align}
\tilde{f}_{3,2} & = 1+\frac{553 x}{90}+\frac{7924298 x^2}{289575}+\frac{16736640773 x^3}{182432250}+\frac{44795102939779 x^4}{180607927500} +\frac{1772758878836270623 x^5}{3143255218228125} + \notag \\
& +\frac{1662997766186668904247953 x^6}{1655489658336388875000}+\frac{191842084036371760940408129 x^7}{148994069250274998750000} + \notag \\
& +\frac{211558979044604598085003381272869x^8}{656043049776761797902314062500}-\frac{7622399141835252594237010120372504573 x^9}{1427788237423243403798308968750000} + \notag \\
& -\frac{333502373626151926062090409882416866407969 x^{10}}{15548613905539120667363584669687500000} + \mathcal{O}(x^{11})
\end{align}

\begin{align}
\tilde{f}_{3,1} & = 1+7 x+\frac{43119 x^2}{1430}+\frac{25868633 x^3}{257400}+\frac{50015328017 x^4}{178378200}+\frac{5683227511324613 x^5}{8278532262000}+\frac{45027198734556529453 x^6}{29660798447280000} + \notag \\
& +\frac{13461109263357330532301 x^7}{4360137371750160000}+\frac{33282286045940516784104263363 x^8}{5688395086888889575200000} + \notag \\
& +\frac{827776812642718129525245705317 x^9}{79405351825142866723200000}+\frac{236590668694363070389235082821742149 x^{10}}{13481837659631881626598512000000}+ \mathcal{O}(x^{11})
\end{align}

\begin{align}
\tilde{f}_{4,4} & = 1+\frac{3753 x}{385}+\frac{18023949 x^2}{275275}+\frac{9876808931021 x^3}{30580889625}+\frac{35556067418327333 x^4}{27052148248125}+\frac{74425041474465790907 x^5}{16961696951574375} + \notag \\
& +\frac{142176034924362610623531901354259 x^6}{12122235997267490151197765625}+\frac{48034592009600594147142098413913 x^7}{1725220606267701307686796875} + \notag \\
& +\frac{2921778795647832282045363249840625783 x^8}{70311365808440166794775406640625}-\frac{970188649919722941796880230833283271602074958177 x^9}{71441880301468810343738993778784316548828125} + \notag \\
& -\frac{12833662638076722274593348623980008443731839555627467 x^{10}}{65163597137751554265136565614903390371005859375}+ \mathcal{O}(x^{11})
\end{align}

\begin{align}
\tilde{f}_{4,3} & = 1+\frac{3411 x}{385}+\frac{2853027 x^2}{55055}+\frac{8689112827201 x^3}{38056218200}+\frac{11665664103047303 x^4}{14427812399000}+\frac{180587241207746167 x^5}{75385319784775} + \notag \\
& +\frac{4560588914814596401660018507787 x^6}{766245040814932710791760000}+\frac{297986951636393568702496721317 x^7}{25165611122765329616400000} + \notag \\
& +\frac{870333136973744852798887232883373 x^8}{52286738262776129845933200000}+\frac{36815391502136834453394781498703459894916515759 x^9}{13647849133475654874774303469970670864000000} + \notag \\
& -\frac{198486736623328222916847333876290264532169758542839 x^{10}}{2196790896100661450781887841634932402960000000} + \mathcal{O}(x^{11})
\end{align}

\begin{align}
\tilde{f}_{4,2} & = 1+\frac{3537 x}{385}+\frac{14205393 x^2}{275275}+\frac{18817349868677 x^3}{85626490950}+\frac{41558767503358199 x^4}{54104296496250}+\frac{390827960831726717963 x^5}{169616969515743750} + \notag \\
& +\frac{589372295742161523579690824044123x^6}{96977887978139921209582125000}+\frac{39390962097053294038969178501413 x^7}{2760352970028322092298875000} + \notag \\
& +\frac{16815868891673891199151819138720699313 x^8}{562490926467521334358203253125000}+\frac{82579184676571932368942126806739871530918534398627 x^9}{1494783957076885877961308177525333392406250000} + \notag \\
& +\frac{90156006023415488161061169196777347026435076854194793 x^{10}}{1042617554204024868242185049838454245936093750000}+ \mathcal{O}(x^{11})
\end{align}

\begin{align}
\tilde{f}_{4,1} & = 1+\frac{3151 x}{385}+\frac{6700637 x^2}{165165}+\frac{7623098572823 x^3}{48929423400}+\frac{593293552751569067 x^4}{1168652804319000}+\frac{54051094394424155111 x^5}{36637265415400650} + \notag \\
& +\frac{244722149644639622697113383628107 x^6}{62065848306009549574132560000}+\frac{2359677722621816352599949134217949 x^7}{238494496610447028774622800000} + \notag \\
& +\frac{1999474342544188944892656769273379653 x^8}{84520375733554510067041323600000}+\frac{6692498757090918922475725488850293312869649041119 x^9}{122830642201280893872968731229736037776000000} + \notag \\
& +\frac{1756557486754788236703118870875147587204872678329212069 x^{10}}{14413145069316439778579966128966791495820560000000} + \mathcal{O}(x^{11})
\end{align}

\begin{align}
\tilde{f}_{5,5} & = 1+\frac{484 x}{39}+\frac{1057496 x^2}{10647}+\frac{25930445123 x^3}{43922424}+\frac{159819541996278523 x^4}{56606528273022}+\frac{415851780776392085 x^5}{36651445537707} + \notag \\
& +\frac{57961576764061338079533425 x^6}{1502317536396079987584}+\frac{200801064446286339933020124762102205 x^7}{1829243946054423359080675639584} + \notag \\
& + \frac{1200988246877357656477324784089081 x^8}{4596613361026354660040992944}+\frac{392399723058789399270205316586732052836493 x^9}{910658092792643141684858597428503552} + \notag \\
& +\frac{468700486175900711022800739936338825880921293591802239 x^{10}}{8856970008936129391951115271468525864824268648192}+ \mathcal{O}(x^{11})
\end{align}

\begin{align}
\tilde{f}_{5,4} & = 1+\frac{47102 x}{4095}+\frac{51590132 x^2}{621075}+\frac{70935355830826 x^3}{158532499125}+\frac{18538514296100710265924 x^4}{9534662105987143125} + \notag \\
& +\frac{39747201415399738623464 x^5}{5612252597961384375} +\frac{23879067623663605533437306761858 x^6}{1087309652501508574998234375} + \notag \\
& +\frac{2922112209221789500287130830396193667292916 x^7}{50549834534643598636313885655398671875} +\frac{1593526410591781186709346301091700236504 x^8}{12830716006120114326123539874609375} + \notag \\
& +\frac{13766420352599877769868251251901093941591640569452 x^9}{72088323259683739298528073061132779919921875} + \notag \\
& +\frac{4224941939644954280315812158713971754960475425647009902071650952 x^{10}}{80310551229633675397391820866119275266375115715278251953125} + \mathcal{O}(x^{11})
\end{align}

\begin{align}
\tilde{f}_{5,3} & = 1+\frac{748 x}{65}+\frac{7055752 x^2}{88725}+\frac{450996994459 x^3}{1098060600}+\frac{876980828878925401 x^4}{508718799839250}+\frac{284658766538640587 x^5}{46277077699125} + \notag \\
& +\frac{664822757657191282281955283 x^6}{34775868898057407120000}+\frac{53159880021120705287637126069347587 x^7}{1014770606723485886684389980000} + \notag \\
& +\frac{10223155954608133840867112672288771 x^8}{80608399288481247545612250000}+\frac{15682166390341161437003285159560323589447271 x^9}{58555690122983741106279488003376000000} + \notag \\
& +\frac{153557644661470371110259045442527260714132429012006157197 x^{10}}{325150950550193780008030658567997967238385060000000}+ \mathcal{O}(x^{11})
\end{align}

\begin{align}
\tilde{f}_{5,2} & = 1+\frac{43307 x}{4095}+\frac{41065937 x^2}{621075}+\frac{98754948996887 x^3}{317064998250}+\frac{23084752481539077182803 x^4}{19069324211974286250} + \notag \\
& +\frac{45789668091314423355523 x^5}{11224505195922768750}+\frac{106887226138061020350063691209869 x^6}{8698477220012068599985875000} + \notag \\
& +\frac{13657289810548211782995994335209665663706453 x^7}{404398676277148789090511085243189375000}+\frac{8822790917318015966631192420724800500767 x^8}{102645728048960914608988318996875000} + \notag \\
& +\frac{12435967951882391906201094707696057093907210920783 x^9}{60705956429207359409286798367269709406250000} + \notag \\
& +\frac{592537805511418875443561952593896015212200633515836434173158740087 x^{10}}{1284968819674138806358269133857908404262001851444452031250000} + \mathcal{O}(x^{11})
\end{align}

\begin{align}
\tilde{f}_{5,1} & = 1+\frac{2156 x}{195}+\frac{2645192 x^2}{38025}+\frac{1075080268009 x^3}{3294181800}+\frac{14091620530884480289 x^4}{11119139482200750}+\frac{2817143441554814803 x^5}{654490098887625} + \notag \\
& +\frac{12418789799338640515546824497 x^6}{938948460247549992240000}+\frac{67577749525412697949213646287824852503 x^7}{1796578875589165799097092146020000} + \notag \\
& +\frac{1032438000617436514089378655019805187 x^8}{10260297680862398794734359250000}+\frac{3646507484757557062242630188589668947420564717 x^9}{14229032699885049088825915584820368000000} + \notag \\
& +\frac{227293485483052250913510946752138318334851724684472293977623 x^{10}}{362450409591880295057523318400869734051588374740000000} + \mathcal{O}(x^{11})
\end{align}

\begin{align}
\tilde{f}_{6,6} & = 1+\frac{1157 x}{77}+\frac{5118789 x^2}{36652}+\frac{234124821879 x^3}{243735800}+\frac{145686247932896729 x^4}{27374503916760}+\frac{577971531705261825521191 x^5}{23358664192171308000} + \notag \\
& +\frac{178773536069908389256841471 x^6}{1811798961869616720000}+\frac{415144797831107890924761330133811 x^7}{1222632740051969146140240000} + \notag \\
& +\frac{38099177830792358368291572999375469013945987 x^8}{38163534638528761871609096131435200000}+\frac{1367892175780997073885390898786293333047540249 x^9}{560157541170003863284600581523344000000} + \notag \\
& +\frac{17714683795689417621512806332630969718877202111529019 x^{10}}{4089314176700591013109526633090797539792000000} + \mathcal{O}(x^{11})
\end{align}

\begin{align}
\tilde{f}_{6,5} & = 1+\frac{26065 x}{1848}+\frac{30355351 x^2}{251328}+\frac{23964796616153 x^3}{31287320064}+\frac{14100775855815374545 x^4}{3604048972812288}+\frac{5313459217145746856999309 x^5}{316320170245788893184} + \notag \\
& +\frac{44004922137350367980484041021 x^6}{710818626409888644071424}+\frac{219598883907493402694652818766394819 x^7}{1112613119899836659464660844544} + \notag \\
& +\frac{459556210913449030487271275301665495221059341 x^8}{850514873947747015190650516546630189056} + \notag \\
& +\frac{6196651026665998684982241790838826707787916952027 x^9}{5063390879089459304865627353275700757921792} + \notag \\
& +\frac{1179939505433223851151564270773290517177114044893592387 x^{10}}{583262122144072996246081346078534521706527064064} + \mathcal{O}(x^{11})
\end{align}

\begin{align}
\tilde{f}_{6,4} & = 1+\frac{3211 x}{231}+\frac{112781903 x^2}{989604}+\frac{758329189033 x^3}{1096811100}+\frac{1690222087839881 x^4}{498725779860}+\frac{949917993609971196308767 x^5}{67573278555924141000} + \notag \\
& +\frac{1915730810998935767102818946 x^6}{37835458008417855410625}+\frac{42259911044368955247597202202269 x^7}{263119723029908845848247500} + \notag \\
& +\frac{2351854276255948140333669285983955807080081749 x^8}{5216478140903900138325568327465548900000}+\frac{42621022523893502521860527569032652334967738919 x^9}{38283266954337451531356920993486041500000} + \notag \\
& +\frac{2667390614692753973844933449414928922854627880673527 x^{10}}{1131494193376036506284227574211737628988500000} + \mathcal{O}(x^{11})
\end{align}

\begin{align}
\tilde{f}_{6,3} & = 1+\frac{120133 x}{9240}+\frac{370459193 x^2}{3769920}+\frac{430895162535553 x^3}{782183001600}+\frac{3395845335003988840423 x^4}{1351518364804608000} + \notag \\
& +\frac{387290038754936980878127753 x^5}{39540021280723611648000} +\frac{164267073690295858717722251224079 x^6}{4886878056567984427991040000} + \notag \\
& +\frac{359875492448536219471467508091258434639 x^7}{3476915999686989560827065139200000} +\frac{769691100175032732089404146006900951085679279537 x^8}{2657858981086709422470782864208219340800000} + \notag \\
& +\frac{1067343482359616523475721064929353496068445861870101 x^9}{1438463317923141847973189588998778624409600000} + \notag \\
& +\frac{90315855792594906802826570877697045972116545389281076193 x^{10}}{51683198441405334024640951034841787722104832000000} + \mathcal{O}(x^{11})
\end{align}

\begin{align}
\tilde{f}_{6,2} & = 1+\frac{3055 x}{231}+\frac{489225763 x^2}{4948020}+\frac{1194907039079 x^3}{2193622200}+\frac{3016315348600170989 x^4}{1231852676254200}+\frac{275896381765976604595103 x^5}{29108489224090399200} + \notag \\
& +\frac{158175329065817204295273074623 x^6}{4842938625077485492560000}+\frac{4349010371379356114619475405883999 x^7}{42442822261804071787439760000} + \notag \\
& +\frac{1909796688132434844042617363396218870829675413 x^8}{6420280788804800170246853326111444800000}+\frac{990929043524015602473040611481965964578850558427x^9}{1225064542538798449003421471791553328000000} + \notag \\
& +\frac{18603058765548549450547167697480724954098534521514461569 x^{10}}{8943330104444192545670534746569574219525104000000} + \mathcal{O}(x^{11})
\end{align}

\begin{align}
\tilde{f}_{6,1} & = 1+\frac{115037 x}{9240}+\frac{7327853 x^2}{83776}+\frac{356554101639401 x^3}{782183001600}+\frac{883178380488410890861 x^4}{450506121601536000} + \notag \\
& +\frac{58193875045423494522821549 x^5}{7908004256144722329600}+\frac{11084159091048065551969800022301 x^6}{444261641506180402544640000} + \notag \\
& +\frac{14303707208776420809074699503001845949 x^7}{182995578930894187411950796800000}+\frac{611178010362329177529580895613352507122419693233 x^8}{2657858981086709422470782864208219340800000} +\notag \\
& +\frac{50857621812689307886025616358860742544553274931326187 x^9}{79115482485772801638525427394932824342528000000} + \notag \\
& +\frac{209368580834681502416171565809626728245345329334178891329 x^{10}}{121512942113348540884600280433028025355526471680000} + \mathcal{O}(x^{11})
\end{align}

\begin{align}
\tilde{f}_{7,7} & = 1+\frac{7785 x}{442}+\frac{1170771595 x^2}{6281704}+\frac{351427951165 x^3}{241955424}+\frac{1098422561526712992659 x^4}{120666290501671200}+\frac{389617496526219452458165529 x^5}{8123254676572505184000} + \notag \\
& +\frac{73713105787670031168841799519614607 x^6}{338572406961151037598026188800}+\frac{208166280728875233728820824528473942157 x^7}{241343323404378634449190642176000} + \notag \\
& +\frac{401362730647593870837169898041740997552182217 x^8}{134253927624891292596652550006016000000} + \notag \\
& +\frac{1091025392133532587237569971948899899375531969934344141723x^9}{121624854363484200405335358837512308800274636800000} + \notag \\
& +\frac{3109795886267670565967910213328619921373128699138806049341 x^{10}}{137868308768244537090696640093276854513259008000000} + \mathcal{O}(x^{11})
\end{align}

\begin{align}
\tilde{f}_{7,6} & = 1+\frac{103225 x}{6188}+\frac{6344008980 x^2}{38475437}+\frac{179710992882133 x^3}{149844338280}+\frac{2031585981981604204555691 x^4}{289719763494512551200} + \notag \\
& +\frac{1362284392438346535587033197 x^5}{39504496918158013491750}+\frac{67419242530982116232019093095687910697 x^6}{461026679474218268777432674608000} + \notag \\
& +\frac{4486160829201350049616412997215930596801757 x^7}{8281525191100508072924575376231232000} + \notag \\
& +\frac{2656966552961215941891705104075719457988629241431 x^8}{1511615578566212789946240658060998130500000} + \notag \\
& +\frac{520212760571259679495217891928382336596265466217020414801411 x^9}{105195506996233637128475303792287134923459451849600000} + \notag \\
& +\frac{139835806063784919915478765442324559843340177920848479019697911 x^{10}}{12019871866826777440612428701793753921163460520460800000} + \mathcal{O}(x^{11})
\end{align}

\begin{align}
\tilde{f}_{7,5} & = 1+\frac{50535 x}{3094}+\frac{47677576435 x^2}{307803496}+\frac{166257824151505 x^3}{154125605088}+\frac{1429658189332616312155 x^4}{236505929383275552} + \notag \\
& +\frac{4460919780449899718124889741 x^5}{156031475827604679574272} +\frac{3814026624002132208159000149285060375 x^6}{32516493964548945650914435172352} + \notag \\
& +\frac{1966437522810913040244499393697178814265 x^7}{4635722555951304810500053854916608} +\frac{539099356688544784106135125445976941346545399 x^8}{395557740466831363184228682298212139008} + \notag \\
& +\frac{2543937386667341264620561570737809997332776362952853034625 x^9}{654127656731865265987990840314262199681989062623232} + \notag \\
& +\frac{353900649397196708561913890376324370437954857894164424621863 x^{10}}{36332953794536451458970746583148000190852116232110080} + \mathcal{O}(x^{11})
\end{align}

\begin{align}
\tilde{f}_{7,4} & = 1+\frac{860425 x}{55692}+\frac{426660846980 x^2}{3116510397}+\frac{2702878604183197 x^3}{3034347850170}+\frac{247009936034237361706287731 x^4}{52801426896874912456200} + \notag \\
& +\frac{8103554970866197929553145376538 x^5}{388783506420052089779057625}+\frac{15342549979104094943291413746203406467257 x^6}{189049752751896628840545986131443000} + \notag \\
& +\frac{8599544768735233157448935372600702396216681213 x^7}{30563486313080868824882703211942376148000} + \notag \\
& +\frac{132391856479983567440961018073189669568740522130427133 x^8}{150625217754151254619091265292611545150821968750} + \notag \\
& +\frac{2175183388499004368049619106251360565337197301638816279408704939 x^9}{873518834899771880456156936291826051107159602350051150000} + \notag \\
& +\frac{5755042877566562029836279844620611711526701295417040427597061890191 x^{10}}{898291789421946951067946679615465146852394931282623001800000} + \mathcal{O}(x^{11})
\end{align}

\begin{align}
\tilde{f}_{7,3} & = 1+\frac{47895 x}{3094}+\frac{372905507995 x^2}{2770231464}+\frac{4881950119315 x^3}{5708355744}+\frac{77908220241168549243433 x^4}{17737944703745666400} + \notag \\
& +\frac{375111514228329401396345835929 x^5}{19503934478450584946784000}+\frac{744513952523350790423953591531254127 x^6}{10035954927329921497195813324800} \notag \\
& +\frac{149273258467553522306568701054070715495237 x^7}{579465319493913101312506731864576000}+\frac{3522731468347286944415840911443190110810289526569 x^8}{4307683726674773368010066142452120832000000} + \notag \\
& +\frac{1519725572510330014700259371983182611273756153398435261573 x^9}{633171489491108132815049322176843471748462796800000} + \notag \\
& +\frac{9206159644443354833252385093276190639693274498464314736931123 x^{10}}{1401734328492918651966463988547376550572998311424000000} + \mathcal{O}(x^{11})
\end{align}

\begin{align}
\tilde{f}_{7,2} & = 1+\frac{819265 x}{55692}+\frac{376667798420 x^2}{3116510397}+\frac{8826939874499197 x^3}{12137391400680}+\frac{752219576377547459363390171 x^4}{211205707587499649824800} + \notag \\
& +\frac{11671923108878048980049416765199 x^5}{777567012840104179558115250}+\frac{170460410374325847033330208250113993854457 x^6}{3024796044030346061448735778103088000} + \notag \\
& +\frac{94326592819053058674071189291982372479050659221 x^7}{489015781009293901198123251391078018368000} + \notag \\
& +\frac{1475197055823876164689951820314812170030701925442926293 x^8}{2410003484066420073905460244681784722413151500000} + \notag \\
& +\frac{20391823003149848112928180049400876594870878273347087173488128847 x^9}{11181041086717080069838808784535373454171642910080654720000} + \notag \\
& +\frac{1481071780820937764653074510818724742058603436352933698130357555601267 x^{10}}{287453372615023024341742937476948846992766378010439360576000000} + \mathcal{O}(x^{11})
\end{align}

\begin{align}
\tilde{f}_{7,1} & = 1+\frac{46575 x}{3094}+\frac{38377086795 x^2}{307803496}+\frac{115788697675225 x^3}{154125605088}+\frac{2416901507277622387531 x^4}{656960914953543200} + \notag \\
& +\frac{162567999562047281768096767 x^5}{10469100632555332768000} +\frac{47613277969241272946195516848536886991 x^6}{812912349113723641272860879308800} + \notag \\
& +\frac{1446900561155184516262261056428295659693 x^7}{7153892833258186435956873232896000} +\frac{1141551959991352461062830181219252954002479515137 x^8}{1754982259015648409189286206184197376000000} + \notag \\
& +\frac{1486108801872753255040846577651688778247704238631539643268697 x^9}{753107499526818562815121033256551874633868986572800000} + \notag \\
& +\frac{2877421101562376966635140717919923180371541929598768729508331 x^{10}}{504624358257450714707927035877055558206279392112640000} + \mathcal{O}(x^{11})
\end{align}

\begin{align}
\tilde{f}_{8,8} & = 1+\frac{3451 x}{171}+\frac{151692071 x^2}{633555}+\frac{3225640796171681 x^3}{1549690394175}+\frac{607527211191466073021 x^4}{41831678347333875} + \notag \\
& +\frac{3698213386973843188333451 x^5}{43396183117524161925}+\frac{6649636233609242135903565338555293 x^6}{15391471896769092395021375625} + \notag \\
& +\frac{85012390917747141863907134470729479964757 x^7}{44222056600768656764674040761865625}+\frac{4773755598385750742019894769665442533655041913 x^8}{630949877241219046010676729466867078125} + \notag \\
& +\frac{6526845214015698532463049936750024448774227891134437 x^9}{247880646724531489182464133770289745815898125} + \notag \\
& +\frac{196890357153877123858256279813979635370526574864045266598435276251 x^{10}}{2457923629640037068345494737599135153176493091328514296875}+ \mathcal{O}(x^{11})
\end{align}

\begin{align}
\tilde{f}_{8,7} & = 1+\frac{65807 x}{3420}+\frac{3638052173 x^2}{16894800}+\frac{11235437299456679 x^3}{6372120744000}+\frac{9841450209953218040167 x^4}{849913464834720000} + \notag \\
& +\frac{268514905754507375039498933 x^5}{4198572516283516800000}+\frac{715003422458468338758257351889158941 x^6}{2342074869649095863870016000000} + \notag \\
& +\frac{169374616699257410613633578571725072607625801 x^7}{132347512503545077478837018338560000000} + \notag \\
& +\frac{75707666804300655232052308199680293349886080739 x^8}{15946261262622260554767191965670400000000} + \notag \\
& +\frac{138592171823229654130712445066192876350439520757518451053 x^9}{8895246568494544461677329924610003624448000000000} + \notag \\
& +\frac{11360804209292324722508643873897415069918943948787813714213070349287 x^{10}}{254177629940025403660662541051839300228355158251520000000000} + \mathcal{O}(x^{11})
\end{align}

\begin{align}
\tilde{f}_{8,6} & = 1+\frac{357 x}{19}+\frac{17086207 x^2}{84474}+\frac{243213102379567 x^3}{153055841400}+\frac{41365335628660127393 x^4}{4131523787391000} + \notag \\
& +\frac{759259470659140290108329 x^5}{14286809256798078000}+\frac{38882220629180460258807095335391 x^6}{158969450721088019572560000} + \notag \\
& +\frac{4461078185894672353861193902861966053173 x^7}{4490661284100581827243698291600000}+\frac{2390790294699208637749232228904762043541650651 x^8}{664704397422848048472153262319004000000} + \notag \\
& +\frac{957465567007020163923600471162643381287992599460137 x^9}{82179684499238498435421094662909495312000000} + \notag \\
& +\frac{121206174287198059795334029605067646004465406163319151094091251 x^{10}}{3603791628200591660698051073941008596987674408080000000}+ \mathcal{O}(x^{11})
\end{align}

\begin{align}
\tilde{f}_{8,5} & = 1+\frac{85765 x}{4788}+\frac{860783219 x^2}{4730544}+\frac{3365516613797827 x^3}{2497871331648}+\frac{3736582445909466620161 x^4}{466432509501294336} + \notag \\
& +\frac{18542203076627325809506615 x^5}{460835319387278803968}+\frac{63322761013750729657102331661195493 x^6}{359892592769758666825722138624} + \notag \\
& +\frac{19442072556877833842417216173793978435984123 x^7}{28471867425882651099991216220367077376} +\frac{8167480956512819288828178677702007109578054841 x^8}{3430512806923421864242844100697886883840} + \notag \\
& +\frac{4023563463150273999928426153123803932603973569527454585 x^9}{535816634130849106622400608409826466003293569024} + \notag \\
& +\frac{20005639436647088777290626810139798218048862009822451961492645263 x^{10}}{931956682100639570395995318741503522851440604791903879168} + \mathcal{O}(x^{11})
\end{align}

\begin{align}
\tilde{f}_{8,4} & = 1+\frac{3043 x}{171}+\frac{52261757 x^2}{295659}+\frac{3949949777516053 x^3}{3099380788350}+\frac{4322691784676697038419 x^4}{585643496862674250} + \notag \\
& +\frac{36677863128465866227855901 x^5}{1012577606075563778250}+\frac{19156959521617456061521636759754723 x^6}{123131775174152739160171005000} + \notag \\
& +\frac{300385565890432585958215123242951866074742759 x^7}{502716339437538090100814495380888425000} + \notag \\
& +\frac{220760033175959556112760329664609404631219510389 x^8}{105999579376524799729793690550433669125000} + \notag \\
& +\frac{8580952313263413694028834799767257761551183789773771039 x^9}{1288979362967563743748813495605506678242670250000} + \notag \\
& +\frac{156936567207894881400910056810741459392678903465516858555524580258171 x^{10}}{7983335949070840397986166907721990977517249560635014436250000} + \mathcal{O}(x^{11})
\end{align}

\begin{align}
\tilde{f}_{8,3} & = 1+\frac{407609 x}{23940}+\frac{2112041699 x^2}{13140400}+\frac{12738857230058133 x^3}{11564219128000}+\frac{197187697209893886235849 x^4}{32391146493145440000} + \notag \\
& +\frac{509825891259953130901413299x^5}{17779140408459830400000}+\frac{1181452213518685671907089242974923643 x^6}{9917675065304196065523648000000} + \notag \\
& +\frac{85766016936441983946702140431340939740898849 x^7}{192037006185537608321403243306240000000} + \notag \\
& +\frac{188260472867049606992031055500421248649721273359 x^8}{122546325121578570253302329843174400000000} + \notag \\
& +\frac{2349004053245805196155829890682977501925479861593315506391 x^9}{478517084378973324815672696982369600554496000000000} + \notag \\
& +\frac{1152217956192499429338282660612878055484149288292268748168496344666817 x^{10}}{78311338531125454922152267506717131622880767711160320000000000} + \mathcal{O}(x^{11})
\end{align}

\begin{align}
\tilde{f}_{8,2} & = 1+\frac{2941 x}{171}+\frac{1435835759 x^2}{8869770}+\frac{13641743576064103 x^3}{12397523153400}+\frac{14119277535187271121391 x^4}{2342573987450697000} + \notag \\
& +\frac{137058169560073210069966633 x^5}{4860372509162706135600}+\frac{230125666321184498511759522196887239 x^6}{1970108402786443826562736080000} + \notag \\
& +\frac{320934292356163353029079832113056970867296431 x^7}{731223766454600858328457447826746800000} + \notag \\
& +\frac{178043960998390777750635124515087825907273703057 x^8}{116965053105130813494944761986685428000000} + \notag \\
& +\frac{81457006445409360118019589660174260059408481581415992949 x^9}{16498935845984815919984812743750485481506179200000} + \notag \\
& +\frac{702676639205494172518921840144112078359383816308789908032170654195737 x^{10}}{46448500067321253224646789281291583869191270170967356720000000} + \mathcal{O}(x^{11})
\end{align}

\begin{align}
\tilde{f}_{8,1} & = 1+\frac{397001 x}{23940}+\frac{17788768091 x^2}{118263600}+\frac{308437579871754359 x^3}{312233916456000}+\frac{1531119983862043674063841 x^4}{291520318438308960000} + \notag \\
& +\frac{3140931189982529675846238121 x^5}{130919124825931478400000}+\frac{549118562449578285789724907710268759581 x^6}{5623321762027479169151908416000000} + \notag \\
& +\frac{807252837361432968995319324132527363394507380383 x^7}{2224364642647082117186813767216177920000000} + \notag \\
& +\frac{335837931521113458232717480651672166854763570585333 x^8}{268008813040892333143972195367022412800000000} + \notag \\
& +\frac{4261185012515345206849484565724839904861817080692305078239933 x^9}{1046516863536814661371876188300442316412682752000000000} + \notag \\
& +\frac{2632100422587109755961072528377034706271736302075036401108585033955896641 x^{10}}{209326207893698341006913011045454892827960292091931535360000000000} + \mathcal{O}(x^{11})
\end{align}
\end{widetext}

\section{Residual phases}
\label{app:residual_phases}
In this Appendix we report the phases associated to the remainder functions $\rho_{\lm}$ and $\tilde{f}_{\lm}$, 
whose PN expansions up to 10 PN are written in the previous Appendix. Since the results for
these quantities, in the comparable-mass case, are usually expressed in terms of $y= (E_{\rm real} \Omega)^{3/2}$
instead of $x = (M \Omega)^{3/2}$, we are going to separate these results for the test-mass scenario, in which
such phases are reported up to $\ell = 8$ and up to 10PN in terms of $x$, from the comparable-mass one,
where we consider only up to the case with $\ell = 5$ for the $\rho_{\lm}$'s and with the full $\nu$ 
dependence accessible in literature
\cite{Pan:2010hz}.

\subsection{Test-mass case}
We have
\begin{widetext}
\begin{align}
& \delta_{22} = -\frac{17 x^{3/2}}{3} -\frac{259 x^{9/2}}{81}-\frac{58940243 x^{15/2}}{3539025} \quad ,\\
& \tilde{\delta}_{22} = \frac{25 x^{3/2}}{3}+\frac{12077 x^{9/2}}{567}+\frac{159283133 x^{15/2}}{694575} \quad , \\
& \delta_{21} =  -\frac{10 x^{3/2}}{3} -\frac{29 x^{9/2}}{81}-\frac{3728003 x^{15/2}}{7078050} \quad ,\\
& \tilde{\delta}_{21} = \frac{25 x^{3/2}}{6}+\frac{12077 x^{9/2}}{4536}+\frac{159283133 x^{15/2}}{22226400} \quad , \\
& \delta_{33} = -\frac{67 x^{3/2}}{10} +\frac{54211 x^{9/2}}{21000}+\frac{87343796839 x^{15/2}}{1886500000} \quad , \\
& \tilde{\delta}_{33} = \frac{77 x^{3/2}}{10}+\frac{370189 x^{9/2}}{21000}+\frac{277058043161 x^{15/2}}{1886500000} \quad , \\
& \delta_{32} =  -\frac{14 x^{3/2}}{3} +\frac{2176 x^{9/2}}{2835}+\frac{139748608 x^{15/2}}{22920975} \quad ,\\
& \tilde{\delta}_{32} = \frac{77 x^{3/2}}{15}+\frac{370189 x^{9/2}}{70875}+\frac{277058043161 x^{15/2}}{14325609375} \quad , \\
& \delta_{31} = -\frac{67 x^{3/2}}{30} +\frac{54211 x^{9/2}}{567000}+\frac{87343796839 x^{15/2}}{458419500000} \quad , \\
&  \tilde{\delta}_{31} = \frac{77 x^{3/2}}{30}+\frac{370189 x^{9/2}}{567000}+\frac{277058043161 x^{15/2}}{458419500000} \quad , \\
& \delta_{44} = -\frac{116 x^{3/2}}{15} +\frac{5474624 x^{9/2}}{779625}+\frac{53161630989358592 x^{15/2}}{608422955765625} \quad , \\
& \tilde{\delta}_{44} = \frac{38 x^{3/2}}{5}+\frac{1195976 x^{9/2}}{70875}+\frac{60849825332512 x^{15/2}}{457117171875} \quad , \\
& \delta_{43} =  -\frac{59 x^{3/2}}{10} +\frac{45627 x^{9/2}}{15400}+\frac{1661300808174373 x^{15/2}}{80121541500000} \quad ,\\
& \tilde{\delta}_{43} = \frac{57 x^{3/2}}{10}+\frac{149497 x^{9/2}}{21000}+\frac{1901557041641 x^{15/2}}{60196500000} \quad , \\
& \delta_{42} = -\frac{58 x^{3/2}}{15} +\frac{684328 x^{9/2}}{779625}+\frac{1661300968417456 x^{15/2}}{608422955765625} \quad , \\
& \tilde{\delta}_{42} = \frac{19 x^{3/2}}{5}+\frac{149497 x^{9/2}}{70875}+\frac{1901557041641 x^{15/2}}{457117171875} \quad , \\
& \delta_{41} =  -\frac{59 x^{3/2}}{30} +\frac{15209 x^{9/2}}{138600}+\frac{1661300808174373 x^{15/2}}{19469534584500000} \quad ,\\
& \tilde{\delta}_{41} = \frac{19 x^{3/2}}{10}+\frac{149497 x^{9/2}}{567000}+\frac{1901557041641 x^{15/2}}{14627749500000} \quad , \\
& \delta_{55} = -\frac{184 x^{3/2}}{21} +\frac{44804315 x^{9/2}}{3972969}+\frac{5089851079303303 x^{15/2}}{39809189109690} \quad , \\
& \tilde{\delta}_{55} = \frac{319 x^{3/2}}{42}+\frac{532744699 x^{9/2}}{31783752}+\frac{248981199300097213 x^{15/2}}{1910841077265120} \quad , \\
& \delta_{54} =  -\frac{106 x^{3/2}}{15} +\frac{58520552 x^{9/2}}{10135125}+\frac{7597678556920544 x^{15/2}}{181346524734375} \quad ,\\
& \tilde{\delta}_{54} = \frac{638 x^{3/2}}{105}+\frac{4261957592 x^{9/2}}{496621125}+\frac{7967398377603110816 x^{15/2}}{186605573951671875} \quad , \\
& \delta_{53} = -\frac{184 x^{3/2}}{35} +\frac{8960863 x^{9/2}}{3678675}+\frac{15269553237909909 x^{15/2}}{1535848345281250} \quad , \\
& \tilde{\delta}_{53} = \frac{319 x^{3/2}}{70}+\frac{532744699 x^{9/2}}{147147000}+\frac{248981199300097213 x^{15/2}}{24573573524500000} \quad , \\
& \delta_{52} =  -\frac{53 x^{3/2}}{15} +\frac{7315069 x^{9/2}}{10135125}+\frac{237427454903767 x^{15/2}}{181346524734375} \quad ,\\
& \tilde{\delta}_{52} = \frac{319 x^{3/2}}{105}+\frac{532744699 x^{9/2}}{496621125}+\frac{248981199300097213 x^{15/2}}{186605573951671875} \quad , \\
& \delta_{51} = -\frac{184 x^{3/2}}{105} +\frac{8960863 x^{9/2}}{99324225}+\frac{5089851079303303 x^{15/2}}{124403715967781250} \quad , \\
& \tilde{\delta}_{51} = \frac{319 x^{3/2}}{210}+\frac{532744699 x^{9/2}}{3972969000}+\frac{248981199300097213 x^{15/2}}{5971378366453500000} \quad ,
\end{align}
\begin{align}
& \delta_{66} = -\frac{137 x^{3/2}}{14} +\frac{13051301 x^{9/2}}{840840}+\frac{4525656132452421 x^{15/2}}{26736047994656} \quad , \\
& \tilde{\delta}_{66} = \frac{533 x^{3/2}}{70}+\frac{225159199 x^{9/2}}{13377000}+\frac{40965008675217057 x^{15/2}}{313862321500000} \quad , \\
& \delta_{65} =  - \frac{172 x^{3/2}}{21} +\frac{71374423 x^{9/2}}{7945938}+\frac{184149419420544737 x^{15/2}}{2707024859458920} \quad ,\\
& \tilde{\delta}_{65} = \frac{533 x^{3/2}}{84}+\frac{225159199 x^{9/2}}{23115456}+\frac{4551667630579673 x^{15/2}}{86776654648320} \quad , \\
& \delta_{64} =  -\frac{137 x^{3/2}}{21} + \frac{13051301 x^{9/2}}{2837835}+\frac{1508552044150807 x^{15/2}}{67675621486473} \quad , \\
& \tilde{\delta}_{64} = \frac{533 x^{3/2}}{105}+\frac{225159199 x^{9/2}}{45147375}+\frac{4551667630579673 x^{15/2}}{264821333765625} \quad , \\
& \delta_{63} =  -\frac{172 x^{3/2}}{35} +\frac{71374423 x^{9/2}}{36786750}+\frac{552448258261634211 x^{15/2}}{104437687479125000} \quad ,\\
& \tilde{\delta}_{63} = \frac{533 x^{3/2}}{140}+\frac{225159199 x^{9/2}}{107016000}+\frac{40965008675217057 x^{15/2}}{10043594288000000} \quad , \\
& \delta_{62} =  -\frac{137 x^{3/2}}{42} + \frac{13051301 x^{9/2}}{22702680}+\frac{1508552044150807 x^{15/2}}{2165619887567136} \quad ,\\
& \tilde{\delta}_{62} = \frac{533 x^{3/2}}{210}+\frac{225159199 x^{9/2}}{361179000}+\frac{4551667630579673 x^{15/2}}{8474282680500000} \quad , \\
& \delta_{61} =  -\frac{172 x^{3/2}}{105} +\frac{71374423 x^{9/2}}{993242250}+\frac{184149419420544737 x^{15/2}}{8459452685809125000} \quad ,\\
& \tilde{\delta}_{61} = \frac{533 x^{3/2}}{420}+\frac{225159199 x^{9/2}}{2889432000}+\frac{4551667630579673 x^{15/2}}{271177045776000000} \quad , \\
& \delta_{77} =  - \frac{389 x^{3/2}}{36} +\frac{612182119 x^{9/2}}{30932928}+\frac{65625416145752217259 x^{15/2}}{310015161999129600} \quad ,\\
& \tilde{\delta}_{77} = \frac{275 x^{3/2}}{36}+\frac{524535701 x^{9/2}}{30932928}+\frac{449219027964262603447 x^{15/2}}{3410166781990425600} \quad , \\
& \delta_{76} =  -\frac{65 x^{3/2}}{7} +\frac{809765 x^{9/2}}{64974}+\frac{16878965057800103 x^{15/2}}{172341800735650} \quad ,\\
& \tilde{\delta}_{76} = \frac{275 x^{3/2}}{42}+\frac{524535701 x^{9/2}}{49120344}+\frac{449219027964262603447 x^{15/2}}{7370714133862279200} \quad , \\
& \delta_{75} =  - \frac{1945 x^{3/2}}{252} +\frac{76522764875 x^{9/2}}{10609994304}+\frac{8203177018219027157375 x^{15/2}}{208416993108774847488} \quad ,\\
& \tilde{\delta}_{75} = \frac{1375 x^{3/2}}{252}+\frac{65566962625 x^{9/2}}{10609994304}+\frac{56152378495532825430875 x^{15/2}}{2292586924196523322368} \quad , \\
& \delta_{74} =  -\frac{130 x^{3/2}}{21} +\frac{3239060 x^{9/2}}{877149}+\frac{270063440924801648 x^{15/2}}{20939528789381475} \quad ,\\
& \tilde{\delta}_{74} = \frac{275 x^{3/2}}{63}+\frac{524535701 x^{9/2}}{165781161}+\frac{449219027964262603447 x^{15/2}}{55971360454016682675} \quad , \\
& \delta_{73} =  -\frac{389 x^{3/2}}{84}+\frac{612182119 x^{9/2}}{392962752}+\frac{65625416145752217259 x^{15/2}}{21442077480326630400} \quad ,\\
& \tilde{\delta}_{73} = \frac{275 x^{3/2}}{84}+\frac{524535701 x^{9/2}}{392962752}+\frac{449219027964262603447 x^{15/2}}{235862852283592934400} \quad , \\
& \delta_{72} =  -\frac{65 x^{3/2}}{21} +\frac{809765 x^{9/2}}{1754298}+\frac{16878965057800103 x^{15/2}}{41879057578762950} \quad ,\\
& \tilde{\delta}_{72} = \frac{275 x^{3/2}}{126}+\frac{524535701 x^{9/2}}{1326249288}+\frac{449219027964262603447 x^{15/2}}{1791083534528533845600} \quad , \\
& \delta_{71} =  -\frac{389 x^{3/2}}{252} +\frac{612182119 x^{9/2}}{10609994304}+\frac{65625416145752217259 x^{15/2}}{5210424827719371187200}  \quad ,\\
& \tilde{\delta}_{71} = \frac{275 x^{3/2}}{252}+\frac{524535701 x^{9/2}}{10609994304}+\frac{449219027964262603447 x^{15/2}}{57314673104913083059200} \quad , \\
& \delta_{88} = -\frac{532 x^{3/2}}{45} +\frac{14889375968 x^{9/2}}{618100875}+\frac{129467292695005941248 x^{15/2}}{507826038729515625} \quad ,\\
& \tilde{\delta}_{88} = \frac{2414 x^{3/2}}{315}+\frac{517746006968 x^{9/2}}{30286942875}+\frac{301716560973469525235168 x^{15/2}}{2264396306694910171875} \quad ,
\end{align}
\begin{align}
& \delta_{87} =  -\frac{373 x^{3/2}}{36} +\frac{3647897461 x^{9/2}}{226048320}+\frac{2220364069714984597 x^{15/2}}{16980044527641600} \quad ,\\
& \tilde{\delta}_{87} = \frac{1207 x^{3/2}}{180}+\frac{64718250871 x^{9/2}}{5651208000}+\frac{9428642530420922663599 x^{15/2}}{137962861787088000000} \quad , \\
& \delta_{86} =  -\frac{133 x^{3/2}}{15} +\frac{465292999 x^{9/2}}{45785250}+\frac{252865806044933479 x^{15/2}}{4179638178843750} \quad ,\\
& \tilde{\delta}_{86} = \frac{1207 x^{3/2}}{210}+\frac{64718250871 x^{9/2}}{8973909000}+\frac{9428642530420922663599 x^{15/2}}{298192106231428500000} \quad , \\
& \delta_{85} =  -\frac{1865 x^{3/2}}{252} +\frac{91197436525 x^{9/2}}{15506914752}+\frac{277545508714373074625 x^{15/2}}{11415344335042894848} \quad ,\\
& \tilde{\delta}_{85} = \frac{1207 x^{3/2}}{252}+\frac{64718250871 x^{9/2}}{15506914752}+\frac{9428642530420922663599 x^{15/2}}{741997381777788165120} \quad , \\
& \delta_{84} =  -\frac{266 x^{3/2}}{45} +\frac{1861171996 x^{9/2}}{618100875}+\frac{4045852896718935664 x^{15/2}}{507826038729515625} \quad ,\\
& \tilde{\delta}_{84} = \frac{1207 x^{3/2}}{315}+\frac{64718250871 x^{9/2}}{30286942875}+\frac{9428642530420922663599 x^{15/2}}{2264396306694910171875} \quad , \\
& \delta_{83} =  -\frac{373 x^{3/2}}{84} +\frac{3647897461 x^{9/2}}{2871650880}+\frac{2220364069714984597 x^{15/2}}{1174418141465318400} \quad ,\\
& \tilde{\delta}_{83} = \frac{1207 x^{3/2}}{420}+\frac{64718250871 x^{9/2}}{71791272000}+\frac{9428642530420922663599 x^{15/2}}{9542147399405712000000} \quad , \\
& \delta_{82} =  -\frac{133 x^{3/2}}{45} +\frac{465292999 x^{9/2}}{1236201750}+\frac{252865806044933479 x^{15/2}}{1015652077459031250} \quad , \\
& \tilde{\delta}_{82} = \frac{1207 x^{3/2}}{630}+\frac{64718250871 x^{9/2}}{242295543000}+\frac{9428642530420922663599 x^{15/2}}{72460681814237125500000} \quad , \\
& \delta_{81} =  -\frac{373 x^{3/2}}{252} +\frac{3647897461 x^{9/2}}{77534573760}+\frac{2220364069714984597 x^{15/2}}{285383608376072371200} \quad ,\\
& \tilde{\delta}_{81} = \frac{1207 x^{3/2}}{1260}+\frac{64718250871 x^{9/2}}{1938364344000}+\frac{9428642530420922663599 x^{15/2}}{2318741818055588016000000} \quad .
\end{align}
\end{widetext}
 
\subsection{Comparable-mass case}
In this scenario we obtain the following results for the phases of the remainder functions $\rho_{\lm}$ 
starting from the results of the PN-expanded waveforms we wrote in Appendix~\ref{app:PN_exp_hhat}
\begin{align}
\delta_{22}&=-\dfrac{17}{3}y^{3/2}-24\nu y^{5/2} \nonumber\\
&+ \left(\dfrac{30995}{1134}\nu+\dfrac{962}{135}\nu^2\right)y^{7/2} -\nu\dfrac{4976}{105}\pi y^4\, \\
\delta_{21} &= -\dfrac{10}{3} y^{3/2} - \dfrac{25}{2} \nu y^{5/2} \ , \\
\delta_{33} &= -\dfrac{67}{10} y^{3/2} - \dfrac{80897}{2430} \nu y^{5/2} \ , \\
\delta_{32} &= -\dfrac{7 (10-39 \nu)}{15 (1-3 \nu)} y^{3/2} \ , \\
\delta_{31} &= -\dfrac{67}{30} y^{3/2} - \dfrac{17}{10} \nu y^{5/2} \ ,
\end{align}
\begin{align} 
\delta_{44} &= -\dfrac{928 - 3339 \nu}{120 (1-3 \nu)} y^{3/2} \ , \\
\delta_{43} &= -\dfrac{4779 - 15491 \nu}{810 (1-2 \nu)} y^{3/2} \ , \\
\delta_{42} &= -\dfrac{58 - 237 \nu}{15 (1-3 \nu)} y^{3/2} \ , \\
\delta_{41} &= -\dfrac{59 - 1651 \nu}{30 (1-2 \nu)} y^{3/2} \ , \\
\delta_{55} &= -\dfrac{575000 - 1675639 \nu}{65625 (1-2 \nu)} y^{3/2} \ , \\
\delta_{53} &= -\dfrac{134136 - 458339 \nu}{25515 (1-2 \nu)} y^{3/2} \ , \\
\delta_{51} &= -\dfrac{184 - 12971 \nu}{105 (1-2 \nu)} y^{3/2} \ .
\end{align}

\section{Residual amplitude and phase corrections for DIN and ILPZ factorizations} 
\label{sec:DIN_ILPZ}
Here we report, for more completeness, the expressions of the residual amplitude corrections $\rho_{\ell m}^{\rm DIN}$ and $\tilde{\rho}_\lm^{\rm ILPZ}$
coming from the DIN or (modified) ILPZ factorizations, keeping into account both the terms $\nu$-dependent and 
the ones that are valid in the test-mass regime. We write also the residual phases.
We only reports the modes up to $\ell = 3$ for simplicity,

\subsection{Residual amplitude corrections}
\begin{widetext}
\begin{align}
\rho_{22}^{\rm DIN}(x)&=1+\left(-\dfrac{43}{42}+\dfrac{55}{84}\nu\right)x+\left(-\dfrac{20555}{10584}-\dfrac{33025}{21168}\nu+\dfrac{19583}{42336}\nu^2\right)x^2\nonumber\\
&+\bigg[\dfrac{1556919113}{122245200}-\frac{428}{105} \, \text{eulerlog}_2(x)
+\left(\frac{41 \pi^2}{192}-\frac{48993925}{9779616}\right)\nu-\frac{6292061}{3259872}\nu^2+\frac{10620745}{39118464}\nu^3\bigg]x^3 \nonumber\\
&+\bigg[-\frac{387216563023}{160190110080}+\frac{9202}{2205}\, \text{eulerlog}_2(x)+\left(-\frac{6718432743163}{145627372800}-\frac{9953 \pi ^2}{21504}+\frac{8819}{441}\, \text{eulerlog}_2(x)\right)\nu\nonumber\\
&+\left(\frac{10815863492353}{640760440320}-\frac{3485 \pi ^2}{5376}\right)\nu^2 -\frac{2088847783}{11650189824}\nu^3+\frac{70134663541}{512608352256} \nu ^4\bigg]x^4 + \left(-\frac{16094530514677}{533967033600} + \notag \right. \\
& \left. + \frac{439877}{55566} \, \text{eulerlog}_2(x) \right) x^5 + \left(\frac{313425353036319023287}{1132319812111488000} - \frac{241777319107}{3208936500} \, \text{eulerlog}_2(x) + \frac{91592}{11025} \, \text{eulerlog}_2(x)^2 + \right. \notag \\
& \left. - \frac{91592 \pi^2}{11025} - \frac{6848 \zeta(3)}{105} \right) x^6 + \left[- \frac{38460677967545998977786359}{411134000579560177920000} + \frac{711515082916633}{21024951948000} \, \text{eulerlog}_2(x) + \right. \notag \\
& \left. - \frac{1969228}{231525} \left( \, \text{eulerlog}_2(x)^2 - \pi^2 \right) + \frac{147232}{2205}\zeta(3)\right] x^7 + \left(- \frac{15305094710902555724554334903}{24377827799070391726080000} + \right. \notag \\
& \left. + \frac{262214117676911}{1557403848000} \, \text{eulerlog}_2(x) - \frac{47066839}{2917215} \, \text{eulerlog}_2(x)^2 + \frac{47066839 \pi^2}{2917215} - \frac{128}{15} \log(2 x) + \frac{3519016}{27783} \zeta(3) \right) x^8 + \notag \\
& + \left[\frac{2029025757272342692216458472843784453}{374311063388772264685516185600000} +\frac{33915179364161}{168469166250} \, \text{eulerlog}_2(x)^2-\frac{39201376}{3472875} \, \text{eulerlog}_2(x)^3 + \right. \notag \\
&-\frac{33915179364161 \pi^2}{168469166250}-\frac{1465472 \pi^4}{165375}-\frac{8128 \log(2 x)}{315}-\frac{1474218276364}{802234125} \zeta(3)+ \notag \\
& +\left. \, \text{eulerlog}_2(x) \left(-\frac{263263957513705951767409}{148616975339632800000}+\frac{39201376 \pi^2}{1157625}+\frac{2930944}{11025} \zeta (3) \right)+\frac{109568}{105} \zeta(5) \right] x^9 + \notag \\
& + \left[-\frac{46049549007414696098131742635031094769709}{116493089147853704215426347282432000000}-\frac{130098015593907827}{1103809977270000}\, \text{eulerlog}_2(x)^2 + \right. \notag \\
& + \frac{842829584}{72930375}\, \text{eulerlog}_2(x)^3+\frac{130098015593907827 \pi^2}{1103809977270000}+\frac{31507648 \pi^4}{3472875}-\frac{474640 \log (2x)}{3969} + \notag \\
& +\, \text{eulerlog}_2(x) \left(\frac{1707101181083798880486383129}{2158453503042690934080000}-\frac{842829584 \pi^2}{24310125}-\frac{63015296}{231525} \zeta(3) \right)+\frac{312387375161861}{262811899350} \zeta(3) + \notag \\
& \left. -\frac{2355712}{2205} \zeta (5) \right] x^{10}\ ,
\end{align}

\begin{align}
\rho_{21}^{\rm DIN}(x)&=1+ \left(-\frac{59}{56}+\frac{23 \nu }{84}\right) x + \left(-\frac{47009}{56448}-\frac{10993 \nu }{14112}+\frac{617 \nu^2}{4704}\right) x^2 + \notag \\
& +x^3 \left[\frac{7613184941}{2607897600}-\frac{107 \, \text{eulerlog}_1(x)}{105}+\left(\frac{1024181}{17385984}-\frac{41 \pi^2}{768}\right) \nu +\frac{622373 \nu^2}{8692992}+\frac{2266171 \nu^3}{39118464}\right] + \nonumber\\
&+\left(-\frac{1168617463883}{911303737344}+\frac{6313 \, \text{eulerlog}_1(x)}{5880}\right) x^4+\left(-\frac{63735873771463}{16569158860800}+\frac{5029963 \, \text{eulerlog}_1(x)}{5927040}\right) x^5 + \notag \\
& + \left(\frac{490932833660765885657}{34355421854878924800}-\frac{1215607983439 \, \text{eulerlog}_1(x)}{273829248000}+\frac{11449 \, \text{eulerlog}_1(x)^2}{22050}-\frac{11449 \pi^2}{22050}-\frac{428 \zeta (3)}{105}\right) x^6 + \notag \\
& + \left(-\frac{3566347735856473936487681203}{249482202425759776112640000}+\frac{6816829896384433 \, \text{eulerlog}_1(x)}{2392172310528000}-\frac{675491 \, \text{eulerlog}_1(x)^2}{1234800} + \right. \notag \\
& \left. +\frac{675491 \pi^2}{1234800}+\frac{6313 \zeta (3)}{1470}\right) x^7 + \left(-\frac{3011026523263793843792391593597}{111768026686740379698462720000} +\frac{8941437217921069 \, \text{eulerlog}_1(x)}{1739761680384000} + \right. \notag \\
& \left. -\frac{538206041 \, \text{eulerlog}_1(x)^2}{1244678400}+\frac{538206041 \pi^2}{1244678400}+\frac{16 \log (2)}{5}+\frac{16 \log (x)}{5}+\frac{5029963 \zeta (3)}{1481760}\right) x^8 + \notag \\
& + \left[\frac{69698391727380944440171591965727282079}{1211399992604769340169382159974400000}+\frac{172976754066437 \, \text{eulerlog}_1(x)^2}{57504142080000}-\frac{1225043 \, \text{eulerlog}_1(x)^3}{6945750} + \right. \notag \\
&-\frac{172976754066437 \pi^2}{57504142080000}-\frac{22898 \pi^4}{165375}+\frac{746 \log(2)}{105}+\frac{746 \log (x)}{105}-\frac{1891753316687 \zeta(3)}{68457312000} + \notag \\
& \left. + \text{eulerlog}_1(x) \left(-\frac{2184537365252992360785787}{90182982369057177600000}+\frac{1225043 \pi^2}{2315250}+\frac{45796 \zeta (3)}{11025}\right)+\frac{1712 \zeta (5)}{105}\right] x^9 + \notag \\
& + \left[-\frac{13722343999227116182528664869983627476666269}{167560846977091695132228940367659008000000}-\frac{1124314064226356987 \, \text{eulerlog}_1(x)^2}{502356185210880000} + \right. \notag \\ 
& +\frac{72277537 \, \text{eulerlog}_1(x)^3}{388962000}+\frac{1124314064226356987 \pi^2}{502356185210880000}+\frac{675491 \pi^4}{4630500}+\frac{298331 \log(2)}{17640}+\frac{298331 \log (x)}{17640} + \notag \\
& +\text{eulerlog}_1(x) \left(\frac{115889995537209277139668976989}{5239126250940955298365440000}-\frac{72277537 \pi^2}{129654000}-\frac{675491 \zeta (3)}{154350}\right)+\frac{521602861743961 \zeta(3)}{23921723105280} + \notag \\
& \left. -\frac{12626 \zeta (5)}{735}\right] x^{10} \ ,
\end{align}

\begin{align}
\rho_{33}^{\rm DIN}(x)&=1+ \left(-\frac{7}{6}+\frac{2 \nu }{3}\right) x + \left(-\frac{6719}{3960}-\frac{1861 \nu }{990}+\frac{149 \nu^2}{330}\right)x^2 + \left[\frac{3203101567}{227026800}-\frac{26 \, \text{eulerlog}_3(x)}{7} + \right. \notag \\
& \left. +\left(-\frac{129509}{25740}+\frac{41 \pi^2}{192}\right) \nu -\frac{274621 \nu^2}{154440}+\frac{12011 \nu^3}{46332}\right] x^3 +\left(-\frac{57566572157}{8562153600}+\frac{13 \, \text{eulerlog}_3(x)}{3}\right) x^4 + \notag \\
 & +\left(-\frac{903823148417327}{30566888352000}+\frac{87347\, \text{eulerlog}_3(x)}{13860}\right) x^5 +\left(\frac{8239014224382011547721}{20928659384400768000}-\frac{4264622767 \, \text{eulerlog}_3(x)}{61122600} + \right. \notag \\
 & \left. +\frac{338 \, \text{eulerlog}_3(x)^2}{49}-\frac{507 \pi^2}{49}-\frac{936 \zeta (3)}{7}\right) x^6 + \left(-\frac{186730759557475960517039}{627859781532023040000}+\frac{104273504957 \, \text{eulerlog}_3(x)}{2305195200} + \right. \notag \\
 & \left. -\frac{169 \, \text{eulerlog}_3(x)^2}{21}+\frac{169 \pi^2}{14}+156 \zeta (3)\right) x^7 + \left(-\frac{54445751267827803613578851861}{69795597941850895994880000} + \right. \notag \\
& \left.+\frac{104211220042957 \, \text{eulerlog}_3(x)}{748140624000}-\frac{1135511 \, \text{eulerlog}_3(x)^2}{97020}+\frac{1135511 \pi^2}{64680}+\frac{87347 \zeta(3)}{385}\right) x^8 + \notag \\
& + \left[\frac{462497649188687912169953452648977167}{84714965366705059820953559040000}+\frac{69239871571 \, \text{eulerlog}_3(x)^2}{427858200}-\frac{8788 \, \text{eulerlog}_3(x)^3}{1029} + \right. \notag \\
& -\frac{69239871571 \pi^2}{285238800}-\frac{6084\pi^4}{245}-\frac{6091647967 \zeta (3)}{1697850}+\text{eulerlog}_3(x) \left(-\frac{5763455760090808884859}{3034036419632064000}+\frac{13182 \pi^2}{343} + \right. \notag \\
& \left. \left.+\frac{24336 \zeta (3)}{49}\right)+\frac{33696 \zeta (5)}{7}\right] x^9 + \left[-\frac{8068812658206523261486971299624727239}{5082897922002303589257213542400000} + \right. \notag \\
& -\frac{1962745690841 \, \text{eulerlog}_3(x)^2}{16136366400}+\frac{4394\, \text{eulerlog}_3(x)^3}{441}+\frac{1962745690841 \pi^2}{10757577600}+\frac{1014 \pi^4}{35}-\frac{64 \log (2)}{35}-\frac{64 \log (x)}{35} + \notag \\
& \left. +\text{eulerlog}_3(x) \left(\frac{244404737248005048311039}{169039171950929280000}-\frac{2197 \pi^2}{49}-\frac{4056 \zeta (3)}{7}\right)+\frac{184662613757 \zeta(3)}{64033200}-5616 \zeta (5)\right] x^{10}\ ,
\end{align}

\begin{align}
\rho_{32}^{\rm DIN}(x)&=1+ \frac{328-1115 \nu + 320 \nu^2}{270 (3 \nu -1)} + \frac{\left(3085640 \nu^4-20338960 \nu^3-4725605 \nu ^2+8050045 \nu -1444528\right) x^2}{1603800 (3 \nu -1)^2} + \notag \\
& +\left(\frac{5849948554}{940355325}-\frac{104 \, \text{eulerlog}_2(x)}{63}\right) x^3+\left(-\frac{10607269449358}{3072140846775}+\frac{17056 \, \text{eulerlog}_2(x)}{8505}\right) x^4 + \notag \\
& +\left(-\frac{1312549797426453052}{176264081083715625}+\frac{18778864 \, \text{eulerlog}_2(x)}{12629925}\right) x^5+ \left(\frac{2121088054187370326502524}{27708335837614916859375} + \right. \notag \\
& \left. -\frac{686761907152 \, \text{eulerlog}_2(x)}{50128172325}+\frac{5408 \, \text{eulerlog}_2(x)^2}{3969}-\frac{2704 \pi^2}{1323}-\frac{1664 \zeta(3)}{63}\right) x^6 + \notag \\
& + \left(-\frac{921336671049791307368884852}{14099280120447898078828125}+\frac{146903645126384 \, \text{eulerlog}_2(x)}{14888067180525}-\frac{886912 \, \text{eulerlog}_2(x)^2}{535815}+\frac{443456 \pi^2}{178605} + \right. \notag \\
& \left. +\frac{272896 \zeta(3)}{8505}\right) x^7 + \left(-\frac{511290545962611000221248395650762}{5959797853272001139128268165625}+\frac{13139758130890608416 \, \text{eulerlog}_2(x)}{854202854482621875} + \right. \notag \\
& \left. -\frac{976500928 \, \text{eulerlog}_2(x)^2}{795685275}+\frac{488250464 \pi^2}{265228425}+\frac{300461824 \zeta (3)}{12629925}\right) x^8 + \notag \\
& + \left[\frac{143808937982050545991412965849898295874336}{312998898582333932825165560622855859375}+\frac{44653873760704 \, \text{eulerlog}_2(x)^2}{3158074856475}-\frac{562432 \, \text{eulerlog}_2(x)^3}{750141} + \right. \notag \\
& -\frac{22326936880352 \pi^2}{1052691618825}-\frac{43264 \pi^4}{19845}-\frac{15723839832832 \zeta(3)}{50128172325}+\text{eulerlog}_2(x) \left(-\frac{583154273043654101594098912}{3542896030266394901859375} + \right. \notag \\ 
& \left. \left. +\frac{281216 \pi^2}{83349}+\frac{173056 \zeta (3)}{3969}\right)+\frac{26624 \zeta(5)}{63}\right] x^9 + \left[-\frac{46804968080981386976552318325118425394324444}{211274256543075404656986753420427705078125} + \right. \notag \\
& -\frac{10865355002211008 \, \text{eulerlog}_2(x)^2}{937948232373075}+\frac{92238848 \, \text{eulerlog}_2(x)^3}{101269035}+\frac{5432677501105504 \pi^2}{312649410791025}+\frac{7095296 \pi^4}{2679075} + \notag \\
& +\frac{1024 \log(2)}{945}+\frac{1024 \log (x)}{945}+\text{eulerlog}_2(x) \left(\frac{871941344832073172195752132256}{6217782533117523052763203125}-\frac{46119424 \pi ^2}{11252115} + \right. \notag \\
& \left. \left. -\frac{28381184 \zeta (3)}{535815}\right)+\frac{4059080596100864 \zeta (3)}{14888067180525}-\frac{4366336 \zeta (5)}{8505}\right] x^{10} \ ,
\end{align}

\begin{align}
\rho_{31}^{\rm DIN}(x)&=1+ \left(-\frac{13}{18}-\frac{2 \nu }{9}\right) x+ \left(\frac{101}{7128}-\frac{1685 \nu }{1782}-\frac{829 \nu^2}{1782}\right) x^2 + \left[\frac{11706720301}{6129723600}-\frac{26 \, \text{eulerlog}_1(x)}{63} + \right. \notag \\
& \left. +\left(-\frac{9688441}{2084940}+\frac{41 \pi^2}{192}\right) \nu +\frac{174535 \nu^2}{75816}-\frac{727247 \nu^3}{1250964}\right]  x^3  + \notag \\
& +\left(\frac{2606097992581}{4854741091200}+\frac{169 \, \text{eulerlog}_1(x)}{567}\right) x^4+\left(\frac{430750057673539}{297110154781440}-\frac{1313 \, \text{eulerlog}_1(x)}{224532}\right) x^5 + \notag \\
&  +  \left(\frac{128318463590300031177841}{15256992691228159872000}-\frac{14891283901 \, \text{eulerlog}_1(x)}{14852791800}+\frac{338 \, \text{eulerlog}_1(x)^2}{3969}-\frac{169 \pi^2}{1323}-\frac{104 \zeta(3)}{63}\right) x^6 + \notag \\
& + \left(\frac{7320626611136687027290387}{1373129342210534388480000}-\frac{784527613381 \, \text{eulerlog}_1(x)}{11763411105600}-\frac{2197 \, \text{eulerlog}_1(x)^2}{35721}+\frac{2197 \pi^2}{23814}+\frac{676 \zeta (3)}{567}\right) x^7 + \notag \\
& + \left(\frac{3450511742152536039994043457511}{238123037410171538883651993600} -\frac{61848173707717 \, \text{eulerlog}_1(x)}{102845822808960}+\frac{17069 \, \text{eulerlog}_1(x)^2}{14145516}-\frac{17069 \pi^2}{9430344} + \right. \notag \\
& \left. -\frac{1313 \zeta (3)}{56133}\right) x^8 + \left[\frac{319497337040080800132992619938404767961}{8337223316564278462279144512921600000}+\frac{234986017513 \, \text{eulerlog}_1(x)^2}{935725883400} + \right. \notag \\
& -\frac{8788 \, \text{eulerlog}_1(x)^3}{750141}-\frac{234986017513 \pi^2}{623817255600}-\frac{676 \pi^4}{19845}+\text{eulerlog}_1(x) \left(-\frac{1072481066956438661850967}{258782068339677634752000}+\frac{4394 \pi^2}{83349} + \right. \notag \\
& \left. \left. +\frac{2704 \zeta (3)}{3969}\right)-\frac{20372359501 \zeta (3)}{3713197950}+\frac{416 \zeta (5)}{63}\right] x^9 +  \left[\frac{1113575895699643740584658755666029532417}{17655296435077295567179364850892800000} + \right. \notag \\
& -\frac{13481555955647 \, \text{eulerlog}_1(x)^2}{741094899652800}+\frac{57122 \, \text{eulerlog}_1(x)^3}{6751269}+\frac{13481555955647 \pi^2}{494063266435200}+\frac{4394 \pi^4}{178605}-\frac{64 \log (2)}{315} + \notag \\
& -\frac{64 \log (x)}{315}+\text{eulerlog}_1(x) \left(-\frac{49486537476392050502065909}{23290386150570987127680000}-\frac{28561 \pi^2}{750141}-\frac{17576 \zeta (3)}{35721}\right) + \notag \\
& \left. +\frac{2350647629819 \zeta (3)}{2940852776400}-\frac{2704 \zeta (5)}{567}\right] x^{10} \ ,
\end{align}

\begin{align}
&\tilde{\rho}_{22}^{\rm ILPZ} =1-\left(\dfrac{43}{42}+\dfrac{55}{84}\nu\right)x+\left(-\dfrac{20555}{10584}-\dfrac{33025}{21168}\nu+\dfrac{19583}{42336}\nu^2\right)x^2 + \nonumber\\
&+ \left[\dfrac{1556919113}{122245200}+\left(\dfrac{41\pi^2}{192}-\dfrac{48993925}{9779616}\right)\nu-\dfrac{6292061}{3259872}\nu^2+\dfrac{10620745}{39118464}\nu^3\right]x^3 + \nonumber\\
&+ \bigg[-\frac{387216563023}{160190110080}+\left(\frac{10815863492353}{640760440320}-\frac{3485 \pi ^2}{5376}\right) \nu ^2-\frac{2088847783 \nu ^3}{11650189824}
+\frac{70134663541 \nu ^4}{512608352256} + \nonumber\\
&+\nu \left(-\frac{6718432743163}{145627372800}-\frac{9953 \pi^2}{21504}+\frac{464}{35}{\rm eulerlog}_2(x)\right)\bigg]x^4
-\dfrac{16094530514677 }{533967033600}x^5 + \nonumber\\
&+\left(\dfrac{230345430821967560887}{1132319812111488000}-\dfrac{91592 \pi^2}{11025}\right)x^6 +\left(-\dfrac{7576963083194058102522359}{411134000579560177920000} + \dfrac{1969228 \pi^2}{231525} \right)x^7 + \nonumber\\
&+\left(-\dfrac{11831416136632492005314654903}{24377827799070391726080000}+\dfrac{47066839\pi^2 }{2917215} -\dfrac{128}{15} \log(2 x) \right)x^8 + \nonumber \\
& + \left(\dfrac{1606998464785272152311147590366114053}{374311063388772264685516185600000}-\dfrac{28672 \,  \text{eulerlog}_2(x)}{1605}-\dfrac{33915179364161 \pi^2}{168469166250}-\dfrac{8128}{315} \log (2 x) + \right. \notag \\
& \left. -\dfrac{313611008 \, \zeta(3)}{496125}\right)x^9 + \left(-\dfrac{2368396146182107522672770370535300593709}{116493089147853704215426347282432000000}+\dfrac{88064 \, \text{eulerlog}_2(x)}{4815} \right. \notag \\
& \left. +\dfrac{130098015593907827 \pi^2}{1103809977270000}-\dfrac{474640 \log (2 x)}{3969}+\dfrac{6742636672 \, \zeta (3)}{10418625}\right) x^{10} \ ,
\end{align}

\begin{align}
&\tilde{\rho}_{21}^{\rm ILPZ} =1+ \left(-\frac{59}{56}+\frac{23 \nu }{84}\right) x+ \left(-\frac{47009}{56448}-\frac{10993 \nu }{14112}+\frac{617 \nu^2}{4704}\right) x^2 + \left[\frac{7613184941}{2607897600}+\left(\frac{1024181}{17385984}-\frac{41 \pi ^2}{768}\right) \nu + \right. \notag \\
& \left. +\frac{622373 \nu ^2}{8692992}+\frac{2266171 \nu^3}{39118464}\right]x^3 -\frac{1168617463883 x^4}{911303737344}-\frac{63735873771463 x^5}{16569158860800}+ \notag\\ 
& + \left(\frac{333388684869106816217}{34355421854878924800}-\frac{11449 \pi^2}{22050}\right) x^6+\left(-\frac{2361005023677528936539393203}{249482202425759776112640000} +\frac{675491 \pi ^2}{1234800}\right) x^7 + \notag \\
& + \left(-\frac{2584194466338595911569649065597}{111768026686740379698462720000}+\frac{538206041 \pi^2}{1244678400}+\frac{16}{5} \log (2 x)\right) x^8 + \notag \\
& + \left(\frac{49827957173414668013436607253148619679}{1211399992604769340169382159974400000}-\frac{448  \, \text{eulerlog}_1(x)}{1605}-\frac{172976754066437 \pi^2}{57504142080000}+\frac{746}{105} \log (2 x) + \right. \notag \\
& \left. -\frac{4900172 \, \zeta(3)}{496125}\right) x^9 + \left(-\frac{12204582364897795934369659654899390447290269}{167560846977091695132228940367659008000000}+\frac{472 \, \text{eulerlog}_1(x)}{1605} + \right. \notag \\
& \left. +\frac{1124314064226356987 \pi^2}{502356185210880000}+\frac{298331 \log (2 x)}{17640}+\frac{72277537 \, \zeta (3)}{6945750}\right) x^{10} \ ,
\end{align}

\begin{align}
&\tilde{\rho}_{33}^{\rm ILPZ} =1+ \left(-\frac{7}{6}+\frac{2 \nu }{3}\right) x+ \left(-\frac{6719}{3960}-\frac{1861 \nu }{990}+\frac{149 \nu^2}{330}\right) x^2 + \left[\frac{3203101567}{227026800}+\left(-\frac{129509}{25740}+\frac{41 \pi ^2}{192}\right) \nu + \right. \notag \\
& \left. -\frac{274621 \nu ^2}{154440}+\frac{12011 \nu^3}{46332}\right]x^3 -\frac{57566572157 x^4}{8562153600}-\frac{903823148417327 x^5}{30566888352000}+ \notag\\ 
& +\left(\frac{4987099197177263643721}{20928659384400768000}-\frac{507 \pi ^2}{49}\right) x^6+\left(-\frac{72913733605309783877039}{627859781532023040000}+\frac{169 \pi^2}{14}\right) x^7 + \notag \\
& + \left(-\frac{36045012217200826661574755861}{69795597941850895994880000} +\frac{1135511 \pi ^2}{64680}\right) x^8 + \notag \\
& + \left(\frac{433945206730860879918922604792644367}{84714965366705059820953559040000}-\frac{69239871571 \pi ^2}{285238800}-\frac{52728 \, \zeta(3)}{49}\right) x^9 + \notag \\
& +  \left(-\frac{13760302405468260039050774267530263239}{5082897922002303589257213542400000}+\frac{1962745690841 \pi^2}{10757577600}-\frac{64}{35} \log (2 x)+\frac{8788 \, \zeta (3)}{7}\right) x^{10} \ ,
\end{align}

\begin{align}
&\tilde{\rho}_{32}^{\rm ILPZ} =1+ \frac{328-1115 \nu + 320 \nu^2}{270 (3 \nu -1)} + \frac{\left(3085640 \nu^4-20338960 \nu^3-4725605 \nu ^2+8050045 \nu -1444528\right) x^2}{1603800 (3 \nu -1)^2} +\frac{5849948554 x^3}{940355325} + \notag \\
& -\frac{10607269449358 x^4}{3072140846775}-\frac{1312549797426453052 x^5}{176264081083715625}+\left(\frac{1270649019848628657252524}{27708335837614916859375}-\frac{2704 \pi ^2}{1323}\right) x^6 + \notag \\
& +\left(-\frac{395634511053935303207884852}{14099280120447898078828125}+\frac{443456 \pi ^2}{178605}\right) x^7+ \notag \\
& + \left(-\frac{346535005732938641595129804518762}{5959797853272001139128268165625}+\frac{488250464 \pi ^2}{265228425}\right) x^8 + \notag \\
& + \left(\frac{135024286976807136011145862629912009874336}{312998898582333932825165560622855859375}-\frac{22326936880352 \pi^2}{1052691618825}-\frac{1124864 \, \zeta (3)}{11907}\right) x^9 + \notag \\
& + \left(-\frac{66218211273663251939996465824084295156824444}{211274256543075404656986753420427705078125}+\frac{5432677501105504 \pi^2}{312649410791025}+\frac{1024}{945} \log (2 x)+\frac{184477696 \, \zeta (3)}{1607445}\right) x^{10} \ ,
\end{align}

\begin{align}
&\tilde{\rho}_{31}^{\rm ILPZ} =1+ \left(-\frac{13}{18} - \frac{2 \nu}{9}\right) x + \left(\frac{101}{7128} - \frac{1685 \nu}{1782} - \frac{829 \nu^2}{1782} \right) x^2 +\left[\frac{11706720301}{6129723600} + \left(-\frac{9688441}{2084940} + \frac{41 \pi^2}{192} \right) \nu + \right. \notag \\
& \left. + \frac{174535 \nu^2}{75816} - \frac{727247 \nu^3}{1250964}\right] x^3 + +\frac{2606097992581 x^4}{4854741091200}+\frac{430750057673539 x^5}{297110154781440} + \notag \\
& + \left(\frac{99051228345457300041841}{15256992691228159872000}-\frac{169 \pi^2}{1323}\right) x^6+\left(\frac{9222996902051464551130387}{1373129342210534388480000}+\frac{2197 \pi^2}{23814}\right) x^7 + \notag \\
& +\left(\frac{3444039304785033167940409511911}{238123037410171538883651993600}-\frac{17069 \pi^2}{9430344}\right) x^8 + \notag \\
& + \left(\frac{310170357782365119359474991341403039961}{8337223316564278462279144512921600000}-\frac{234986017513 \pi^2}{623817255600}-\frac{17576 \zeta(3)}{11907}\right) x^9 + \notag \\
& + \left(\frac{1062945274148287957798165336967073500417}{17655296435077295567179364850892800000}+\frac{13481555955647 \pi^2}{494063266435200}-\frac{64}{315} \log (2 x)+\frac{114244 \zeta (3)}{107163}\right)  x^{10} \ .
\end{align}
\end{widetext}

\subsection{Phases}
Here instead we report the phases of the same residual functions we have just 
considered with the two methods in the comparable-mass case, at the PN order
which is accessible in the literature.
\begin{widetext}
\begin{align}
& \delta_{22}^{\rm DIN} = \frac{7}{3}y^{3/2} - 24 \nu y^{5/2} + \frac{428 \pi}{105} y^3 + \left(\frac{30995 \nu}{1134} + \frac{962 \nu^2}{135}\right)y^{7/2}  - \frac{5536 \pi y^4}{105} \quad ,\\
& \delta_{21}^{\rm DIN} = \frac{2}{3}y^{3/2} - \frac{25}{2} \nu y^{5/2} + \frac{107 \pi}{105} y^3  \label{eq:delta21_DIN} \quad ,\\
& \delta_{33}^{\rm DIN} = \frac{13}{10}y^{3/2} - \frac{80897}{2430} \nu y^{5/2} + \frac{39 \pi}{7} y^3 \quad ,\\
& \delta_{32}^{\rm DIN} = \frac{10+33 \nu}{15(1-3 \nu)} y^{3/2}\quad ,\\
& \delta_{31}^{\rm DIN} = \frac{13}{30}y^{3/2}  - \frac{17}{10} \nu y^{5/2} + \frac{13 \pi}{21} y^3 \quad ,\\
& \notag \\
& \tilde{\delta}_{22}^{\rm ILPZ} = \frac{7}{3} y^{3/2}-24 \nu y^{5/2} + \left(\frac{30995}{1134} \nu + \frac{962}{135} \nu^2 \right) y^{7/2} - \frac{4976}{105} \pi \nu y^4 \quad , \\
& \tilde{\delta}_{21}^{\rm ILPZ} = \frac{2}{3} y^{3/2} - \frac{25}{2} \nu y^{5/2} \quad , \\
& \tilde{\delta}_{33}^{\rm ILPZ} = \frac{13}{10} y^{3/2} - \frac{80897}{2430} \nu y^{5/2} \quad , \\
& \tilde{\delta}_{32}^{\rm ILPZ} = \frac{10+33 \nu}{15(1-3 \nu)} y^{3/2} \quad , \\
& \tilde{\delta}_{31}^{\rm ILPZ} = \frac{13}{30} y^{3/2} - \frac{17}{10} \nu y^{5/2} \quad . \\
\end{align}
\end{widetext}
We outline that the coefficient of the term $y^{3/2}$ of $\delta_{21}^{\rm DIN}$ in \eqref{eq:delta21_DIN} is
different with respect the one of~\cite{Damour:2008gu} (see Eq. (21) of that reference), but it is the same
of the more recent paper ~\cite{Pompili:2023tna} (see Eq. (B2c) of that reference).

\bibliographystyle{apsrev4-1}
\bibliography{refs20260603.bib, local.bib}

\end{document}